\documentclass[12pt,righttag]{article}
\usepackage{latexsym}
\usepackage{amsfonts}
\usepackage{amsmath}
\usepackage{amssymb}
\usepackage{epsfig}
\usepackage{graphicx}

\newcommand\wt{\widetilde}
\newcommand\wh{\widehat}
\newcommand\dt{{|\det|}}
\newcommand\diy{\displaystyle}
\newcommand\DT{\big|{\rm{Det}}\big|}
\newcommand\TN{{\big|{\rm{Tan}}\big|}}
\newcommand\C{{\mathbb C}}
\newcommand\Hb{{\mathbb H}}
\newcommand\oHb{{\overline\Hb}}

\newcommand\Eb{{\mathbb E}}

\newcommand\R{{\mathbb R}}

\newcommand\Z{{\mathbb Z}}
\newcommand\Lam{{\Lambda}}
\renewcommand\H{{\cal H}}
\newcommand\pSigma{\partial\Sigma}

\newcommand\oLoop{\overline{Loop}}
\newcommand\bA{{\mathbf A}}
\newcommand\cA{{\mathcal A}}
\newcommand\bB{{\mathbf B}}

\newcommand\cD{{\mathcal D}}
\newcommand\cI{{\mathcal I}}
\newcommand\cJ{{\mathcal J}}
\newcommand\cK{{\mathcal K}}

\newcommand\cL{{\mathcal L}}
\newcommand\ocL{\overline{\mathcal L}}
\newcommand\cM{{\mathcal M}}
\newcommand\cS{{\mathcal S}}

\newcommand\cU{{\mathcal U}}

\newcommand\bW{{\mathbf W}}

\newcommand\rLL{{\rm L}_{\rm{Liouv}}}
\newcommand\rSL{{\rm S}_{\rm{Liouv}}}
\newcommand\ma{{\mathfrak a}}
\newcommand\mC{{\mathfrak C}}

\newcommand\mE{{\mathfrak E}}
\newcommand\mF{{\mathfrak F}}
\newcommand\mf{{\mathfrak f}}

\newcommand\mS{{\mathfrak S}}
\newcommand\ms{{\mathfrak s}}
\newcommand\mU{{\mathfrak U}}
\newcommand\mv{{\mathfrak v}}
\newcommand\mX{{\mathfrak X}}

\newcommand\obU{\overline{\mathbf U}}

\newcommand\obV{\overline{\mathbf V}}

\newcommand\oa{{\overline a}}
\newcommand\ob{{\overline b}}

\newcommand\ux{{\underline x}}
\newcommand\uy{{\underline y}}
\newcommand\uz{{\underline 0}}
\newcommand\urho{{\underline\rho}}

\begin{document}

\centerline{\Large{\bf{On Malliavin measures, SLE and CFT}}}
\bigskip

\centerline{\Large{{\bf Maxim Kontsevich}}}

\centerline{{\sl IHES, 35 Route de Chartres,}}

\centerline{{\sl 91440 Bures-sur-Yvette, France}}
\bigskip

\centerline{\Large{{\bf Yuri Suhov}}}

\centerline{{\sl Statistical Laboratory, DPMMS/CMS,}}

\centerline{{\sl University of Cambridge, Wilberforce Road}}

\centerline{{\sl Cambridge, CB3 0WB, UK}}
\bigskip

{\bf Abstract.} This paper is motivated by emerging connections 
between the conformal field theory (CFT) on the one hand 
and stochastic L\"owner 
evolution (SLE) processes and measures that play the r\^ole of the
Haar measures for the diffeomorphism
group of a circle, on the other hand. We attempt to build a framework
for widely spread beliefs that SLE-processes would provide
a picture of phase separation in a small massive perturbation 
of the CFT.  

\bigskip

{\bf Table of Contents}

{\bf 1. Introduction}

{\bf 2. Malliavin measures}

\quad {\bf 2.1. The space of simple loops}

\quad {\bf 2.2. Determinant lines}

\quad {\bf 2.3. Determinant bundles on loops}

\quad {\bf 2.4. The covariance property and the main conjecture}

\quad {\bf 2.5. Reduction to $\C^*$}

{\bf 3. Properties of determinant lines}

\quad {\bf 3.1. A preliminary: metrics with pole singularities}

\quad {\bf 3.2. The canonical vector for the special four-sphere
neutral collection}

\quad {\bf 3.3. A variation formula for the special neutral 
collection}

\quad {\bf 3.4. The limit formula for degenerating neutral 
collections}

{\bf 4. The SLE measures, I}

\quad {\bf 4.1. Spaces of intervals and associated line
bundles}

\quad {\bf 4.2. Reduction to $\oHb$}

\quad {\bf 4.3. A reminder on SLE processes}

{\bf 5. The SLE measures, II}

\quad {\bf 5.1. The restriction matringale}
 
\quad {\bf 5.2. End of proof of Theorem 1}

\quad {\bf 5.3. Concluding remarks}

{\bf 6. Applications to statistical physics}

\quad {\bf 6.1. Phase boundaries} 

\quad {\bf 6.2. The Malliavin measures and the CFT}
 
\quad {\bf 6.3. A proposal by Friedrich and the SLE measures}

\quad {\bf 6.4. Operadic structure and quadratic identities
for partition functions}

\quad {\bf References}
\bigskip 

\section{Introduction}

The main object of study in this paper are certain natural
determinant bundles, on spaces of simple Jordan curves in surfaces. 
We consider two classes of such curves: (i) loops
in an open surface $\Sigma$ and (ii) intervals in a
surface $\Sigma$ with a non-empty boundary $\pSigma$,
joining two distinct points $x,y\in\pSigma$. In case (i), 
we put forward, in chapter 2 of the paper, a conjecture of existence and uniqueness 
(up to a positive scalar factor) 
of a one-parameter family of (locally) conformally covariant assignments 
$$\Sigma\mapsto{\mbox{\boldmath${\lambda}$}}_{\Sigma}.\eqno (1.1)$$ 
Here ${\mbox{\boldmath${\lambda}$}}_{\Sigma}$
is a measure
on the space of loops in $\Sigma$ (called a Malliavin measure), 
with values in a given power of a determinant bundle.
In case (ii), our aim is to prove a theorem of
existence of a one-parameter family of (locally) 
conformally covariant assignments 
$$(\Sigma,x,y)\mapsto{\mbox{\boldmath${\lambda}$}}_{\Sigma,x,y}.\eqno (1.2)$$ 
Here ${\mbox{\boldmath${\lambda}$}}_{\Sigma,x,y}$
is a measure
on the space of intervals in $\Sigma$ joining distinct points $x\in\pSigma$ and
$y\in\pSigma$ (called an SLE-measure), 
with values in a given product of determinant bundles. 
To this end, in chapter 3 of the paper we develop 
some useful geometric techniques. Next, in section 4 we 
state and the aforementioned existence theorem.
The proof is carried in sections 4 and 5
and is based on the so-called restriction 
covariance property introduced and verified for (scalar)
probability measures on intervals, in a special situation where
the surface $\Sigma$ (with a boundary) is a closed disk,
$x,y\in\pSigma$ are the endpoints of a diameter, and the
measure is generated by the (chordal) SLE$_\kappa$
process, with 
$$0<\kappa\leq 4,\eqno (1.3)$$  
Condition (1.4) is necessary and sufficient for 
an SLE$_\kappa$-process to generate simple Jordan
curves; a similar condition is introduced in
the conjecture in case (i).

In the concluding chapter 6 we discuss possible 
applications of Malliavin and SLE-measures to
the problem of describing probability distributions 
on phase-separating curves (domain walls) in two-dimensional
Gibbs random fields just below the critical temperatures. 

The paper contains 6 chapters numbered from 1 to 6. Chapters 2--6 are divided
into sections numbered by 2.1, 2.2, and so on. Most of the sections 
contain subsections labeled by triple numbers: 2.2.1, 2.2.2, and so on.
Throughout the paper, symbol $\Box$
marks the end of a proof. Symbol $\blacksquare$ 
is used to mark the end of a definition or a remark. 

The results of this paper have been announced in [K2]. 

\section{Malliavin measures}

\subsection{The space of simple loops}

Throughout this paper all surfaces are supposed to be open, paracompact 
and endowed with a conformal structure. Correspondingly, an embedding of 
a surface into another surface will mean a conformal embedding. In 
chapters 2 and 3 the surfaces are assumed open, whereas in 
chapters 4 and 5 they should have a non-empty boundary. 
A basic examples of 
an (oriented) compact surface repeatedly mentioned below  is 
a Riemann sphere; other examples are a torus (also oriented)
or a Klein bottle (non-oriented). Basic examples of non-compact
surfaces repeatedly mentioned throughout the paper are an open disk, an open 
annulus and a punctured plane (oriented), or an open 
M\"obius strip (non-oriented). 

Speaking about a metric on a surface $\Sigma$, we always 
have in mind a Riemannian metric compatible with the conformal structure.

A standard way of producing a compact surface from a non-compact
one is to pass to a Schottki double (or briefly a double).
Suppose that $\Sigma$ is a non-compact surface of finite 
topological type (i.e. with finite Betti numbers).
Then there exists a unique oriented compact surface 
$\Sigma_{\rm{double}}$, 
with an orientation-reversing conformal involution $\sigma$ and an
embedding $\eta:\;\Sigma\hookrightarrow\Sigma_{\rm{double}}$,
such that 

\begin{enumerate}
\item
$\sigma [\eta (\Sigma )]\cap \eta (\Sigma )=\emptyset\,$,  

\item  the complement
$\Sigma_{\rm{double}}\setminus
\left(\sigma [\eta (\Sigma )]\cup \eta (\Sigma )\right)$ 
is a disjoint union of finitely many isolated points 
and closed loops.
\end{enumerate}

Given a surface $\Sigma$, we  
denote by $Comp\,(\Sigma)$ the space of compact subsets of $\Sigma$ 
equipped with the standard topology. $Comp\,(\Sigma)$ 
is a locally compact Hausdorff space. 

By definition, a simple closed Jordan loop on $\Sigma$
(or, shortly, a loop)
is a compact subset of $\Sigma$ homeomorphic to $S^1$.
The space of loops on $\Sigma$ is denoted
by $Loop\,(\Sigma)$; it 
is a Borel subset of $Comp\,(\Sigma)$, but not closed and not locally 
compact.\footnote{In [K2] it was wrongly stated that $Loop\,(\Sigma)$
is locally compact.}  An embedding 
of surfaces $\beta:\Sigma_1\hookrightarrow \Sigma_2$ gives rise 
to an open embedding of corresponding spaces of loops
$\beta_*:Loop\,(\Sigma_1)\hookrightarrow Loop\,(\Sigma_2)$.

An important special case is where surface $\Sigma$ is an 
annulus $A$. In this case we denote 
by $Loop^1(A)$ the component of 
$Loop\,(A)$ consisting of single-winding loops ${\cL}\subset A$.
For a general oriented surface $\Sigma$ and a loop
${\cL}\in Loop\,(\Sigma)$ there exists a 
fundamental system of neighborhoods of 
${\cL}$ in\\ $Loop\,(\Sigma)$ formed by the images of 
$Loop^1(A)$ under embeddings
$A\hookrightarrow \Sigma$ of $A$ in $\Sigma$.
In the case of a non-oriented surface $\Sigma$, a similar role 
is played by a M\"obius strip $M$. 

As was said earlier, our goal in this paper is to study
measures on spaces of loops (and intervals) which are not locally compact.
In this situation, a natural analog of a sigma-finite 
measure on a non locally compact space ${\mathfrak X}$ is a 
{\it locally finite measure} ${\mbox{\boldmath${\mu}$}}$, with
the property that  
every point $x\in {\mX}$ has a neighborhood ${\mU}$ of finite measure:
${\mbox{\boldmath${\mu}$}}(\mU)<\infty$. More generally, if ${\Lam}$  
is a continuous oriented real line bundle on ${\mX}$, then one can 
speak of locally finite measures with values in ${\Lam}$. For any 
local trivialisation $s$ of ${\Lam}$ around point $x\in {\mX}$ 
(i.e., positive section of the dual bundle ${\Lam}^*$), every such 
${\Lam}$-valued measure ${\mbox{\boldmath${\mu}$}}$ gives an ordinary
locally finite measure ${\mbox{\boldmath${\mu}$}}_s$ on a neighborhood 
of $x$. Further, for every two local trivialisations ${\ms}$ and 
${\ms}'$, the Radon--Nikodym derivative 
$\diy{\frac{{\rm d}{\mbox{\boldmath${\mu}$}}_{\ms}
}{{\rm d}{\mbox{\boldmath${\mu}$}}_{{\ms}'}}}$ is
a continuous strictly positive function equal to ${\ms}/{\ms}'$ 
in a neighborhood of $x$. 

In what follows, speaking of a measure with values in a continuous 
oriented real line bundle, we always mean a locally finite measure.
The same agreement is applied to scalar measures. 

Note that space $Loop\,(\Sigma)$ depends only on the 
{\it topological} structure on surface $\Sigma$. However,
the continuous oriented real line bundle $\DT_{\Sigma}$ 
on $Loop\,(\Sigma)$ which is introduced below depends 
non-trivially on the choice of the {\it conformal} structure.  

\subsection{Determinant lines}

{\bf 2.2.1. Liouville action.}
Let $\Sigma$ be a compact surface. Given a pair of metrics, $g_1$
and $g_2$, on $\Sigma$ (compatible with the conformal structure), we 
define the Liouville action $\rSL  (g_1,g_2)\in\R$ by 
$$\begin{array}{cl}
\rSL (g_1,g_2)&=\diy{\frac{1}{48 \pi{\rm i}}}
\diy{\int\limits_\Sigma}{\rLL}(g_1,g_2)\\
\;&:= \diy{\frac{1}{48 \pi{\rm i}}}
\diy{\int\limits_\Sigma}
(\varphi_1-\varphi_2)\partial\overline{\partial}(\varphi_1
+\varphi_2).\end{array}\eqno (2.1)$$
Here we use the representation $g_i=\exp\;(\varphi_i) |{\rm{d}}z|^2$ where 
$z$ is an arbitrary local complex coordinate on 
$\Sigma$. Equation (2.1) gives a natural definition, 
as the density ${\rLL}(g_1,g_2)$ 
does not depend on the choice of coordinate $z$; it also shows 
a `local character' of the Liouville action, where the integrand
depends on the values of functions
$\varphi_1$, $\varphi_2$ and their derivatives at a single point.

The main property of Liouville action 
is the following well-known cocycle identity:

{\bf Lemma 2.1.} {\sl
$$\rSL (g_1,g_3)=\rSL (g_1,g_2)
+\rSL (g_2,g_3)\,.\eqno (2.2)$$}
\medskip

{\it Proof} : Obviously, 
$\rSL (g_1,g_2)$ is antisymmetric in $g_1,g_2$.
A straightforward calculation shows that
$${\rLL}(g_1,g_2)+{\rLL}(g_2,g_3)+{\rLL}(g_3,g_1)
={\rm d}\alpha(g_1,g_2,g_3).\eqno (2.3)$$
Here
$$\alpha(g_1,g_2,g_3)=\frac{-1}{6}
\sum_{1\le i,j,k\le 3}\epsilon^{ijk} \log \frac {g_i}{g_j}
\left(\partial-\overline{\partial}\right)\log\frac{g_j}{g_k}\,,
\eqno (2.4)$$
where $\epsilon^{ijk}$ is the standard fully antisymmetric tensor.
$\Box$
\medskip

Form $\alpha(g_1,g_2,g_3)$ introduced in (2.4) 
also satisfies a useful cocycle identity:
\medskip

{\bf Lemma 2.2.} {\sl For any collection of four 
metrics $(g_i)_{1\le i\le 4}$ on $\Sigma$, the following equation 
holds:
$$\sum_{i=1}^4 (-1)^i \alpha(g_1,\dots,\hat{g_i},\dots, g_4)=0.
\eqno (2.5)$$}
\medskip

{\it Proof} : It is easy to see that if we write 
$g_i=\exp\;(\phi_i)|{\rm{d}}z|^2$ in local coordinate $z$, then
$$\alpha(g_1,g_2,g_3)=\frac{-1}{2}
\sum_{1\le i,j,k\le 3}\epsilon^{ijk}\phi_i
\left(\partial-\overline{\partial}\right)\phi_j\,\,.$$
Each term in this formula depends only on {\it two} functions $\phi_i$.
It is easy to see that it leads to the assertion of Lemma 2.2.
$\Box$
\medskip

In what follows we repeatedly use the following straightforward
assertion
\medskip

{\bf Lemma 2.3.} {\sl The Liouville density 
${\rLL}(g_1,g_2)$ 
vanishes at points where both metrics $g_1$, $g_2$ are flat.}
\medskip

{\it Proof} : The curvature of metric $\exp\;(\phi )|{\rm d}z|^2$
equals $(-2)\exp\;(-\phi )\diy{\frac{\partial^2\phi}{\partial z 
{\partial}\overline{z}}}$. This implies the statement of Lemma 
2.3. $\Box$ 
\medskip

{\bf Remark 2.1.} In the physical literature (see, e.g.,
the contributions by K. Gawedzki and E. D'Hoker in [20]), the Liouville
action (with the cosmological constant zero)
is written as a functional of $\sigma\in{\rm C}^\infty (\Sigma )$,
depending on a background metric $g$:
$${\rm S}_{{\rm{Liouv}},g}(\sigma )=
\diy{\frac{1}{12\pi}}\diy{
\int\limits_\Sigma}\left(\frac{1}{2}\big|{\rm{grad}}_g\sigma\big|^2+
R_g\cdot\sigma\right){\rm{area}}_g.\eqno (2.6)$$
However, one can check that the following identity holds:
$${\rm S}_{{\rm{Liouv}},g}(\sigma )=-\rSL(g,e^{2\sigma}g),
\eqno (2.7)$$
establishing the connection between the two forms of 
the action. Thus our choice of the local density in Eqn 
(2.1) differs from that in (2.6) by a total derivative; 
an advantage being the property stated in Lemma 2.3. $\blacksquare$ 
\medskip

{\bf 2.2.2. Determinant lines for compact surfaces.}
For a compact surface $\Sigma$, 
we define the determinant line 
$\dt_{\Sigma}$, an oriented one-dimensional
vector space over $\R$, as follows.
Any smooth metric $g$ on $\Sigma$ 
compatible with conformal structure gives a positive
point (a base vector) in 
$\dt_{\Sigma}$, denoted by $[g]$.
For two such metrics, $g_1$, $g_2$, the ratio 
of corresponding vectors is given by            
$$[g_2]/[g_1]:=\exp\;\big[\rSL (g_1,g_2)\big].\eqno (2.8)$$
Cocycle identity (2.2) ensures that $\dt_{\Sigma}$ is correctly 
defined. Obviously, for any finite collection $(\Sigma_i)_{i=1,n}$ 
of compact surfaces we have a canonical isomorphism
$$\dt_{\sqcup_{i=1}^n \Sigma_i}\simeq 
\operatornamewithlimits{\otimes}_{i=1}^n\dt_{\Sigma_i}\,.\eqno (2.9)$$

For any real ${\rm c}$ we define the ${\rm c}$-th
tensor power $\left(\dt_{\Sigma}\right)^{\otimes c}$ 
using the homomorphism
$$\lambda\mapsto\lambda^c,\,\,\,\lambda\in\R_{>0}^\times\,.
\eqno (2.10)$$

If $\Sigma$ is a connected orientable compact
surface of genus zero, then there is a canonical
vector $v_\Sigma \in \dt_\Sigma$. 
Namely, let us choose a conformal isomorphism between $\Sigma$ and
$\C P^1$. Then the round 
metric on $\C P^1$ (the standard metric
on the unit sphere $S^2\subset\R^3$) 
gives rise to an element $v_\Sigma$ of $\dt_{\Sigma}$.
Vector $v_\Sigma$ does not depend on the choice
of the aforementioned conformal isomorphism because there is no 
non-trivial homomorphism from the group $\Z_2\ltimes PSL(2,\C)$ 
of conformal automorphisms of $\C P^1$ to $\R_{>0}^\times$.  
\medskip

{\bf 2.2.3. Determinant lines for non-compact surfaces.}
For a non-compact surface $\Sigma$ of finite 
type, we define the oriented line $\dt_{\Sigma}$ as 
$$\left(\dt_{\Sigma_{\rm{double}}}\right)^{\otimes (1/2)}\,.
\eqno (2.11)$$ 

We say that $\Sigma$ is puncture-free if the complement
$\Sigma_{\rm{double}}\setminus
\left(\sigma [i(\Sigma )]\cup i(\Sigma )\right)$
does not contain isolated points. For a puncture-free surface
$\Sigma$, we can define line $\dt_{\Sigma}$
in an alternative way. Namely, we say metric $g$ is well-behaving 
at infinity if there exists a relatively compact open
subset $U\subset\Sigma$ such that the set $\Sigma\setminus U$ with
metric $g$ is isometric to a finite disjoint union 
$\sqcup_i(0,\epsilon_i]\times S^1$ of
semi-open flat cylinders $(0,\epsilon_i]\times S^1$, 
where $\epsilon_i$ is a positive number, and $S^1$ is the standard circle
of length $2\pi$. Such a metric compatible with the conformal
structure exists iff $\Sigma$ is puncture-free.

The Liouville action $\rSL (g_1,g_2)$ is
then alternatively defined as in Eqn (2.1), for any 
two metrics $g_1,g_2$ 
well-behaving at infinity. Moreover, metrics $g_1$ and $g_2$ 
well-behaving at infinity can be extended to $\sigma$-invariant 
metrics ${\wt g}_1$ and ${\wt g}_2$ on $\Sigma_{\rm{double}}$, with
the property that $\rSL (g_1,g_2)=\diy{\frac{1}{2}}
\rSL \big({\wt g}_1,{\wt g}_2\big)$. Therefore, the 
alternative definition of $\dt_{\Sigma}$
coincides with the original one.
\medskip
 
{\bf 2.2.4. Canonical vectors.} Recall, for a compact orientable   
surface $\Sigma$ of genus $0$, we defined
a canonical vector $v_\Sigma$ in $\dt_{\Sigma}$ by using a 
conformal isomorphism of $\sigma$ and $S^2$. Now, 
if $\Sigma$ is a non-compact puncture-free conformal surface
homeomorphic to an open disk, we define a canonical element
$v_\Sigma$ in $\dt_{\Sigma}$ by using the fact that 
$\Sigma_{\rm{double}}$ is homeomorphic to a sphere.

Next, if $\Sigma$ is a puncture-free conformal surface
homeomorphic to an annulus or to a M\"obius strip, then
we define a canonical element $v_\Sigma$ in $\dt_{\Sigma}$ 
by using the unique flat metric on $\Sigma$ well-behaving at 
infinity. Moreover, any multiple of this metric gives the same 
vector $v_\Sigma$, as can be seen from Eqn (2.7).
Later on (see Lemma 2.4 below), it will be convenient to use 
a special {\it normalised} metric $g_\Sigma^{\rm{norm}}$,
in the case where $\Sigma$ is an annulus.  
Metric $g_\Sigma^{\rm{norm}}$ is simply a mutliple of the 
flat metric, 
specified by the condition that the height of $\Sigma$ is equal to 1.
This metric can also be specified as follows.
Consider a harmonic function $h$ on $\Sigma$ which tends to 0 
at one component of the boundary
of $\Sigma$ and to 1 at the other component (such $h$ is defined 
uniquely up to the involution
$h\mapsto 1-h$). The normalised metric is given by 
$$g_\Sigma^{\rm can}=\left|2\frac{\partial h}{\partial z}\right|^2  
|{\rm{d}}z|^2\eqno (2.12)$$
in any local coordinate $z$.
\medskip

{\bf 2.2.5. Neutral collections.}
In this subsection, we describe a construction of special vectors in
tensor products of determinant lines for several compact
surfaces with conformal structures. Informally, we will deal with 
two-dimensional non-Hausdorff ``manifolds'' where some points are 
non-separable. A particular example is where such a ``manifold'' 
is the Cartesian product $N=S^1\times T^1$ of a circle $S^1$ and 
a ``train track'' $T^1$, the quotient of $\R^1\times \{1,2\}$
 by the equivalence relation
 $$(x,1)\sim (x,2),\,\,\,x<0$$

%\vskip 5 truemm
%\begin{center}
%\includegraphics[width=.3\textwidth]{Fig1.eps}
%\end{center}
\vskip 5 truemm

\begin{figure}[ht]
\centering
\includegraphics[height=60mm]{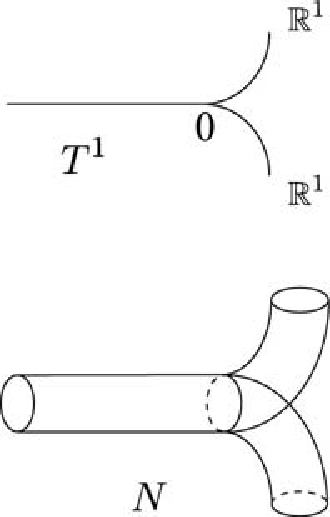}
\caption{Train track $T^1$ and its product with $S^1$}
\end{figure}
\vskip 5 truemm

In general, we define
an admissible (non-Hausdorff) surface $\Sigma^{\rm{nH}}$
as a topological space with countable base,
 such that any point $x\in\Sigma^{\rm{nH}}$
has a neighbourhood homeomorphic to an open disk and
such that there exists a neighbourhood of the set of 
non-separable points homeomorphic to a disjoint union
of finitely many copies of space $M$ defined above. 
Clearly, such surfaces can be 
endowed with smooth or even conformal structures. We will
assume that inseparable points form smooth curves on $\Sigma^{\rm{nH}}$.

Further, consider a finite collection $\Sigma_1,\dots,\Sigma_n$ 
of compact surfaces.
Assume also that for all  
$i=1,\ldots ,n$ a  `weight' $m_i\in \Z$ is given. Next, let 
us fix  an admissible surface $\Sigma^{\rm{nH}}$ with a conformal
structure, and an $n$-tuple of embeddings 
$\phi_i:\;\Sigma_i\to\Sigma^{\rm{nH}}$.

We call the collection 
$${\mC}=\left(\{(\Sigma_i,\phi_i,m_i)\};
\Sigma^{\rm{nH}}\right)$$ 
{\it neutral} if for any separable point $x\in\Sigma^{\rm{nH}}$, 
the sum of weights of surfaces $\Sigma_i$ whose images $\phi_i(\Sigma_i)$ 
contain $x$ equals $0$. Given a neutral collection ${\mC}$, we define 
a vector $v_{\mC}\in
\operatornamewithlimits{\otimes}_{i}\left(\dt_{\Sigma_i}
\right)^{\otimes m_i}$ by
$$v_{\mC}=\operatornamewithlimits{\otimes}_{i}
\big([\phi^*_i g]\big)^{\otimes m_i},\eqno (2.13)$$
where $\phi^*_i g$ is the pullback image of a metric $g$ on
$\Sigma^{\rm{nH}}$ compatible with the given conformal structure. 
It follows from the locality of Liouville action that this definition does 
not depend on the choice of $g$.\footnote{More generally,
one can allow  maps $\phi_i$ to be only immersions (local 
homeomorphisms). In the definition of 
neutrality one should count each weight $w_i$ with the multiplicity
equal to the number of  points in $\phi_i^{-1}(x)$.}

A working example of a neutral collection is where we take the 
non-Hausdorff surface 
$$\R\times\{1,2\}/\,(x,1)\sim (x,2)\,\mbox{ for }x\in (0,1)\,,$$
mulitply it by the unit circle $S^1$ and `compactify' the product  by 
adding four `caps' (closed disks) to four ends; this gives an 
admissible surface which we will denote by $S^{\rm{nH}}$. There are four 
distinct embeddings of sphere $\C P^1$ in $S^{\rm{nH}}$;
we deem them $\phi_i$, $i=1,2,3,4$, and 
denote by $S_1$, $S_2$, $S_3$ and $S_4\simeq S^2$ the 
images $\phi_1(\C P^1 )$, $\phi_2(\C P^1 )$, $\phi_3(\C P^1 )$,  
and $\phi_4(\C P^1 )$. Then take any pair of embeddings
covering together all four end caps and assign to them  
multiplicities $+1$. The remaining pair of embeddings 
gets multiplicities $-1$. We denote these basic multiplicities
by $\mu_i$, $i=1,2,3,4$. This gives a neutral collection which
we will denote by $\mF$. 

%\vskip 5 truemm
%\begin{center}
%\includegraphics[width=.5\textwidth]{Fig2.eps}
%\end{center}
\vskip 5 truemm

\begin{figure}[ht]
\centering
\includegraphics[height=45mm]{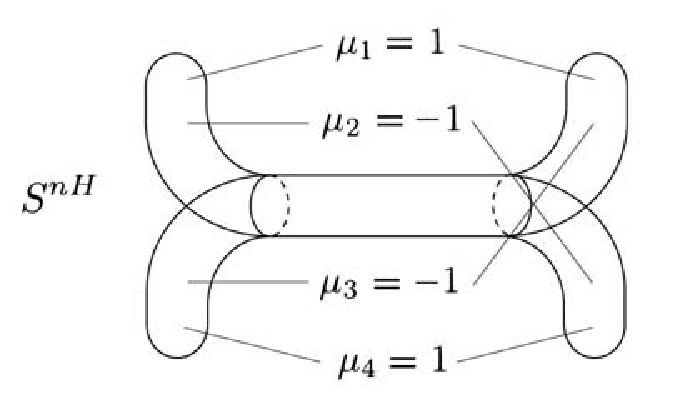}
\caption{Non-Hausdorff surface $S^{\rm{nH}}$}
\end{figure}

Other working examples are a neutral collection 
${\mE}$ of eight spheres in subsection 2.5.1
and a neutral collection ${\mS}$ of six spheres in
subsection 4.2.3.

A useful fact is as follows. Suppose we have a neutral collection 
${\mC}=\left(\{(\Sigma_i,\phi_i,m_i)\};
\Sigma^{\rm{nH}}\right)$, and a continuous map 
$\psi:\;\Sigma^{\rm{nH}}\to {\Sigma'}^{\rm{nH}}$ where
${\Sigma'}^{\rm{nH}}$ is another admissible
(non-Hausdorff) surface with a conformal structure, and $\psi$ is
locally a conformal homeomorphism. 
Then the compositions $\phi'_i=\psi\circ\phi_i$ give a new
neutral collection ${\mC}'=\left(\{(\Sigma_i,\phi'_i,m_i)\};
{\Sigma'}^{\;{\rm{nH}}}\right)$, and vectors $v_{\mC}$
and $v_{\mC'}$ coincide. This follows from the fact that
we can choose a metric on  $\Sigma^{\rm{nH}}$
which is a pullback image of the chosen metric 
on ${\Sigma'}^{\;{\rm{nH}}}$.

Informally, it means that we can ``move'' sets of nonseparable 
points in a zip-like fashion. 

\subsection{Determinant bundles on loops}

{\bf 2.3.1. Determinant line for an individual loop.}
Suppose we are given a surface $\Sigma$ and a loop 
${\cL}\in Loop\,(\Sigma )$. Next, let ${D}\subset\Sigma$ be a 
puncture-free domain that is a surface of a finite topological type containing 
${\cL}$ and contractible to ${\cL}$). We 
define the oriented line $\dt_{{\cL},\Sigma}$ as the quotient:
$$\dt_{{\cL},\Sigma}=\frac{\dt_{{D}}}{\dt_{{D}
\setminus{\cL}}}:\simeq \dt_{{D}}\otimes\left(\dt_{{D}
\setminus{\cL}}\right)^{\otimes (-1)}.\eqno (2.14)$$

%\vskip 5 truemm
%\begin{center}
%\includegraphics[width=.4\textwidth]{Fig3.eps}
%\end{center}
\vskip 2 truemm

\begin{figure}[ht]
\centering
\includegraphics[height=65mm]{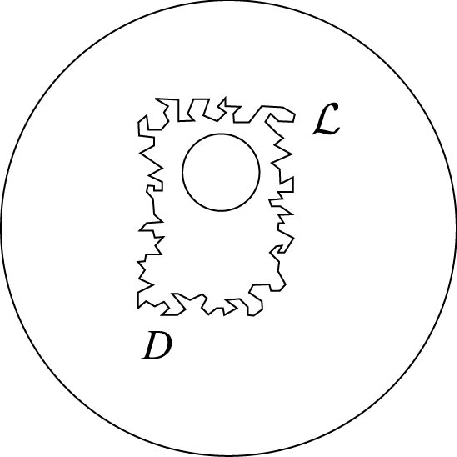}
\caption{Loop $\cL$ in an annulus domain $D$}
\end{figure}
\vskip 3 truemm

To make this definition independent
of ${D}$, we construct for every pair of domains ${D}_1,
{D}_2\supset{\cL}$, of the same kind as above, an isomorphism 
$$i_{{D}_1,{D}_2}:
\dt_{{D}_1}\left/\dt_{{D}_1\setminus{\cL}}
\right.\to
\dt_{{D}_2}\left/\dt_{{D}_2\setminus{\cL}}\right.\eqno (2.15)$$
satisfying the corresponding cocycle identity
$$i_{{D}_2,{D}_3}\circ i_{{D}_1,{D}_2}
=i_{{D}_1,{D}_3}.$$

%\vskip 5 truemm
%\begin{center}
%\includegraphics[width=.4\textwidth]{Fig4.eps}
%\end{center}
\vskip 5 truemm

\begin{figure}[ht]
\centering
\includegraphics[height=65mm]{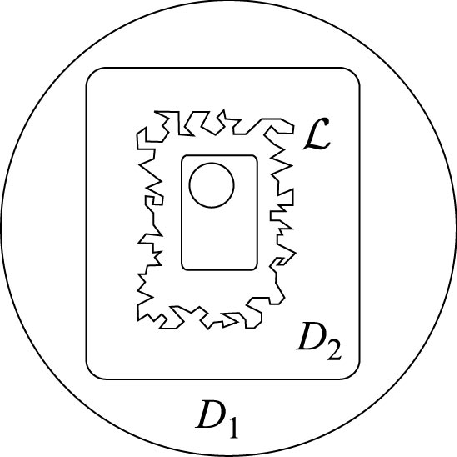}
\caption{Loop in two domains $D_2\subset D_1$}
\end{figure}
\vskip 5 truemm

The construction of isomorphism $i_{{D}_1,{D}_2}$ is as
follows. First, assume that ${D}_2$ is relatively compact  
in ${D}_1$, and the boundary $\partial{D}_2$ consists of two 
real analytic loops. Next, choose a metric $g_{1\setminus 2}$ 
on ${D}_1\setminus{\overline D}_2$ well-behaving at infinity. Then  
$\exists$ metrics $g_1$ on ${D}_1$ and $g_{1,{\cL}}$ on ${D}_1
\setminus{\cL}$ well-behaving at infinity, such that their restrictions on 
${D}_1\setminus{\overline D}_2$ coincide with $g_{1\setminus 2}$. 
Further, define metrics $g_2$ on ${D}_2$ and $g_{2,{\cL}}$
on ${D}_2\setminus{\cL}$ as the restrictions of 
$g_1$ and $g_{1,{\cL}}$, respectively. The isomorphism 
$i_{{D}_1,{D}_2}$ is then determined by
$$i_{{D}_1,{D}_2}\left(\frac{[g_1]}{[g_{1,{\cL}}]}\right)=
\frac{[g_2]}{[g_{2,{\cL}}]}\,.\eqno (2.16)$$

In general, we choose a domain ${D}_3\subset{D}_1\cap{D}_2$
which is relatively compact in ${D}_1\cap{D}_2$ and
with boundary $\partial{D}_3$ consisting of two 
real analytic loops. Then define $i_{{D}_1,{D}_2}$  by 
$$i_{{D}_1,{D}_2}=i_{{D}_2,{D}_3}^{-1}\circ
i_{{D}_1,{D}_3}\,,\eqno (2.17)$$
to guarantee the cocycle identity. The independence 
of the choice
of ${D}_3$ follows from the locality of the Liuoville action. 
\vskip 5 truemm

{\bf Remark 2.2.} It is instructive to give an alternative definition 
of isomorphism $i_{{D}_1,{D}_2}$, by using the construction of 
vector $v_{\mC}$ for a particular neutral collection
$\mC$ described as follows. Consider four compact surfaces
$$\Sigma_1=({D}_1)_{\rm{double}}, \,\,
\Sigma_2=({D}_1\setminus{\cL})_{\rm{double}},\,\,
\Sigma_3=({D}_2)_{\rm{double}},\,\,  
\Sigma_4=({D}_2\setminus{\cL})_{\rm{double}}\,.\eqno (2.18)$$
Assign to them multiplicities $(m_1,m_2,m_3,m_4)=(-1,+1,+1,-1)$. 
Next, fix two relatively compact open neighbourhoods ${\cU}_1$ 
and ${\cU}_2$ of loop ${\cL}$, with smooth boundaries, such that 
${D}_1\cap{D}_2\supset{\bar{\cU}_1}$ and ${\cU}_1\supset{\bar{\cU}_2}$. 

As a non-Hausdorff `manifold' $\Sigma^{\rm{nH}}$, we take
the union $\Sigma_1\sqcup_{\cU}\Sigma_4$ where the 
set ${\cU}$ (along which $\Sigma_1$ and $\Sigma_4$ are identified)
is the formed by the pullback images
of ${\cU}_1\setminus{\bar{\cU}_2}$. It is easy to see
that $\Sigma_2$ and $\Sigma_3$ are naturally embedded
in $\Sigma_1\sqcup_{\cU}\Sigma_4$; this yields the
neutral collection $\mC$ under consideration.   
Then $i_{{D}_1,{D}_2}$ is given by multiplication
by the vector  
$$\big(v_{\mC}\big)^{\otimes 1/2}\in 
\left(\dt_{{D}_1}\right)^{\otimes (-1)}
\otimes\dt_{{D}_1\setminus{\cL}}\otimes
\dt_{{D}_2}\otimes\left(\dt_{{D}_2
\setminus{\cL}}\right)^{\otimes (-1)}\,.\quad \blacksquare \eqno (2.19)$$

{\bf 2.3.2. Topology on the determinant bundle.}
Our goal in this subsection is to define a continuous real line 
bundle $\DT_{\Sigma}$ on the space $Loop\,(\Sigma)$ whose fiber 
at each point ${\cL}\in Loop\,(\Sigma)$ is canonically identified 
with $\dt_{{\cL},\Sigma}$. Here we will assume for simplicity
that $\Sigma$ is orientable  near ${\cL}$;  the non-orientable 
case follows by passing to the double cover.

To start with, assume that surface $\Sigma$ is a 
puncture-free annulus $A$. Recall (see subsection 2.2.4)
that in this case we have the 
canonical vector $v_A \in \dt_A$. Hence we have a canonical
vector $v_{{\cL},A}\in \dt_{{\cL},A}$ for an arbitrary
non-contractible loop ${\cL}\in Loop^1(A)$, namely:
$$v_{{\cL},{A}}=\frac{v_{A}}{v_{{A}_1}\otimes v_{{A}_2}}\,.\eqno (2.20)$$
Here ${A}_1$ and ${A}_2$ are two annuli forming the 
connected components of\\ ${A}\setminus{\cL}$. This yields a trivialisation
of the bundle $\DT_A$ on the space $Loop^1(A)$.
We then define the continuous structure on $\DT_{A}$
by declaring that the map $L\mapsto (v_{{\cL},{A}})$ is continuous.

%\vskip 5 truemm
%\begin{center}
%\includegraphics[width=.4\textwidth]{Fig5.eps}
%\end{center}
%\vskip 5 truemm

\vskip 5 truemm
\begin{figure}[ht]
\centering
\includegraphics[height=65mm]{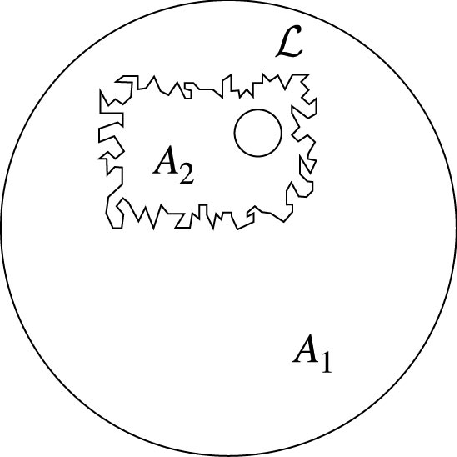}
\caption{Loop $\cL$ and two annuli $A_1,A_2$, connected components of $A\setminus \cL$}
\end{figure}
\vskip 5 truemm

Next, this construction is extended to the case of
a loop ${\cL}$ on a general surface
$\Sigma$ orientable near ${\cL}$. Here, we use 
the fact that there exists 
a fundamental system of neighborhoods
of ${\cL}\in Loop\,(\Sigma)$ consisting of sets 
$i_*[Loop^1(A)]$ where $i:A\hookrightarrow \Sigma$ 
is an embedding of an annulus $A$ in $\Sigma$.
To justify the  correctness of the definition, 
we have to check that for any two annuli, $A\subset\Sigma$ 
and $A'\subset\Sigma$, such that ${\cL}\in Loop^1(A)\cap
Loop^1(A')$, the ratio $v_{{\cL},A}/v_{{\cL},A'}$ is a 
continuous function
in a neighborhood of ${\cL}$. In order to calculate this ratio, 
we have to introduce certain interpolations
between six flat metrics: the normalised metrics on annuli 
$A$ and $A'$, and the normalised metrics on 
the connected components
of $A\setminus{\cL}$ and $A'\setminus{\cL}$. The continuity of 
the ratio follows from Lemma 2.3 and the following 
assertion.
\medskip

{\bf Lemma 2.4.} {\sl For any annulus $A$, the normalised 
metrics (see Eqn {\rm{(2.12)}}) on both connected components
of $A\setminus{\cL}$ depend continuously on ${\cL}\in Loop\,(A)$ 
on compacts in $A\setminus{\cL}$.}
\medskip

{\it Proof} : The harmonic function $h$ used in the definition of 
the normalised metric coincides with the 
probability of hitting a component of the boundary $\partial A$
in the Brownian motion. Hence it
depends continuously on the boundary curve. The expression for the 
normalised metric 
includes the first derivative of $h$ which can be replaced by 
a suitable contour integral because $h$ is harmonic.
$\Box$ 
\vskip 5 truemm

{\bf Remark 2.3.} The concept of line $\dt_{{\cL},\Sigma}$ 
associated with loop ${\cL}\in Loop (\Sigma)$ seems novel and 
is central for this paper. It is easy 
to see that the restriction of $\DT_\Sigma$ to the subspace of 
$Loop (\Sigma )$ formed by sufficiently smooth curves (e.g.,
curves of class C$^2$) is canonically trivialised. In a sense, 
one can interpret a non-smooth loop ${\cL}$ as an infinitesimally 
tiny open subset of $\Sigma$, and $\dt_\Sigma$ can be seen as
an analog of the determinant line for such an `open surface'.
$\blacksquare$ 

\subsection{The covariance property and the main conjecture}

Given an embedding $\xi:\;\Sigma \hookrightarrow
\Sigma'$, we have an associated open embedding
$$\xi_*:\;Loop\,(\Sigma )\hookrightarrow
Loop\,(\Sigma').$$
Further, it generates the canonical isomorphism of line bundles
$$\xi_{\rm{det}}:\;(\xi_*)^*\DT_{\Sigma'}\to \DT_{\Sigma}.$$
(We can use here any annulus ${A}$ containing a given loop 
${\cL}\in Loop (\Sigma )$.
\medskip

{\bf Definition 2.1.} Fix a real number ${\rm c}$ and assume that 
for every surface $\Sigma$  we are given a measure 
${\mbox{\boldmath${\lambda}$}}_{\Sigma}$  
on $Loop\,(\Sigma )$ with values in $\left(\DT_{\Sigma}\right)^{\otimes c}$. 
We say that the assignment 
$\Sigma\mapsto{\mbox{\boldmath${\lambda}$}}_{\Sigma}$ is
{\it locally conformally covariant}, with parameter ${\rm c}$
(briefly: ${\rm c}$-LCC, or, simply, LCC) 
if for any embedding $\xi:\;\Sigma \hookrightarrow
\Sigma'$ we have 
$$\xi^*\big({\mbox{\boldmath${\lambda}$}}_{\Sigma'}\big)
={\mbox{\boldmath${\lambda}$}}_{\Sigma},\eqno (2.21)$$
where we use the obvious identification of the line bundles 
via isomorphism $\xi$. $\blacksquare$ 
\medskip 

{\bf Conjecture 1.} {\sl For any ${\rm c}\in (-\infty, 1]$, 
there exists a unique (up to a positive constant factor) non-zero
${\rm c}$-LCC assignment}   
$\Sigma\mapsto{\mbox{\boldmath${\lambda}$}}_{\Sigma}$. 
\medskip 

The bound ${\rm c}\leq 1$ is motivated by properties of 
the family of random SLE$_\kappa$-processes (see chapter 4).
A well-known fact is that a trajectory of an SLE$_\kappa$-process
remains `simple' (dividing a unit disk or a half-plane
into two domains) for $\kappa\in (0,4]$ which implies 
the above bound on ${\rm c}$. 

We will call measures ${\mbox{\boldmath${\lambda}$}}_{\Sigma}$ 
figuring in Conjecture 1 {\it Malliavin measures}.
The relation between Conjecture 1 and a series
of papers by Malliavin and his followers initiated 
in [M] and [AM] is discussed in subsection 2.5.2. 

Observe that if Conjecture 1 has been verified when $\Sigma$
is an arbitrary annulus $A=A_{r_1,r_2}=\{z\in\C:\;\;r_1<|z|<r_2\}$,
$0<r_1<r_2<+\infty$, and  
${\mbox{\boldmath${\lambda}$}}_{A}$ is a measure on
the space $Loop^1(A)$ of single-winding loops in $A$ and
$\xi$ is an embedding $A_{r_1,r_2}\to A_{r'_1,r'_2}$, then it will be verified
in full generality, for all orientable surfaces $\Sigma$.
\medskip

Similarly, to establish Conjecture 1 for non-orientable surfaces,
it is enough to check the conjecture when $\Sigma$ is an arbitrary M\"obius strip $M$,
${\mbox{\boldmath${\lambda}$}}_{M}$ is a measure on
$Loop^1 (M)$ and $\xi$ is an embedding $M\to M'$.
In our view, the first step in proving Conjecture 1 would be
a construction of such measures ${\mbox{\boldmath${\lambda}$}}_{A}$ 
and ${\mbox{\boldmath${\lambda}$}}_{M}$.

Next, in the orientable case, in section 2.5 we provide
a further reduction, which we believe is equivalent to
the initial problem of constructing an LCC assignment
$\Sigma\mapsto{\mbox{\boldmath${\lambda}$}}_{\Sigma}$.
It will be stated in terms of scalar measures on the space
of single-winding loops on a punctured plane. 

Parameter ${\rm c}$ is interpreted as the {\it central charge}
(in the corresponding conformal field theory; see section 6.2). 

In the case ${\rm c}=0$, Conjecture 1 was recently  
established (in the case of an orientable surface $\Sigma$) 
by Werner [W4]. Unfortunately, the method in [W4] works
(for both existence and uniqueness) specifically for ${\rm c}=0$;
it seems that an extension to other values of ${\rm c}$ requires
new ideas. 

In chapter 4 we will define a space of intervals
$Int_{x,y}(\Sigma)$, an analog of 
space $Loop\,(\Sigma )$ for a surface $\Sigma$ with a 
boundary, and natural 
determinant line bundles on $Int_{x,y}(\Sigma)$. 
The main result of chapter 4 is the verification that
an SLE$_\kappa$ process, with $0<\kappa \leq 4$, gives 
rise to an LCC assignment
$(\Sigma,x,y)\mapsto{\mbox{\boldmath${\lambda}$}}_{\Sigma,x,y}$. 
Here ${\mbox{\boldmath${\lambda}$}}_{\Sigma,x,y}$ is
a measure on $Int_{x,y}(\Sigma)$ with values
in a tensor product of the aforementioned
determinant line bundles. This will extend
the LCC property that was previously
established in [W4] for SLE$_{4/3}$ by direct methods.
\medskip

\subsection{ Reduction to $\C^*$}

{\bf 2.5.1. Restriction covariance property for measures
on $Loop^1(\C^*)$.}
Fix ${\rm c}\in (-\infty, 1]$. Under an additional assumption
of strong local finiteness (see below), we will reduce the problem 
of constructing an LCC assignment 
$\Sigma\mapsto{\mbox{\boldmath${\lambda}$}}_{\Sigma}$ 
to a simpler problem (in the orientable case), of constructing
a scalar measure on the set $Loop^1 (\C^*)$ of single-winding loops
in $\C^*$ satisfying a restriction covariance
property. Here, and below,
$$\C^*=\C\setminus\{0\}$$ 
is a punctured plane.
First, if 
$\Sigma$ is a sphere $\C P^1$, we have a scalar-valued measure
$${\mbox{\boldmath${\nu}$}}_{\C P^1}^{\rm c}=
{\mbox{\boldmath${\lambda}$}}_{\C P^1}\otimes
\left(\diy{\frac{v_{\C P^1}}{v_{D_L}\otimes v_{D_R}}}\right)^{\otimes(-{\rm c})}
\eqno (2.22)$$
on $Loop (\C P^1)$, where $D_{\rm L}\big(=D_{{\rm L},{\cL}}\big)\subset\C P^1$ 
and $D_{\rm R}\big(=D_{{\rm R},{\cL}}\big)\subset\C P^1$ are two open
disks, to the left and to the right of ${\cL}\in Loop (\C P^1)$, 
respectively. Measure ${\mbox{\boldmath${\nu}$}}_{\C P^1}^{\rm c}$ is 
invariant under the action of $PSL (2,\C )$ on $Loop (\C P^1)$.  

%\vskip 5 truemm
%\begin{center}
%\includegraphics[width=.6\textwidth]{Fig6.eps}
%\end{center}
%\vskip 5 truemm

\vskip 5 truemm
\begin{figure}[ht]
\centering
\includegraphics[height=2cm]{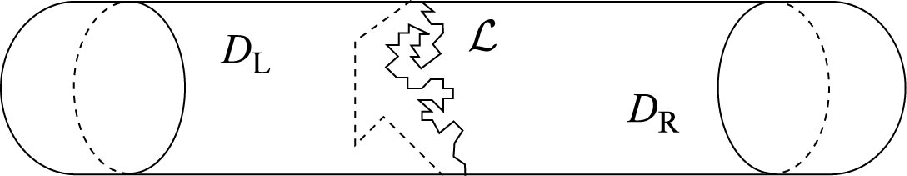}
\caption{Loop $\cL$ on the complement to two caps $D_L,D_R$}
\end{figure}
\vskip 5 truemm

Next, denote by ${\mbox{\boldmath${\nu}$}}^{\rm c}$ the restriction of
measure ${\mbox{\boldmath${\nu}$}}_{\C P^1}^{\rm c}$ to $Loop^1$ 
where 
$$Loop^1:=Loop^1 (\C^* )\eqno (2.23)$$
is an open subset in $Loop (\C P^1)$. 
Measure ${\mbox{\boldmath${\nu}$}}^{\rm c}$ is invariant under dilations 
$z\mapsto t z$, $z\in \C^*$, for any fixed $t\in\C^*$.

%\vskip 5 truemm
%\begin{center}
%\includegraphics[width=.6\textwidth]{Fig7.eps}
%\end{center}
%%\vskip 5 truemm

\vskip 5 truemm
\begin{figure}[ht]
\centering
\includegraphics[height=2cm]{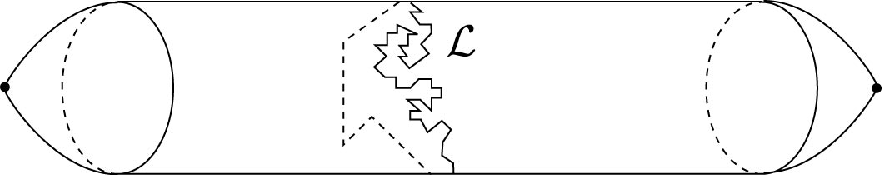}
\caption{A single-winding loop on $\C^*$}
\end{figure}
\vskip 5 truemm

In what follows, we assume that measure ${\mbox{\boldmath${\nu}$}}^{\rm c}$ 
satisfies the following
\medskip

{\bf Strong local finiteness condition.}  For any annulus 
$$A_{r_1,r_2}=\{z\in\C^*:\;\;r_1<|z|<r_2\},$$ 
the set $Loop^1 (A_{r_1,r_2})\subset Loop^1$ of simple loops in 
$A_{r_1,r_2}$ has a finite ${\mbox{\boldmath${\nu}$}}^{\rm c}$-measure:
$${\mbox{\boldmath${\nu}$}}^{\rm c}\left( Loop^1(A_{r_1,r_2})\right)
<\infty,\;\;0<r_1<r_2<\infty .\eqno (2.24)$$
\medskip

%\vskip 5 truemm
%\begin{center}
%\includegraphics[width=.4\textwidth]{Fig8.eps}
%\end{center}
%\vskip 5 truemm

\vskip 5 truemm
\begin{figure}[ht]
\centering
\includegraphics[height=65mm]{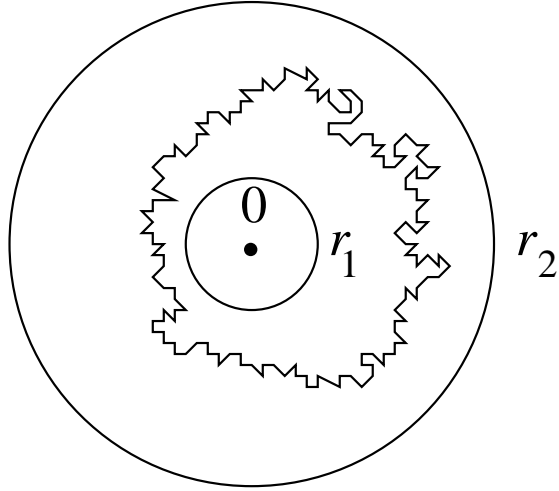}
\caption{Loop in the annulus $A_{r_1,r_2}$}
\end{figure}
\vskip 5 truemm

Observe that the condition of local finiteness of 
${\mbox{\boldmath${\nu}$}}^{\rm c}$ 
implies only that the volume of the above set is finite 
when $|r_2/r_1-1|<\delta$
for some $\delta >0$. It is not clear whether the strong local finiteness
condition would hold $\forall$ ${\rm c}\in (-\infty ,1]$. 
However, we will assume that this property holds true. 
(It holds for ${\rm c}=0$; see [W4].)

Let $A\subset\C^*$  be a relatively compact annulus 
and $\alpha\;$ be 
an embedding $\;A\hookrightarrow\C^*$. Assume that both annuli
$A$ and $\alpha (A)$ surround the origin. 
Then $\alpha$ induces an open embedding 
$$\alpha_*:\;Loop^1(A)\hookrightarrow
Loop^1(\alpha (A))\,.\eqno (2.25)$$
Given $A$ and $\alpha$ as above, there is defined a positive 
continuous function, $q^{\rm{det}}_\alpha ({\cL} )$, 
${\cL}\in Loop^1$. In terms of this function we will state
a condition on a scalar measure ${\mbox{\boldmath${\nu}$}}$
on $Loop^1$ called restriction covariance and
guaranteeing that ${\mbox{\boldmath${\nu}$}}$
obtained from an LCC assignment. In fact, the
assignment will be reconstructed from
a scalar measure ${\mbox{\boldmath${\nu}$}}$ satisfying
the restriction covariance condition.  

%\vskip 5 truemm
%\begin{center}
%\includegraphics[width=.65\textwidth]{Fig9.eps}
%\end{center}
%\vskip 5 truemm

\vskip 5 truemm
\begin{figure}[ht]
\centering
\includegraphics[height=2cm]{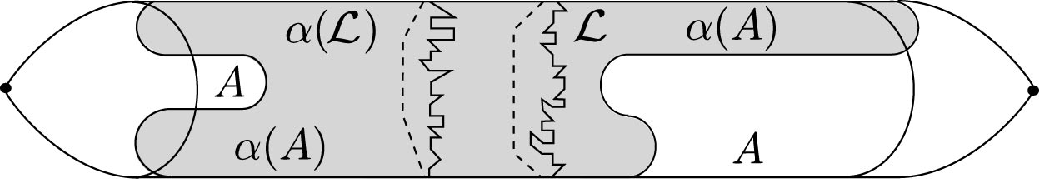}
\caption{Loop in an annulus, and their images under the embedding $\alpha$}
\end{figure}
\vskip 5 truemm

To define function $q^{\rm{det}}_\alpha ({\cL} )$, we construct 
a neutral collection $\mE_{\alpha,{\cL}}$ associated with loop
${\cL}\in Loop^1$ (more precisely, with the corresponding 
loop in $\C P^1$
which we denote by the same symbol ${\cL}$). Collection 
$\mE_{\alpha,{\cL}}$ 
consists of eight spheres $S_i$, $1\leq i\leq 8$. 
Spheres $\Sigma_1$, $\Sigma_2$, $\Sigma_3$, 
and $\Sigma_4$ in the collection are the doubles of 
four open disks $D_1$, $D_2$, $D_3$, and $D_4$, respectively.
In turn, the disks are identified as follows:
$$D_1=D_{{\rm L},{\cL}},\;D_2=D_{{\rm R},{\cL}},\;
D_3= D_{{\rm L},\alpha({\cL})},\;D_4= D_{{\rm R},\alpha({\cL})}.$$

%\vskip 5 truemm
%\begin{center}
%\includegraphics[width=.5\textwidth]{Fig10.eps}
%\end{center}
%\vskip 5 truemm

\vskip 5 truemm
\begin{figure}[ht]
\centering
\includegraphics[height=65mm]{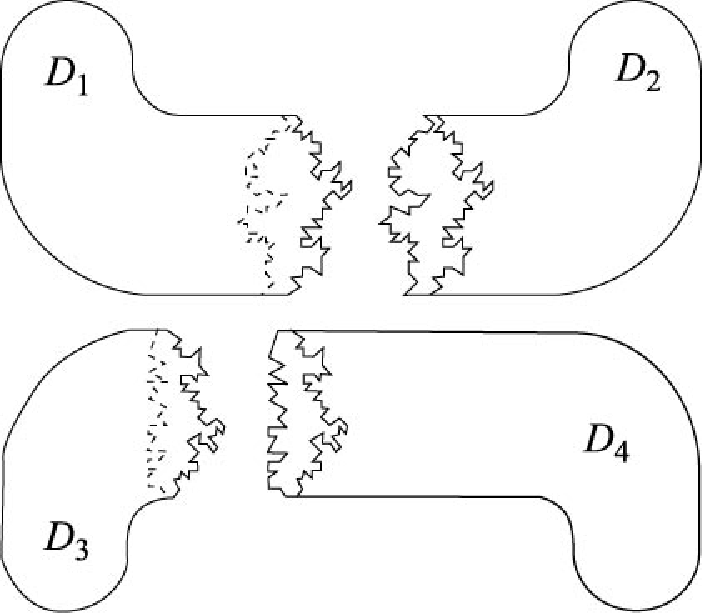}
\caption{Four discs $D_1,\dots,D_4$ in $\C P^1$}
\end{figure}
\vskip 5 truemm

Geometrically, disks $D_{{\rm L},{\cL}}$ and $D_{{\rm R},{\cL}}$ are
domains in $\C P^1$ to the left and to the right of 
loop ${\cL}$, respectively, and disks
$D_{{\rm L},{\cL}}$ and $D_{{\rm R},{\cL}}$ are
domains in $\C P^1$ to the left and to the right of 
loop $\alpha ({\cL})$, respectively.

Further, spheres $\Sigma_5$ and $\Sigma_6$ in the collection
constitute the double of sphere $S_{\rm{in}}=\overline{D_1\cup D_2}$, and spheres 
$\Sigma_7$ and $\Sigma_8$ the doubles of sphere $S_{\rm{fin}}=
\overline{D_3\cup D_4}$, 
correspondingly. (Subscript ${\rm{in}}$ stands for initial
and ${\rm{fin}}$ for final.) Formally, 
$$S_{\rm{in}}=\C P^1,\;\;S_{\rm{fin}}=\C P^1.$$

The non-Hausdorf surface $S^{\rm{nH}}$ for collection 
$\mE_{\alpha,{\cL}}$ is formed by glueing the above eight spheres
$S_1$, $\ldots$, $S_8$. 
These spheres are  
glued all along the domains that are the pullback to the double covering
of the union of two thin strips on the left and on the right of loop 
${\cL}$ in $\C^*$. In the figure below, each sphere $S_i$ is identified
by a pair of half-spherical 
caps which have value $i$ among the pair of indeces attached to
them. (So, spheres $S_1$, $S_2$, $S_3$, and $S_4$ `live' on both
levels while $S_5$, $S_6$, $S_7$ and $S_8$ on a single level.
Spheres $S_6$ and $S_5$ are drown horizontal.)

The weights $\mu_i$ are: $\mu_1=1$, $\mu_2=1$, $\mu_3=-1$, $\mu_4=-1$, 
$\mu_5=-1$, $\mu_6=-1$, $\mu_7=1$ and $\mu_8=1$.  
  
%\vskip 5 truemm
%\begin{center}
%\includegraphics[width=.75\textwidth]{Fig11.eps}
%\end{center}
%\vskip 5 truemm

\vskip 5 truemm
\begin{figure}[ht]
\centering
\includegraphics[height=65mm]{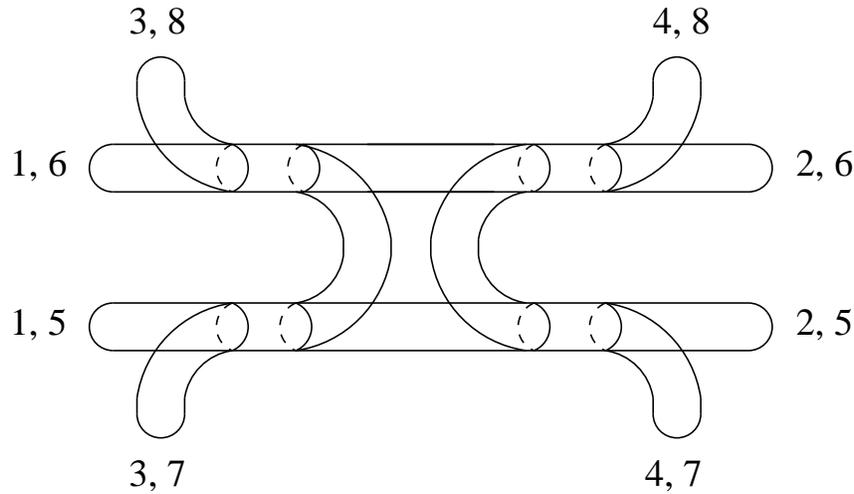}
\caption{Neutral collection of 8 spheres}
\end{figure}
\vskip 5 truemm

Value $q_\alpha^{\rm{det}}({\cL})$ is then defined as follows:
$$q_\alpha^{\rm{det}}({\cL})=\left(v_{\mE_{\alpha,{{\cL}}}}\left/
\operatornamewithlimits{\otimes}_{k=1}^8v_{S_k}^{\otimes \mu_k}
\right.\right)^{1/2}.\eqno (2.26)$$ 
\medskip

{\bf Definition 2.2.} We say that a scalar
measure ${\mbox{\boldmath${\nu}$}}$ on $Loop^1$ is 
${\rm c}$-{\it restriction covariant} (${\rm c}$-RC, or briefly, RC) if, 
for each relatively compact annulus 
$A\subset\C^*$ and an embedding $\alpha:\;A\hookrightarrow\C^*$,
the pullback $\alpha^*\big({\mbox{\boldmath${\nu}$}}
\big|_{\alpha_*(Loop^1(A))}\big)$ of the restriction 
${\mbox{\boldmath${\nu}$}}\big|_{\alpha_*(Loop^1(A))}$
of measure ${\mbox{\boldmath${\nu}$}}$ to the  
image $\alpha_*(Loop^1)$ (which is an open subset in $Loop^1$)
is absolutely continuous with respect to 
${\mbox{\boldmath${\nu}$}}$ and has the Radon-Nikodym 
derivative
$$\frac{{\rm d}\big[\alpha^*\big(
{\mbox{\boldmath${\nu}$}}\big)\big]\big|_{\alpha_*(Loop^1(A)}}{
{\rm d}{\mbox{\boldmath${\nu}$}}}\;({\cL} )=
\left(q^{\rm{det}}_\alpha ({\cL} )
\right)^{\rm c},\;\;{\cL}\in Loop^1(A).\quad\blacksquare\eqno (2.27)$$
\medskip

As follows from definitions, if 
$\Sigma\mapsto{\mbox{\boldmath${\lambda}$}}_{\Sigma}$
is an ${\rm c}$-LCC assignment, then scalar measure
${\mbox{\boldmath${\nu}$}}^{\rm c}$ on $Loop^1$ is ${\rm c}$-RC.
Moreover, any  ${\rm c}$-RC measure $\nu$ on
$Loop^1$ gives rise to a unique ${\rm c}$-LCC assignment.
In fact, it suffices to define measures 
${\mbox{\boldmath${\lambda}$}}_{\Sigma}$ when $\Sigma$
is an arbitrary annulus (and restrict the measures
on $Loop^1 (\Sigma )$). Further, an annulus can be embedded in 
$\C^*$. Hence, the RC property of $\nu$ is necessary
and sufficient for constructing an LCC assignment.
\medskip

{\bf 2.5.2. Infinitesimal restriction covariance property 
for measures on $Loop^1 (\C^*)$.}
Perhaps a simpler task is to check the RC property in an 
infinitesimal form, where embedding $\alpha$ is close to
identity. To this end, observe that the Lie algebra (over $\R$) 
$${\mv}=\C [z,z^{-1}]\diy{\frac{\partial}{\partial z}}\eqno (2.28)$$
acts on $Loop^1$. The basis of algebra ${\mv}$
consists of elements 
$$L_n=-z^{n+1}\diy{
\frac{\partial}{\partial z}}\;\hbox{ and }\;L'_n={\rm i}z^{n+1}\diy{
\frac{\partial}{\partial z}},\;\;n\in\Z.\eqno (2.29)$$ 
Formally speaking, the infinitesimal RC 
property is that
$${\rm{div}}_{{\mbox{\boldmath${\nu}$}}^{\rm c}}L_n={\rm c}P_n,\;\;
{\rm{div}}_{{\mbox{\boldmath${\nu}$}}^{\rm c}}L'_n={\rm c}P'_n,\;\;n\in\Z,$$
where $P_n$, $P'_n$ are certain explicit functions on $Loop^1$
related to so-called Neretin polynomials, and 
${\rm{div}}_{{\mbox{\boldmath${\nu}$}}^{\rm c}}$ stands for the divergence 
relative to measure ${\mbox{\boldmath${\nu}$}}^{\rm c}$ (see [AM]).
In reality, it is enough to check this property when $|n|\leq 2$,
because algebra ${\mv}$ is generated by $L_n$ and $L'_n$ with
$n=-2,-1,0,1,2$.

It looks plausible that the property of 
restriction covariance can be deduced from that of 
infinitesimal restriction covariance. 
However, in this paper we do not offer
a formal proof of this fact. We consider this as 
an interesting open question.  

Explicit formulas for $P_n$ and $P'_n$ for $n=-2,-1,0,1,2$, are
given below. First,
$$P_n=P'_n=0,\;\;n=-1,0,1,\eqno (2.30)$$
which follows from invariance of measure 
${\mbox{\boldmath${\nu}$}}^{\rm c}$ under the action of 
$PSL (2,\C)$. Next,
$$P_{-2}({\cL})=\frac{1}{12}{\rm{Re}}\;{\cS}_{\phi_{{\rm L},{\cL}}}
(0),\;\;P'_{-2}({\cL})=\frac{1}{12}{\rm{Im}}\;
{\cS}_{\phi_{{\rm L},{\cL}}}(0).\eqno (2.31)$$
Here, ${\cS}_f$ stands for the Schwartzian derivative
of function $f$:
$${\cS}_f(z)=\frac{f'''(z)}{f'(z)}-\frac{3 (f''(z))^2}{2(f'(z))^2}\,.
\eqno (2.32)$$
Next, $\phi_{{\rm L},{\cL}}:\;U\to\C$ is an embedding of the open 
unit disk $U=\{t\in\C:\;\;|t|<1\}$, with the image 
$\phi_{{\rm L},{\cL}}(U)=D_{{\rm L},{\cL}}$, normalised so that 
$\phi_{{\rm L},{\cL}}(0)=0$. For $P_2$ and $P'_2$, the formulas 
are similar to (2.31); the only change is that one uses embedding
$\phi_{{\rm R},{\cL}}:\;U\to\C$ with the image $\phi_{{\rm R},{\cL}}(U)=
\varpi [D_{{\rm R},{\cL}}]$ where $\varpi :\;\C P^1\to\C P^1$ is 
the involution $z\mapsto 1/z$.

The justification of formula (2.31) will not be given here:
we refer the reader to section 3.3 where a similar argument 
is used in a slightly different situation.

We will consider two coordinates on $Loop^1$: 
$$\{A,a_1,a_2,\ldots \}\;\hbox{ and }\;\{B,b_1,b_2,\ldots \}.$$ 
Here, the components are as follows:
$$\hbox{$A,B$ are real positive numbers, and
$a_k,b_k$ complex numbers, $k\geq 1$,}$$
identified from the representations
$$\begin{array}{l}\phi_{{\rm L},{\cL}}(t)=A(t+a_1t^2+a_2t^3+\ldots ),\\
\phi_{{\rm R},{\cL}}(t)=B(t+b_1t^2+b_2t^3+\ldots ),\end{array}\;\;
t\in U.\eqno (2.33)$$
One can show that the inequality
$$0<AB\leq 1$$
holds, with equality only when the loop ${\cL}\in Loop^1$ is a circle
$\{z\in\C^*:\;\;|z|=r\}$, 
for some $r\in (0,+\infty )$. In fact, we may assume that 
$AB<1$, as equality $AB=1$ holds on a set of  
${\mbox{\boldmath${\nu}$}}^{\rm c}$-measure $0$.
\medskip

{\bf Definition 2.3.} It is convenient to introduce a set ${\bB}$
of functions $F:\;Loop^1\to\C$ written as finite sums 
over quadruples of multi-indeces $(I,I',J,J')$:
$$F({\cL})=\sum\limits_{I,I'J,J'} f_{I,I',J,J'}(A,B)
a^I{\oa}^{I'}b^J{\ob}^{J'}.\eqno (2.34)$$
Here $a^I=a_1^{i_1}a_2^{i_2}\ldots$ is a monomial
in $a_1$, $a_2$, $\ldots$ associated with an integer-valued multi-index
$I=(i_1,i_2,\ldots )$, of a finite total degree 
($|I|=|i_1|+|i_2|+\ldots <+\infty$). Similarly, 
${\oa}^{I'}$, $b^J$ and ${\ob}^{J'}$
are monomials in the corresponding variables associated with
finite-degree multi-indeces $I'=(i'_1,i'_2,\ldots )$, 
$J=(j_1,j_2,\ldots )$ and $J'=(j'_1,j'_2,\ldots )$.   
Further,
$f_{I,I',J,J'}$ is a function with compact support on 
$\{(A,B)\in\R^2:\;\;A,B>0,\;AB<1\}$. Then    
${\bB}$ is a non-unital commutative $*$-algebra, separating
points of $Loop^1$. Thus, 
measure ${\mbox{\boldmath${\nu}$}}^{\rm c}$ is uniquely
determined by its integrals for functions from ${\bB}$ 
(generalised moments). $\blacksquare$
\medskip

{\bf Remark 2.4.} The Bieberbach conjecture established by L. De Branges
implies that, $\forall$ ${\cL}\in Loop^1$,
$$|a_k|,|b_k|\leq k+1,\;\;k\geq 1.\quad \blacksquare$$
\medskip
 
Note that the Lie algebra 
$\C [z,z^{-1}]\diy{\frac{\partial}{\partial z}}$
and the operators of multiplication by $P_n$, $P'_n$, $n\in\Z$,
preserve ${\bB}$.
\medskip

{\bf Remark 2.5.} Coordinates $a_1$, $a_2$, $\ldots$ were used
in paper [AM], in an attempt to identify a `natural' measure on
the quotient space $Loop^1/\R$. In our context, it is 
not enough to use a single
coordinate, say $(A,a_1,a_2,\ldots )$. The reason is that for 
a non-zero function $f_{I,I'}\in C^\infty_0(0,1)$, the integral
$$\int_{Loop^1}f_{I,I'}(A)a^I{\oa}^{I'}{\mbox{\boldmath${\nu}$}}^{\rm c}$$ 
diverges. Hence, there is no obvious algebra of functions in variables 
$A$, $a_1$, $a_2$, $\ldots$ for which the generalised moments are finite.
Therefore, the second coordinate $(B,b_1,b_2,\ldots )$ is
needed. (An indication of this fact can be found in [AMT].)
$\blacksquare$ 
\medskip

{\bf Definition 2.4.} We say that a strongly locally finite measure 
${\mbox{\boldmath${\nu}$}}$ on $Loop^1$ is   
{\it infinitesimally ${\rm c}$-restriction covariant} 
(${\rm c}$-IRC, or briefly, IRC) if,
$\forall$ $F\in{\bB}$ and $\forall$ $n\in\Z$, 
$$\int\limits_{Loop^1}\big[(L_n+{\rm c}P_n)F\big]
\;{\mbox{\boldmath${\nu}$}}=0\quad\;\;\eqno (2.35)$$
and
$$\int\limits_{Loop^1}\big[(L'_n+{\rm c}P'_n)F\big]
\;{\mbox{\boldmath${\nu}$}}=0.\quad \blacksquare\eqno (2.36)$$
\medskip

In fact, as we mentioned earlier,
in order to check that ${\mbox{\boldmath${\nu}$}}$ 
is ${\rm c}$-IRC, it suffices to verify the above equations 
for $|n|\leq 2$.
 
We note that the equations for $n\leq 0$ coincide with 
conditions (2.3.3) from paper [AM]; the case $n=1$
was considered in article [AMT]. 

It also possible to consider a larger space $AHull^1$
formed by `annular hulls' (called `bubbles' in [LW]). An annular hull is a
closed compact in $\C^*$ homotopically equivalent to 
$S^1$ and separating $0$ from $\infty$ on $\C P^1$. Coordinates
$(A,a_1,a_2,\ldots )$ and $(B,b_1,b_2,\ldots )$, and thus algebra $\bB$,
can be extended to $AHull^1$. In turn, it allows us to define 
the IRC property for a measure on $AHull^1$. The domain
$$\{A\geq A_0,\;\;B\geq B_0\}$$
is a compact in $AHull^1$, in the topology generated jointly by 
the pair of coordinates $(A,a_1,a_2,\ldots )$ and $(B,b_1,b_2,\ldots )$.
Then measures on $AHull^1$ are identified with positive
functionals on ${\bB}$. 
\medskip

{\bf Remark 2.6.} A (Borel) measure ${\mbox{\boldmath${\pi}$}}$ on $AHull^1$ 
invariant under
the action of $\R_+^\times$ gives rise to a countable
collection of distributions (generalised functions)
$M_{I,I',J,J'}$
on $(0,1)$ (more precisely, on the test-function space $C^\infty_0(0,1)$), 
labeled by quadruples of integer-valued multi-indeces
$I,I',J,J'$ of finite total degree. Namely, 
$$\int\limits_{AHull^1} f_{I,I',J,J'}(A,B)a^I{\oa}^{I'}b^J{\ob}^{J'}
 {\mbox{\boldmath${\pi}$}}=\int\limits_0^{+\infty}
\int\limits_0^{+\infty} f(A,B)
M_{I,I',J,J'}(AB)\frac{{\rm d}A\times{\rm d}B}{AB}.\eqno (2.37)$$
The fact that ${\mbox{\boldmath${\pi}$}}$ is IRC
gives rise to a countable system of
differential equations involving distributions $M_{I,I',J,J'}$.
One can show that any solution to this system of differential
equations can be uniquely 
reconstructed from distribution $M_{\uz}:=M_{0,0,0,0}$. The  
latter can be arbitrary, provided that it satisfies a countable system
of inequalities, depending on ${\rm c}$ (which follow from non-negativity of
measure ${\mbox{\boldmath${\pi}$}}$). In particular, $M_{\uz}$ 
is a (non-negative) measure on $(0,1)$. $\blacksquare$ 
\medskip

In relation to measure $M_{\uz}$, we put forward the following
comment.
\medskip

{\bf Remark 2.7.} It is plausible that the measures 
$M_{\uz}$ associated with IRC measures on $AHull^1$
form an infinite-dimensional cone, with a continuum
of extremal rays. We expect that $\forall$ $r\in (0,1)$, 
there is a `canonical' extremal measure $M^{(r)}_{\uz}$,
unique up to a scalar factor, and the associated 
IRC measure
${\mbox{\boldmath${\pi}$}}^{(r)}$ on $AHull^1$ 
admits the following description.
Consider the measure on the Cartesian product
$Loop^1\times Loop^1$ which is the product 
${\mbox{\boldmath${\nu}$}}^{\rm c}\times{\mbox{\boldmath${\nu}$}}^{\rm c}$ 
of two copies of the (hypothetic) ${\rm c}$-IRC measure
${\mbox{\boldmath${\nu}$}}^{\rm c}$. Consider the restriction of
${\mbox{\boldmath${\nu}$}}^{\rm c}\times
{\mbox{\boldmath${\nu}$}}^{\rm c}$ on the open subset 
$(Loop^1)^{\times 2}_{\rm{dis}}\subset
Loop^1\times Loop^1$ consisting of pairs of disjoint loops.
With each pair of disjoint loops there is associated an
annular hull which is the set bounded by these loops. The  
conformal parameter of this annular hull generates a map
$\Upsilon$: $(Loop^1)^{\times 2}_{\rm{dis}}\to (0,1)$. We
conjecture that, 
$\forall$ $r\in (0,1)$, ${\mbox{\boldmath${\pi}$}}^{(r)}$ is the measure, 
on the pullback image of $\Upsilon^{-1}r$, induced by the above restriction
$\big({\mbox{\boldmath${\nu}$}}^{\rm c}\times
{\mbox{\boldmath${\nu}$}}^{\rm c}\big)\big|_{(Loop^1)^{\times 2}_{\rm{dis}}}$.

Finally, we conjecture that for $r=1$, the limiting measure 
$\lim\limits_{r\to 1}{\mbox{\boldmath${\pi}$}}^{(r)}$ is supported by
$Loop^1$ and coincides with ${\mbox{\boldmath${\nu}$}}^{\rm c}$. 
$\blacksquare$
\medskip

There are two open problems related to IRC measures
on $AHull^1$.
\medskip

1. {\sl Write explicitly the system of inequalities upon 
measure $M_{\uz}$ associated with an IRC measure 
${\mbox{\boldmath${\pi}$}}$ on $AHull^1$.} 

2. {\sl Calculate, in a closed form, measure $M_{\uz}$
associated with a (hypothetic) IRC measure 
${\mbox{\boldmath${\nu}$}}^{\rm c}$ on $Loop^1$.}
\medskip

We expect that the latter measure $M_{\uz}$ has a real analytic
density relative to Lebesgue's measure on $(0,1)$, and
the Radon-Nikodym derivative $\diy{\frac{{\rm d}M_{\uz}}{{\rm d}\log\;r}}$
is a kind of indefinite $\theta$-series, presumably related
to Kac' character formulas for representations of the Virasoro
algebra.   

\section{Properties of determinant lines}

In this chapter we prove some useful results relating 
the determinant lines of various surfaces. These results (Propositions 
1 and 2) will be used in chapter 5. In a sense, the results 
of this chapter are not new and have been known to specialists
in a somewhat different form.

\subsection{A preliminary: metrics with pole singularities.}
 
In this subsection we spell out some general concepts needed
in the context of subsequent parts of the paper. 
Assume that $\Sigma$ is a compact surface and 
$\cD=\sum_{i=1}^n k_i p_i$ is a divisor on $\Sigma$, i.e. 
a formal linear combination of distinct points $p_i\in \Sigma$ with 
integral weights $k_i\in \Z$. We define a metric on $\Sigma$ with 
singularities given by $\cD$ as a metric $g$ on non-compact 
surface $\Sigma\setminus \{p_1,\dots,p_n\}$ such 
that near each point $p_i$ there exists a local holomorphic 
coordinate $z_i$ in which metric $g$ has form
$$g=|z_i^{k_i} {\rm{d}}z|^2\,.$$

We claim that such a metric defines a positive vector $[g]$ 
in the tensor product
$$\big|\det\big|_{\Sigma}\otimes \left(
\diy{\operatornamewithlimits{\otimes}_{i=1}^n}
{\big|\det T_{p_i}\Sigma\big|}^{\otimes k_i/24}\right)\,.\eqno (3.1)$$
Here and below, $\det T_p\Sigma$ stands for the wedge 
square $\wedge^2 T_p\Sigma$. Next, given a one-dimensional 
real vector space $V$, we denote by $|V|$ 
the oriented one-dimensional real vector space associated
with the homomorphism 
$$GL (1,\R )\to\R^\times_{>0},\quad x\in GL (1,\R)\mapsto |x|.
\eqno (3.2)$$ 

In order to define $[g]$, it suffices to define the ratio 
$$[g]/[g_0]\in\diy{\operatornamewithlimits{\otimes}_{i=1}^n}   
{\big|\det T_{p_i}\Sigma\big|}^{\otimes k_i/24},\eqno (3.3)$$ 
for any non-singular metric $g_0$ on $\Sigma$. Furthermore, 
we can assume that
$g_0$ is flat near each point $p_i$. In this case we set 
$$[g]/[g_0]:=\exp\;\left[\frac{1}{48 \pi{\rm i}}
\int\limits_{\Sigma \setminus\{p_1,\dots,p_n\}} {\rLL}
(g_0,g)\right]
\diy{\operatornamewithlimits{\otimes}_{i=1}^n}   
[g_0]_{p_i}^{\otimes k_i/24}\,.\eqno (3.4)$$
Here $[g_0]_p\in {\big|\det T_p\Sigma\big|}$ is the inverse
to the natural volume element on $\det T_p\Sigma$ 
generated by metric $g_0$.
Notice that the integral in the above expression is absolutely 
convergent as the density ${\rLL}(g_0,g)$ vanishes near points 
$p_i$ (because both metrics $g_0$ and $g$ are flat there).

The consistency of the above definition is guaranteed by 
Eqn (2.3) and the following general lemma that is valid for any 
surface $\Sigma$.
\medskip

{\bf Lemma 3.1.} {\sl Let $\Sigma$ be a surface with a 
marked point $p$ and $z,w_1,w_2$ be local coordinates near 
point $p$, vanishing at $p$. Let $k$ be an integer. Consider 
the $1$-form $\alpha$ defined in Eqn {\rm{(2.4)}}.
Then the integral, over a small, anticlock-wise oriented, circle around $p$, 
of the closed $1$-form 
$$\alpha\Big(| z^k {\rm{d}}z|^2,|{\rm{d}}w_1|^2, |{\rm{d}}w_2|^2\Big)$$
is equal to $2\pi{\rm i} k\log| ({\rm{d}}w_1/{\rm{d}}w_2)(p)|^2$.}
\medskip

{\it Proof} : Observe that for any three flat metrics $g_1$, $g_2$
and $g_3$ on $\Sigma$, the form  
$\alpha(g_1,g_2,g_3)$ is 
closed. Furthermore, after rescaling one of the metrics 
as $g_i\to bg_i$, $b>0$, the above integral
increases by the amount $\log\;b$ times the difference 
of the rotation numbers of the two other metrics. Next, let 
us consider the integral of $\alpha(g_1,g_2,g_3)$
over the unit circle in coordinate $\tilde{z}:=tz$, for 
real $t\to +\infty$, where
$$g_1=|z^k {\rm{d}}z|^2/t^{2(k+1)}=|\tilde{z}^k {\rm{d}}\tilde{z}|^2,$$
and
$$g_2=|{\rm{d}}w_1|^2/|({\rm{d}}w_1/{\rm{d}}\tilde{z})(p)|^2,
\,g_3=|{\rm{d}}w_2|^2/|({\rm{d}}w_2/{\rm{d}}\tilde{z})(p)|^2\,\,.$$
This integral tends to zero as $t\to \infty$ because $g_2$ 
and $g_3$ become close
to $|{\rm{d}}\tilde{z}|^2$, and form $\alpha(g_1,g_2,g_3)$ is 
antisymmetric in indices $1,2,3$. By the above remark on rescaling,
the difference of the integral in the statement of Lemma 3.1 
and the integral of $\alpha (g_1,g_2,g_3)$ is equal to $2\pi{\rm i} k\log\;
\left|\diy{\frac{{\rm d}w_1}{{\rm d}w_2}(p)}\right|^2$. 
The assertion of Lemma 3.1 then follows. $\Box$
\medskip

Later on, we will also need
\medskip

{\bf Lemma 3.2.} {\sl Let $\Sigma$ be a surface with a marked 
point $p$ and $z_1,z_2,w$ be local coordinates near point 
$p$, vanishing at $p$ and such that
$\diy{\frac{{\rm d}z_1}{{\rm d}z_2}}(p)=1$. Given an integer $k$, 
the integral, over a small circle around $p$, of
the $1$-form 
$$\alpha\Big(|z_1^k {\rm{d}}z_1|^2,|z_2^k {\rm{d}}z_2|^2, 
|{\rm{d}}w|^2\Big)$$
equals zero.}
\medskip

The proof of Lemma 3.2 is similar to that of Lemma 3.1, 
and we omit it.
\medskip

\subsection{The canonical vector for the special four-sphere neutral collection}

The central result of section 3.2 is a formula (see Eqn (3.8))
for the ratio between the canonical vector $v_\mF$ and the 
product of canonical vectors $(v_{S_i})^{\otimes m_i}$. Here and 
throughout the rest of the paper, $\mF$ stands for the neutral collection 
$\left(\{(S_i,\phi_i,\mu_i)\}_{i=1}^4;S^{\rm{nH}}\right)$ consisting of 
four spheres introduced in subsection 2.2.5.
\medskip

{\bf 3.2.1. A formula for the canonical vector for general metrics}
Assume that the common part of spheres $S_1$, $S_2$, $S_3$ and $S_4$
contains a closed cylinder $C$. Moreover, we assume that
$$\begin{array}{l}
S_1=S_{11}=S_{1,L}\cup C\cup S_{1,R},\;\;
S_2=S_{12}=S_{1,L}\cup C\cup S_{2,R},\\
S_3=S_{21}=S_{2,L}\cup C\cup S_{1,R},\;\;
S_4=S_{22}=S_{2,L}\cup C\cup S_{2,R},\end{array}\eqno (3.5)$$
where $S_{iL},S_{iR}$, $i=1,2,3,4$, are half-spheres whose boundary 
circle is identified with the corresponding boundary circle of $C$ (left 
or right, respectively). 

From now on we will use the pair of lower indices $(ij)$,
$1\le i,j\le 2$, instead of a single index $i,\,1\le i\le 4$.
The weights will be
$$\mu_{11}=+1,\;\;\mu_{12}=-1,\;\;
\mu_{21}=-1,\;\;\mu_{22}=+1.\eqno (3.6)$$

Suppose that $g_{ij}$ are metrics on surfaces $S_{ij}$, 
$1\le i,j\le 2$. Lemma 3.3 below gives
an expression for the logarithm of the ratio $\diy{
v_\mF\left/\left(\diy{\operatornamewithlimits{\otimes}_{1\le i,j\le 2}} 
[g_{ij}]^{\otimes \mu_{ij}}\right)\right.}$:
\medskip

{\bf Lemma 3.3.} {\sl 
$$\begin{array}{l}\log\;\left[\diy{v_\mF\left/\left(\diy{
\operatornamewithlimits{\otimes}_{1\le i,j\le 2}} 
[g_{ij}]^{\otimes \mu_{ij}}\right)\right.}\right]\\
\quad =\diy{\frac{1}{48 \pi{\rm i}}}\left[
\diy{\int\limits_{S_{1,L}}}{\rLL}(g_{11},g_{12})
+\diy{\int\limits_{S_{2,L}}}{\rLL}(g_{22},g_{21})\right.\\
\quad +\diy{\int\limits_{S_{1,R}\cup C}}{\rLL}(g_{11},g_{21})
+\diy{\int\limits_{S_{2,R}\cup C}}{\rLL}(g_{22},g_{12})
\\
\quad 
\left.+\diy{\int\limits_L}\alpha(g_{11},g_{12},g_{21})
-\diy{\int\limits_{L_{\,}}^{\,}}\alpha(g_{22},g_{12},g_{21})
\right]\,.\end{array}\eqno (3.7)$$ 
Here $L$ is the left boundary circle of cylinder $C$ endowed
with the standard orientation on $\partial C$.}
\medskip

{\it Proof} : First, assume that all metrics $g_{ij}$ are 
restrictions of a metric on the
non-Hausdorff surface associated with collection $\mF$. In this case,
the LHS in (3.7) vanishes. On the other hand, 
every term in the sum in the RHS also
vanishes. Hence, in this case Eqn (3.7) holds. 

Thus, we should check
that, after the change of metric $g_{ij}$ for some $(i,j)$, both the 
LHS and the RHS of (3.7) increase by same 
amount. This follows directly from Lemmas 2.1 and 2.2   
and the Stokes formula. $\Box$
\medskip

{\bf 3.2.2. The residue formula.}
Now let us apply results from section 3.1 to the 
special neutral four-sphere collection ${\mF}$.
Choose points $p_{1,L},p_{2,L}$ on pieces $S_{1,L}$ 
and $S_{2,L}$ respectively, and
fix a holomorphic parametrisation $z_{ij}$ of each surface 
$S_{ij}$ by $\C P^1$
such that $z_{ij}(p_{i,L})=\infty$. Then $|{\rm{d}}z_{ij}|^2$ is 
a metric with  singularities on $S_{ij}$ at divisor $-2 p_{i,L}$. 
Combining the results from section 3.1 with the 
explicit formula for form $\alpha$ in Eqn (2.4), 
we obtain the following assertion for the logarithm of the ratio 
$v_\mF\left/\left(\diy{\operatornamewithlimits{\otimes}_{1\le i,j\le 2}} 
v_{S_{ij}}^{\otimes \mu_{ij}}\right)\right.$:
\medskip

{\bf Lemma 3.4.} {\sl In the above notation, the following formula
holds true:
$$\log\;\left[\diy{v_\mF\left/\left(
\operatornamewithlimits{\otimes}_{1\le i,j\le 2} 
v_{S_{ij}}^{\otimes \mu_{ij}}\right)\right.}\right]
=\diy{\frac{-1}{24 \pi}}
{\rm{Im}}
\diy{\int\limits_L}\log\left(\diy{\frac{{\rm{d}}z_{11}}{{\rm{d}}z_{22}}}
\right) 
{\rm d}\log\left(\diy{\frac{{\rm{d}}z_{12}}{{\rm{d}}z_{21}}}\right)\,.
\eqno (3.8)$$} 
\medskip

{\it Proof} : Without loss of generality, assume that for $i=1,2$,
$$\frac{{\rm d}z^{-1}_{i1}}{{\rm d}z^{-1}_{i2}}\big(p_{i,L}
\big)=1;\eqno (3.9)$$
this can be achieved by rescaling coordinates $z_{ij}$. 
Set ${\wt g}_{ij}=\left|{\rm d}z_{ij}\right|^2$, and denote by
$g_{ij}$ the round metric on $\sigma_{ij}$ determined by the 
stereographic projection in coordinate $z_{ij}$. The LHS in 
(3.8) is equal by definition to
$$\log\;\left[\diy{v_\mF\left/\left(
\operatornamewithlimits{\otimes}_{1\le i,j\le 2} 
\big[g_{ij}\big]^{\otimes \mu_{ij}}\right)\right.}\right].$$
Owing to Lemma 3.3, this expression coincides with a certain sum of 
integrals over pieces of $\Sigma$ and over contour ${\cL}$. For 
singular metrics ${\wt g}_{ij}$, the expression
$$\log\;\left[\diy{v_\mF\left/\left(
\operatornamewithlimits{\otimes}_{1\le i,j\le 2} 
\left[{\wt g}_{ij}\right]^{\otimes \mu_{ij}}\right)\right.}\right]$$
also makes sense, because terms taking values in 
$\left|\wedge^2 T_{p_{i,L}}S_{i,L}\right|$ vanish. Further, for 
metrics ${\wt g}_{ij}$, the RHS in (3.7) is well-defined.
 
Next, we claim that the assertion of Lemma 3.3 remains valid 
for metrics ${\wt g}_{ij}$. The reason is as follows. Take the 
difference of the LHSs in (3.7) for metrics $g_{ij}$ and 
${\wt g}_{ij}$. It is equal to
$$\frac{-1}{48\pi{\rm i}}\sum_{i,j=1,2}\mu_{ij}\int\limits_{S_{ij}}
{\rLL}\left(g_{ij},{\wt g}_{ij}\right).\eqno (3.10)$$ 
On the other hand, the difference of the RHSs in (3.7) for 
metrics $g_{ij}$ and ${\wt g}_{ij}$ coincides with (3.8) modulo
possible boundary terms around points $p_{i,L}$. This is because the
proof of Eqn (3.7) for smooth metrics is based on combination of 
Eqn (2.4) and Lemma 2.2; hence it works for singular metrics, too. 

Near each point $p_{i,L}$ we have four metrics, two smooth and two 
singular. The integral of $1$-form $\alpha$ over a small circle surrounding $p_{i,L}$
vanishes for any choice of
three of them, by virtue of Lemmas 3.1 and 3.2. 
Therefore, we have
$$\log\;\left[\diy{v_\mF\left/\left(
\operatornamewithlimits{\otimes}_{1\le i,j\le 2} 
\left[{\wt g}_{ij}\right]^{\otimes \mu_{ij}}\right)\right.}\right]=
\frac{1}{48\pi {\rm i}}\int_L\Big[\alpha\big({\wt g}_{11},{\wt g}_{12},
{\wt g}_{21}\big)
-\alpha\big({\wt g}_{22},{\wt g}_{12},{\wt g}_{21}\big)\Big].
\eqno (3.11)$$
The assertion of Lemma 3.4 then follows, as the expression in 
(3.11) coincides with the RHS of (3.8) by a straightforward
calculation. $\Box$.
 
\subsection{A variation formula for the special neutral collection}

{\bf 3.3.1. Schiffer variation.} Let $\Sigma$ be a surface with a conformal 
structure and $z$ be a local holomorphic coordinate on $\Sigma$ 
defined in a neighborhood $U_p$ of  point $p\in \Sigma$, such that $z(p)=0$.
We associate with the triple $(\Sigma,p,z)$ the germ of a one-parameter 
family $(\Sigma_t)_{0\le t< \epsilon}$ of new surfaces with conformal 
structures such that $\Sigma_0$ is canonically identified with 
$\Sigma$. Moreover, on each $\Sigma_t$ for $t\ne 0$ there will be an 
open part identified conformally with $\Sigma\setminus U_p$.

Namely, we define $\Sigma_t$ for  $t\in [0,\epsilon)$ as the result of 
glueing of 
$$\Sigma\setminus \{p'\in U:\;\;|z(p')|\le \delta_1\}$$ 
with the disk $\{w\in \C:\;\;|w|\le \delta_2\}$, by the correspondence
$$z(p')=\sqrt{w^2+t}\,\,.\eqno (3.12)$$
Here  $\epsilon^2/\delta_1$, $\delta_1/\delta_2$ and 
$\delta_2$ are small enough:
$$0\ll\epsilon^2\ll\delta_1\ll\delta_2\ll 1.$$
Family $(\Sigma_t)_{0\le t< \epsilon}$ is called the {\it Schiffer variation}
(of the complex structure on $\Sigma$). Informally, this 
construction describes the following modification of the surface. 
We cut a segment 
$$\{ p':\;\;z(p')\in [-\sqrt{t},\sqrt{t}]\subset \R\}\eqno (3.13)$$
from our surface. The resulting surface has the boundary which consists
of two copies of interval $[-\sqrt{t},\sqrt{t}]$. Then the boundary 
is glued with itself in a different manner. More precisely, we glue 
together the sides of the two cuts with the same number $i=1,2,3,4$; 
see the figure below.

%\vskip 5 truemm
%\begin{center}
%\includegraphics[width=.5\textwidth]{Fig12.eps}
%\end{center}
%\vskip 5 truemm

\vskip 5 truemm
\begin{figure}[ht]
\centering
\includegraphics[height=45mm ]{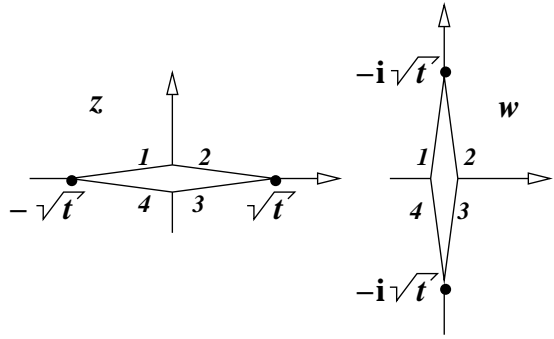}
\caption{Coordinate planes $z$ and $w$ with cuts}
\end{figure}
\vskip 5 truemm

Transformation $w\mapsto \sqrt{w^2+t}$ is the exponential map 
(at time $t$) of the meromorphic vector field 
$$\dot{w}=\frac{1}{2w}\eqno (3.14)$$
in a certain domain in $\C$.

Let us assume that $\Sigma=\Sigma_0$ is a sphere. Let 
$x:\Sigma\to \C P^1$ be a holomorphic parametrisation of $\Sigma$
such that $x(p)=0$ and $({\rm{d}}z/dx)(p)=1$. Denote by 
$q\in \Sigma$ the point corresponding to $\infty\in \C P^1$ 
in coordinate $x$. There exists a unique family 
$x_t:\Sigma_t\to \C P^1$ of holomorphic parametrisations of 
$\Sigma_t$, depending smoothly on $t$ outside of $U_p$ and such that 
$$x_t(q)=\infty,\,\, \left(\frac{d(1/x_t)}{d(1/x)}\right)(q)=1,\,\,
\left(\frac{d}{d(1/x)}\right)^2(1/x_t) (q)=0\,\,.$$

In other words,  near point $q$ we have $x_t=x+O(x^{-1})$.
\medskip

{\bf Lemma 3.5.} {\sl On $\Sigma\setminus U_p$, one has:
$$\frac{\partial x_t}{\partial t} _{|t=0} = -\frac{1}{2x}\,.
\eqno (3.15)$$}
\medskip

{\it Proof} : First, observe that
$$ x_t=x+\frac{c_{-1}(t)}{x}+\frac{c_{-2}(t)}{x^2}+\dots\,.$$
Hence $\left.\diy{\frac{\partial x_t}{\partial t}}\right|_{t=0}$ is a Laurent 
series in $x$ consisting of strictly negative powers of $x$.
For small $t$ function $w=\sqrt{z^2-t}$ is a convergent series 
in non-negative powers of $x_t$:
$$\sqrt{z^2-t}=\sum_{i\ge 0} a_i(t) x_t^i=:f_t(x_t)\,\,.$$
Expanding this identity in $t$ up to $t^1$ we get
$$z-\frac{t}{2z} +O(t^2)=f_0(x)+t\left( 
\left.\frac{\partial f_t}{\partial t}\right|_{t=0}(x)+ 
f_0'(x)\left.\frac{\partial x_t}{\partial t}\right|_{t=0}
\right)+O(t^2)\,\,.$$
Comparing coefficients in front of $t^1$ we conclude that
$\left.\diy{\frac{\partial x_t}{\partial t}}\right|_{t=0}$ 
is the negative power part of the series 
$$-\frac{1}{2 f_0(x) f_0'(x)}= -\frac{1}{2x}+O(x)\,.\quad\Box$$
\medskip

{\bf Remark 3.1.} The Schiffer variation corresponds, up to a 
scalar factor, to the action of the generator $L_2=-z^{-1}{\rm d}
\big/{\rm d}z$ (see (2.29)), in the so-called Virasoro uniformisation 
of moduli spaces. Cf. [BS], [K1].
\medskip

{\bf 3.3.2. Connection with the Schwarzian derivative.} In this 
subsection we continue to work with special neutral four-sphere 
collection $\mF$. Such a collection gives rise to a real number
$$\rho_\mF:=\log\left[v_\mF\left/\left(
\operatornamewithlimits{\otimes}_{1\leq i,j\leq 2} 
v_{S_{ij}}^{\otimes \mu_{ij}}\right)\right.\right]\,.$$
Let us assume that a point $p\in S_{1,L}$ is given, 
together with a germ of local coordinate $z$ at $p$.
Then we can perform Schiffer variations of surfaces $S_1$ and 
$S_2$ and obtain a one-parameter family of  neutral collections 
$\mF_t$. Our goal here is to calculate the first derivative 
$\left.\diy{\frac{\partial \rho_{\mF_t}}{\partial t}}\right|_{t=0}$.

It follows easily from the definitions that the expression
in question coincides with the limit, as $t\to 0$, 
of the value $\diy{\frac{1}{t}}\rho_{\wt{\mF}_t}$. Here
$\wt{\mF}_t$ is a `perturbed' neutral four-sphere collection 
consisting of $S_1$, $S_2$, $S_{1,t}$ and $S_{2,t}$,
with multiplicities $(-1,+1,+1,-1)$.

Let us choose parametrisations $x_1,x_2,x_{1,t},x_{2,t}$ of  
spheres  $S_1$, $S_2$, $S_{1,t}$ and $S_{2,t}$ by $\C P^1$
such that $x_1(p)=x_2(p)=0$ and 
$x_{1,t}=x_1+O(1/x_1)$ at $x_1=\infty$, and a similar condition for $x_{2,t}$.
Moreover, we can assume that $x_1=z+O(z^2),\,x_2=z+O(z^2)$ near $p$.
From subsection 3.3.1, we know that
$x_{i,t}=x_i-\diy{\frac{t}{2x_i}}+O(t^2)$.

The application of the residue formula (3.8) from 
subsection 3.2.2 yields  the following integral
$$\begin{array}{r}\diy{\frac{-1}{24\pi}}
{\rm{Im}}\diy{\int\limits_L} 
\log\left[\diy{\frac{{\rm d}x^{-1}_1}{{\rm d}\left(x_2-\diy{\frac{t}{2 x_2}}
+O(t^2)\right)^{-1}}}\right]\qquad\qquad{}\qquad{}\qquad{}\\
\times{\rm d}\log\left[\diy{\frac{{\rm d}x^{-1}_2}{{\rm d}\left(x_1
-\diy{\frac{t}{2 x_1}}+O(t^2)\right)^{-1}}}\right]\,.
\end{array}\eqno (3.16)$$

A straightforward calculation then shows that the above expression is 
equal to
$$\frac{t}{12}{\rm{Re}}\;{\cS}_f (0) +O(t^2),\eqno (3.17)$$
where function $f$ is defined by $f(x_1)=x_2$ and
its Schwarzian derivative ${\cS}_f$ is given by the standard formula
$${\cS}_f=\frac{f'''}{f'}-\frac{3 (f'')^2}{2(f')^2}\,.
\eqno (3.18)$$

Thus, we have proved the following
\medskip

{\bf Proposition 1.} {\sl In the above notation,
$$\left.\frac{\partial\rho_{\mF_t}}{\partial t}\right|_{t=0}
=\frac{1}{12}{\rm{Re}}\;{\cS}_f(0).\eqno (3.19)$$}

\subsection{The limit formula for degenerating neutral collections}

Let $\Sigma$ be a compact surface with two marked points 
$p_1,p_2$, and $(\Sigma_t)_{t\in [t_0,+\infty)}$
be a one-parameter family of compact surfaces which approach in a 
certain sense the singular surface
$\Sigma_\infty:=\Sigma/\{p_1=p_2\}$, the result of identification 
of points $p_1$ and $p_2$ on $\Sigma$.
More precisely, we assume that for each $t$ an open part 
$U_t\subset \Sigma_t$ is identified
with an open domain $U_t'\subset \Sigma\setminus \{p_1,p_2\}$, and 
$U_t$ is the complement to a closed cylinder in $\Sigma_t$, $U_t'$ 
is the complement to the union of two small closed
$\epsilon(t)$-neighborhoods of points $p_1$ and $p_2$ in a certain 
metric on $\Sigma$, such that $\epsilon(t)\to 0$ as $t\to +\infty$.

First, we will define a determinant line $\DT_{\Sigma_\infty}$ 
and the notion of convergence of points $v_t\in \DT_{\Sigma_t}$ to 
a point in $\DT_{\Sigma_\infty}$ as $t\to +\infty$.
Namely, we define an admissible metric $g$ on $\Sigma_\infty$ as a 
singular metric on $\Sigma$ with divisor $-(p_1+p_2)$.

By definition, there will be a vector $[g]\in \DT_{\Sigma_\infty}$ for 
every admissible metric $g$. For any two admissible metrics $g_1,g_2$ 
we define the ratio of corresponding vectors
by the same formula as in the non-singular case:
$$[g_2]/[g_1]:=\exp\;\big[\rSL (g_1,g_2)\big]\eqno (3.20)$$
where 
$$\rSL (g_1,g_2):=\frac{1}{48 \pi{\rm i}}
\int\limits_{\Sigma \setminus\{p_1,p_2\}} {\rLL}
(g_1,g_2)\,.\eqno (3.21)$$
The cocycle identity for singular metrics
$$\rSL (g_1,g_3)=\rSL (g_1,g_2)
+\rSL (g_2,g_3)\eqno (3.22)$$
again follows from (2.3); the argument here is similar 
to the one used in Lemmas 3.1 and 3.2.

Next, from results in section 3.1 it follows that there is 
a canonical isomorphism
$$i_{\Sigma_\infty}:\DT_\Sigma\otimes
{\big|\det T_{p_1}\Sigma\big|}^{\otimes 1/24}\otimes 
{\big|\det T_{p_2}\Sigma\big|}^{\otimes 1/24}\to \DT_{\Sigma_\infty}\,.
\eqno (3.23)$$

Further, we say that a family of metrics $(g_t)_{t\in [t_0,+\infty)}$ 
on surfaces $\Sigma_t$ is convergent to an admissible
metric $g_\infty$ on $\Sigma_\infty$ if the following holds.
There exists a pair of closed geodesics $L_1$, $L_2$, in metric 
$g_\infty$, surrounding points $p_1, p_2$, such that, on the cylinders 
$C_t\subset \Sigma_t$ bounded by curves $L_1$, $L_2$, metric $g_t$ is 
flat, and both curves $L_1$, $L_2$ are geodesics of length $2\pi$ 
in metric $g_t$.We can also assume that metrics $g_t$ converge 
uniformly to $g_\infty$ on the part of $\Sigma$ lying outside to 
punctured disks bounded by $L_1$ and $L_2$. Indeed, such families 
of metrics exist because of the following result:
\medskip

{\bf Lemma 3.6.} {\sl Given $s\in [0,1)$, set
$$A_s=\{z\in\C:\;\;s<|z|\leq 1\}.$$
Assume that a positive function 
$r(t)$ is given, where
$t>0$, such that $r(t)\to 0$ as $t\to\infty$. Let 
$\phi_t$ be a holomorphic embedding $A_{r(t)}\to A_0$
mapping the boundary circle $|z|=1$ to itself. Denote by  
$g_t$ the pullback by $\phi_t$ of the flat metric $\big|{\rm d}z/z\big|^2$.
Then, as $t\to\infty$, the 
metrics $g_t$ converge, uniformly in the
$C^\infty$ topology on compacts in the punctured disk 
$A_0=$, to metric $\big|{\rm d}z/z\big|^2$.}
\medskip

{\it Proof} : The main part of the proof of Lemma 3.6 
is the following fact 
[SS]. Given $s\in (0,1)$, consider an embedding 
$\phi$: $A_s\to A_0$ such that $|\phi (z)|=1$ for $|z|=1$. Then, as 
$s\to 0$, the image $\phi\big(A_s\big)$ contains the annulus
$\{z\in\C:\;\;(4+o(s))s<|z|<1\}$. The assertion of Lemma 3.6 is
then deduced by means of a straightforward argument using the potential
theory. $\Box$
\medskip

Next, assume that $(g_t)_{t\in [t_0,+\infty)}$ and 
$(g_t')_{t\in [t_0,+\infty)}$ are two families
of metrics converging, respectively, to admissible metrics 
$g_\infty$ and $g_\infty'$ on $\Sigma_\infty$.
Then we have that 
$$\lim_{t\to\infty} [g_t]/[g_t']=[g_\infty]/[g_\infty']\,.\eqno (3.24)$$
It allows us to define a topology near $+\infty$, on the line bundle 
over $[t_0,+\infty]$ with fibers $\DT_{\Sigma_t}$.

Further, we are going to introduce a map
$$dist:[t_0,+\infty)\to {\big|\det T_{p_1}\Sigma\big|} \otimes 
{\big|\det T_{p_2}\Sigma\big|}\eqno (3.25)$$
defined up to a mutliplication by a positive function $f(t)$ such 
that $\lim\limits_{t\to \infty} f(t)=1$.
Let us choose two local coordinates $z_1$ and $z_2$ near points 
$p_1,p_2$. These coordinates
give an identification of each line $\big|\det T_{p_i}\Sigma\big|$,
$i=1,2$, with $\R$. Hence, to define map $dist$, it suffices
to fix a real-valued function of $t$. We choose this function to 
be equal to the conformal parameter of the cylinder on $\Sigma_t$ 
bounded by circles $|z_1|=1$ and
$|z_2|=1$. Here, the conformal parameter of a cylinder $C$ is a  number 
$t\in (0,1)$
such that $C$ is conformally equivalent to $\{z\in \C:\;\;t<|z|<1\}$.
The fact that map $dist$ is defined up to a mutliplication by 
a positive function $f(t)$ such that $\lim_{t\to \infty} f(t)=1$, 
for different choices of pairs of local coordinates $z_1,z_2$, follows
easily from arguments similar to those used earlier in this subsection.

Now assume that $\Sigma$ is a disjoint union of two 
spheres, and that points $p_1,p_2$ belong to 
different components. Then each surface $\Sigma_t,\,t\in[t_0,\infty)$,  
is a  sphere. Therefore, we have a canonical vector 
$v_{\Sigma_t}\in \DT_{\Sigma_t}$, and also a canonical vector 
$v_{\Sigma_\infty}\in \DT_{\Sigma_\infty}\otimes 
{\big|\det T_{p_1}\Sigma\big|^{\otimes 1/24}}\otimes 
{\big|\det T_{p_2}\Sigma\big|^{\otimes 1/24}}$
(the tensor product of the canonical vectors of two connected components).
Our goal in this subsection is to understand the behavior, when 
$t\to \infty$, of vectors
$v_{\Sigma_t}\in \DT_{\Sigma_t}$ in relation to 
$v_{\Sigma_\infty}\in \DT_{\Sigma_\infty}$.

%\vskip 5 truemm
%\begin{center}
%\includegraphics[width=.5\textwidth]{Fig13.eps}
%\end{center}
%\vskip 5 truemm

\vskip 5 truemm
\begin{figure}[ht]
\centering
\includegraphics[height=65mm]{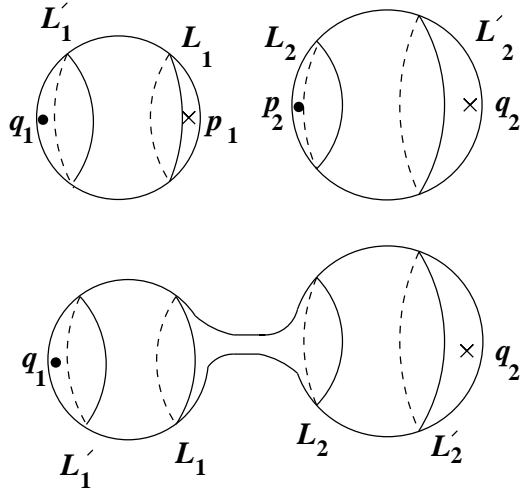}
\caption{Two surfaces, disjoint, and with a small connecting tube}
\end{figure}
\vskip 5 truemm

{\bf Proposition 2.} {\sl In the case where $\Sigma$ is a disjoint 
union of two spheres and points $p_1,p_2$ belong to different 
components as above, one has 
$$\lim_{t\to \infty}\left[v_{\Sigma_\infty}\otimes 
dist(t)^{\otimes -1/24}\right]/v_{\Sigma_t}=1\,.\eqno (3.26)$$
Here we use an identification of lines $\DT_{\Sigma_t}$ with 
$\DT_{\Sigma_\infty}$ compatible with the topology at $t=\infty$ 
introduced above.}
\medskip

{\it Proof} : It is convenient here to use singular metrics with two 
simple poles. We choose two points $q_1,q_2$ on two corresponding 
components of $\Sigma\setminus \{p_1,p_2\}$.
Surface  $\Sigma\setminus \{p_1,p_2,q_1,q_2\}$ is represented as
a disjoint union of two copies of $\C^*=\C\setminus\{0\}$. Thus,
we have on  $\Sigma\setminus \{p_1,p_2,q_1,q_2\}$ a unique 
flat metric $g_\Sigma$ with singularity at divisor 
$-(p_1+p_2+q_1+q_2)$. Similarly, on surface $\Sigma_t, \,t\ge t_0$, 
we have a unique flat metric $g_{\Sigma_t}$ with singularity at
$-(q_1+q_2)$. Also let us choose positive elements $d_i\in 
\big|\det T_{q_i}\Sigma\big|$.

Owing to results in section 3.1, metric $g_\Sigma$, together 
with pair $d_1$, $d_2$, gives a vector 
$\delta_\infty \in \DT_{\Sigma_\infty}$. Also for any $t\in [t_0,+\infty)$ 
metric $g_{\Sigma_t}$ together with pair $d_1,d_2$
yields a vector $\delta_t\in \DT_{\Sigma_t}$. It follows from the 
above definitions that 
$$\lim_{t\to \infty} \delta_t=\delta_\infty\,\,.$$
Now choose positive elements  $d_i'\in\big|\det T_{p_i}\Sigma\big|$. 
They can be represented as closed geodesics $L_i$, $i=1,2$, in metric 
$g_\Sigma$. For large $t$ circles $L_i$ are close to geodesics in metric 
$g_{\Sigma_t}$.

It is easy to see that function $dist(t)$ is equal, 
asymptotically as $t\to +\infty$, to the conformal parameter of the 
cylinder on $\Sigma_t$ bounded by circles $L_1$ and $L_2$, in the 
trivialisation of real  line $\big|\det T_{p_1}\Sigma\big|\otimes 
\big|\det T_{p_2}\Sigma\big|$ given by $d_1'\otimes d_2'$.
Finally, we should compare our vectors with the canonical vectors in 
the determinant line of spheres $\Sigma,\Sigma_1$ and $\Sigma_2$.
This can be done using the following straightforward fact which we 
give without proof:
\medskip

{\bf Lemma 3.7.} {\sl 
Let $d_0$ be a vector from $\big|\det T_0 \C P^1\big|$ and 
$d_\infty$ be a vector from $\big|\det T_\infty \C P^1\big|$. 
Then 
$$v_{\C P^1}\big/(d_0\otimes d_\infty)=
{\rm{const}}\cdot h_{C_{d_0,d_\infty}}^{1/24}.\eqno (3.27)$$
Here $h_{C_{d_0,d_\infty}}$ is
the conformal parameter of the cylinder $C_{d_0,d_\infty}$ bounded 
by two circles corresponding to $d_0$ and $d_\infty$ and ${\rm{const}}>0$
is an absolute constant.}
\medskip

Let $L_i'$,$i=1,2$, denote circles (in metric $g_\Sigma$) surrounding 
points $q_1,q_2$, corresponding to vectors $d_1,d_2$. The assertion 
of Proposition 2 can be restated as follows:
\medskip

{\bf Lemma 3.8.} {\sl Let $h_1$ and $h_2$ be conformal 
parameters of cylinders 
$C_{L_1,L_1'}$ and $C_{L_2,L_2'}$ bounded by pairs of circles
$(L_1,L_1')$ and $(L_2,L_2')$ respectively. Let $h_{\rm{in}}(t)$ be 
the conformal 
parameter of cylinder $C_{L_1,L_2}$ in $\Sigma_t$ bounded by $(L_1,L_2)$, and 
$h_{\rm{out}}(t)$ the conformal paramater
of cylinder $C_{L_1',L_2'}$ in $\Sigma_t$ bounded by $(L_1',L_2')$.
Then one has
$$\lim_{t\to \infty} h_{out}(t)/h_{in}(t)=h_1 h_2\,.\eqno (3.28)$$} 

%\vskip 5 truemm
%\begin{center}
%\includegraphics[width=.7\textwidth]{Fig14.eps}
%\end{center}
%\vskip 5 truemm
%\medskip

\vskip 5 truemm
\begin{figure}[ht]
\centering
\includegraphics[height=30mm]{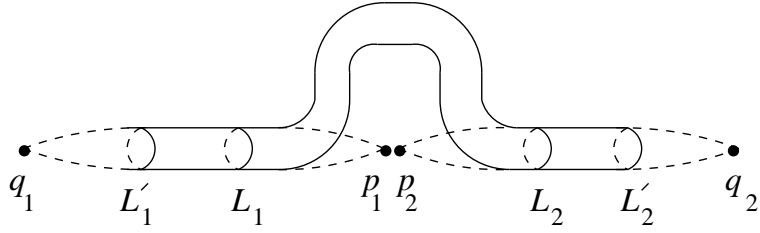}
\caption{Another picture of glued surfaces and of tube $C_{L_1,L_2}$}
\end{figure}
\vskip 5 truemm
\medskip

{\it Proof of Lemma} 3.8: Let $g_{\rm{int}}$ be the unique flat metric with
geodesic boundaries of length $2\pi$ on the cylinder $C_{L_1,L_2}$.
Let us glue two flat cylinders with conformal parameters $h_1$ and $h_2$
to the ends of $C_{L_1,L_2}$. We obtain a flat metric on a cylinder $C'$ 
embedded into $\Sigma_t$ such that the geodesic boundaries 
of $C'$ are close to to lines $L'_1$, $L'_2$. This follows from Lemma 
3.6 and the reflection principle. The conformal parameter of $C'$ will 
be close to that of cylinder $C_{L'_1,L'_2}$, owing to monotonicity
of the conformal parameter with respect to embeddings of 
annuli. By construction, the conformal parameter of $C'$
is equal to $h_1h_2h_{\rm{in}}(t)$.

%\vskip 5 truemm
%\begin{center}
%\includegraphics[width=.6\textwidth]{Fig15.eps}
%\end{center}
%\vskip 5 truemm

\vskip 5 truemm
\begin{figure}[ht]
\centering
\includegraphics[height=50mm]{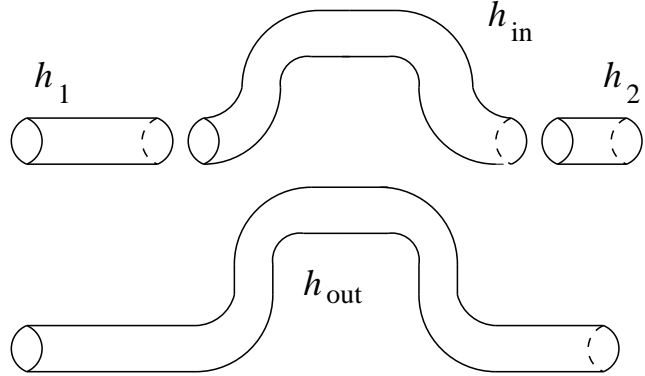}
\caption{Various tubes and their conformal parameters}
\end{figure}
\vskip 5 truemm

\noindent This completes the proof of Lemma 3.8 and that of 
Proposition 2. $\Box$

\section{The SLE-measures, I}

\subsection{Spaces of intervals and associated line bundles}

In chapters 4 and 5 we work with a surface $\Sigma$ with a 
non-empty boundary  $\pSigma\subset\Sigma$, and with a conformal 
structure that 
is smooth everywhere including $\pSigma$, and a pair of distinct 
points $x,y\in \pSigma$. Note that it is not meant that $\Sigma$ 
should be necessarily closed; a working example
of a surface with a boundary is a semi-open rectangle 
$(a,b)\times [c,d]$ where $a<b$ and $c<d$ are real numbers. Here, 
the boundary $\pSigma$ is $(a,b)\times\{c,d\}$. The above conditions 
on surface $\Sigma$ and points $x,y$  are assumed in this and the 
following chapter without stressing them every time again. 

We define an (oriented) {\it interval} ${\cI}$ in $\Sigma$ with 
endpoints $x$ and $y$ as an equivalence class of homeomorphic 
embeddings of the unit segment 
$$\iota:\;[0,1]\hookrightarrow{\Sigma},\;\;\hbox{with
$\iota(0)=x,\,\iota(1)=y$ and $\iota((0,1))\subset \Sigma\setminus\pSigma$,} 
\eqno (4.1)$$
modulo the action of the group of orientation-preserving homeomorphisms 
$[0,1]\to [0,1]$ acting by re-parametrisations. 

The space of intervals in $\Sigma$ with endpoints $x$ and $y$ is denoted by
$Int_{x,y}(\Sigma)$ and is endowed with the topology induced from
$Comp\,(\Sigma)$. Like $Loop\,(\Sigma )$, space $Int_{x,y}(\Sigma)$
is not closed in $Comp\,(\Sigma)$ and not locally compact. 
\medskip

Assume that 
${\cI}\in Int_{x,y}(\Sigma )$ is an interval. First, we introduce 
line $\dt_{{\cI},\Sigma}$. Suppose we are given an open subset 
$U\subset{\Sigma}$ containing ${\cI}$ and such that $U$ as a surface
is of finite type. (A surface
with boundary is called of finite type iff it has finite Betti numbers 
and its boundary has finitely many components. An example is the union 
of an open disk $U=\{z\in\C:\;\;|z|<1\}$ with a finite number 
of open disjoint arcs lying in the 
circle $\{z\in\C:\;\;|z|=1\}$).  
Following (2.14), we set:
$$\dt_{{\cI},\Sigma}=\frac{\dt_{\,{U}\setminus \pSigma}}{
\dt_{\,{U}\setminus ({\cI}\cup\pSigma )}}
:\simeq \dt_{\,{U}\setminus \pSigma}\otimes
\left(\dt_{{U}\setminus ({\cI}\cup\pSigma )}\right)^{\otimes (-1)}.
\eqno (4.2)$$

%\vskip 5 truemm
%\begin{center}
%\includegraphics[width=.45\textwidth]{Fig16.eps}
%\end{center}
%\vskip 5 truemm

\vskip 5 truemm
\begin{figure}[ht]
\centering
\includegraphics[height=45mm]{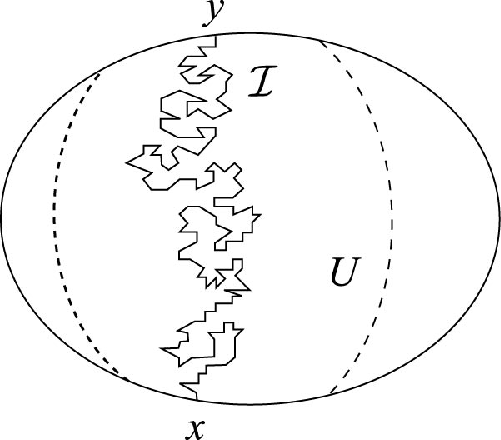}
\caption{An interval and its open neighborhood}
\end{figure}
\vskip 5 truemm

The identification of lines defined as above for different 
subsets $U\supset{\cI}$ is given in the same manner as for the 
case of loops; cf (2.15)--(2.17).
Next, we define the continuous line bundle $\DT_{\Sigma, x,y}$ on 
the space of intervals 
$Int_{x,y}(\Sigma )$, similarly to the analogous bundle for loops.
\medskip

{\bf Remark 4.1.} We would like to warn the reader of a possible caveat.
Namely, one may try to define a determinant line bundle on 
$Int_{x,y}(\Sigma)$ using the following observation. On surface 
$$\Sigma':=\left(\Sigma\setminus \partial\Sigma\right)_{\rm{double}}$$ 
we have involution $\sigma$
that exchanges the copies of $\Sigma$. Obviously, any 
interval ${\cI}\in Int_{x,y}(\Sigma)$ gives a loop ${\cI}'$ 
on $\Sigma'$ invariant
under involution $\sigma$. An alternative approach to the 
definition of the determinant line of ${\cI}$ would be as
$|\det|_{{\cI}',\Sigma'}^{\otimes 1/2}$.
This line is {\it not} isomorphic to our $|\det|_{{\cI},\Sigma}$, the
ratio is certain line bundle on $Int_{x,y}(\Sigma)$  depending on ${\cI}$
is only via the germ of ${\cI}$ near its endpoints. $\blacksquare$ 
\medskip

We will also need another {\it trivial} line bundle 
$\TN_{\Sigma, x,y}$ on $Int_{x,y}(\Sigma )$. The fiber of $\TN_{\Sigma, x,y}$
at any point ${\cI}\in Int_{x,y}(\Sigma)$ is the product
$$|T_x\pSigma |\otimes |T_y\pSigma |\,.\eqno (4.3)$$
Here we use the notation $|V|$, where $V$ is a non-oriented 
one-dimensional real vector space, introduced in subsection 3.1.2.
\medskip

{\bf Definition 4.1.} 
Fix real numbers ${\rm c}$ and ${\rm h}$ and assume that 
for every surface $\Sigma$ 
and pair of points $x,y\in\pSigma$ we are given a measure 
${\mbox{\boldmath${\lambda}$}}_{\Sigma, x,y}$  
on $Int_{x,y}(\Sigma )$ with values in 
$$\TN^{\otimes (-{\rm h})}_{\Sigma ,x,y}\otimes\DT^{\otimes{\rm c}}_{\Sigma ,x,y}.
\eqno (4.4)$$
We say that the (measure-valued) assignment 
$(\Sigma,x,y)\mapsto{\mbox{\boldmath${\lambda}$}}_{\Sigma, x,y}$ is
$({\rm c},{\rm h})$-LCC (or briefly, LCC) if for any 
embedding $\xi:\;\Sigma \hookrightarrow\Sigma'$ we have 
$$\xi^*\big({\mbox{\boldmath${\lambda}$}}_{\Sigma',\xi (x),\xi (y)}\big)
={\mbox{\boldmath${\lambda}$}}_{\Sigma,x,y},\eqno (4.5)$$
where we again use the obvious identification of the line bundles,
associated with $\xi$. $\blacksquare$
\medskip

Now consider a family of values ${\rm c}(\theta)$ and 
${\rm h}(\theta)$ parametrised by $\theta\in (0,1]$:
$${\rm c}=(3-2\theta )\left(3-\frac{2}{\theta}\right),\;\;
{\rm h}=\frac{3/\theta -2}{4}\,.\eqno (4.6)$$
Note that the correspondence between $\theta$ and ${\rm c}$ 
and between $\theta$ and ${\rm h}$ is one-to-one,  
the range for ${\rm c}(\theta)$ is $(-\infty, 1]$ and 
the range for ${\rm h}(\theta)$ is $[1/4,+\infty )$. 
\medskip

{\bf Theorem 1.} {\sl 
For any $0<\theta\leq 1$ there exists a non-zero $({\rm c},{\rm h})$-LCC
assignment
$(\Sigma,x,y)\mapsto{\mbox{\boldmath${\lambda}$}}_{\Sigma,x,y}$.
Here ${\rm c}={\rm c}(\theta )$, and ${\rm h}={\rm h}(\theta )$
are given by {\rm{(4.6)}}.} 
\medskip

We also put forward
\medskip

{\bf Conjecture 2.} {\sl For any $0<\theta\leq 1$,
the LCC assignment in Theorem {\rm 1} is unique, up to a scalar
factor.}
\medskip

Sections 4.2--5.2 aim at the proof of Theorem 1. 
In fact, we will prove that an LCC assignment is generated 
by the chordal SLE$_\kappa$ processes (see section 4.3), 
with $\kappa\in (0,4]$. The key property here is that paths 
produced by the process SLE$_\kappa$ are simple 
Jordan curves precisely for $\kappa\in (0,4]$. The relation between
$\kappa$ and $\theta$ is straightforward: $\kappa =4\theta$. 
As to uniqueness, it can be verified for ${\rm c}=0$;
see Remark 5.2 in section 5.3.

Measures ${\mbox{\boldmath${\lambda}$}}_{\Sigma,x,y}$ will be called
the {\it SLE measures}, in analogy with the Malliavin measures.

In what follows, we use relation (4.6) between $\theta$
and pair $({\rm c},{\rm h})$ without specifying it every time again.

Exponents ${\rm c}$ and ${\rm h}$ come from highest vectors in level 2 
degenerate Virasoro modules, see section 6.3. 

\subsection{Reduction to $\oHb$}

{\bf 4.2.1. Space $Int_{0,\infty}$.} The first (obvious) step 
of the proof of Theorem 1 is that it suffices to construct 
measures ${\mbox{\boldmath${\lambda}$}}_{\Sigma,x,y}$ on 
$Int_{x,y}(\Sigma )$ in the case where $\Sigma$ is
a semi-open rectangle $R_\epsilon=(-\epsilon ,\epsilon )\times [0,1]$, 
with the boundary
$\partial R_\epsilon =(-\epsilon ,\epsilon )\times \{0,1\}$, 
where $x =(0;0)$, and $y=(0;1)$,  
such that property (4.5) holds for all embeddings
$\xi:\;R_\epsilon \hookrightarrow R_\epsilon$, with $\xi (x)=x$, 
$\xi (y)=y$. 
 
%\vskip 5 truemm
%\begin{center}
%\includegraphics[width=.25\textwidth]{Fig17.eps}
%\end{center}
%\vskip 5 truemm

\vskip 5 truemm
\begin{figure}[ht]
\centering
\includegraphics[height=45mm]{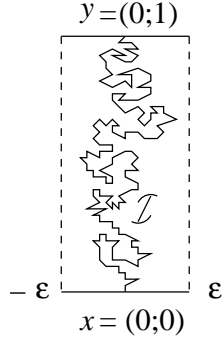}
\caption{An interval in a rectangle}
\end{figure}
\vskip 5 truemm

Next, a semi-open rectangle $R_\epsilon$ can be embedded in a closed
disk identified with the compactified upper half-plane 
${\oHb}$, so that $x$ is taken to $0$ and $y$ to $\infty$. Formally:
$${\oHb}=\Hb\sqcup\R P^1\eqno (4.7)$$
where $\Hb$ is the open upper half-plane and $\R P^1=\partial\oHb$ is the
extended real line
$$\Hb=\{z\in \C:\;\;\hbox{Im}\;z>0\}\,,\;\R P^1=\R\cup\{\infty\}.
\eqno (4.8)$$
Note that with every such embedding we have $R_\epsilon\subset{\overline R}_\epsilon
=\oHb$. Thus, we can associate with $R_\epsilon$  a natural isomorphism 
$$Int_{x,y}(R_\epsilon)\simeq Int_{0,\infty}(\oHb )\,,$$
and the corresponding identification of line bundles
$$\DT_{R_\epsilon,x,y}\simeq\DT_{\oHb,0,\infty},\;\;
\TN_{R_\epsilon,x,y}\simeq\TN_{\oHb,0,\infty}\,.$$ 
The reason is that bundles $\DT_{\Sigma,x,y}$ and 
$\TN_{\Sigma,x,y}$ (for a general surface $\Sigma$)
do not change if we modify $\pSigma$ without
changing neighbourhoods $U_x,U_y\subset\pSigma$ of points $x$ and $y$ in
$\pSigma$ and the interior $\Sigma\setminus\pSigma$. 

Therefore, the assertion of Theorem 1 follows
if, $\forall$ $\theta\in (0,1]$, we construct a measure 
${\mbox{\boldmath${\lambda}$}}_{\oHb,0,\infty}$ on  
$$Int_{0,\infty}:=Int_{0,\infty}({\oHb}),\eqno (4.9)$$

%\vskip 5 truemm
%\begin{center}
%\includegraphics[width=.4\textwidth]{Fig18.eps}
%\end{center}
%\vskip 5 truemm

\vskip 3 truemm
\begin{figure}[ht]
\centering
\includegraphics[height=55mm]{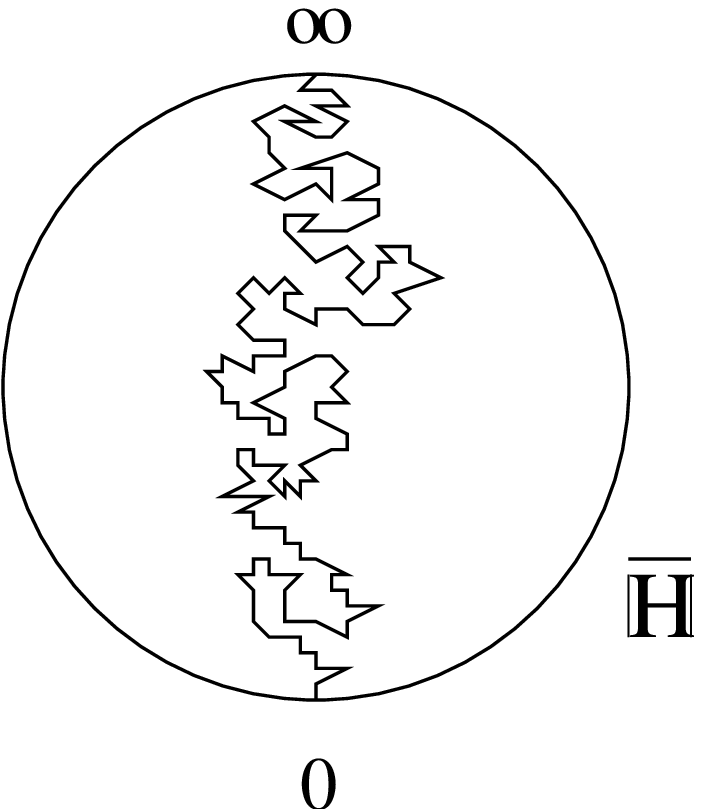}
\caption{Interval in $\oHb$ connecting $0$ and $\infty$}
\end{figure}

\noindent with values in the bundle
$$\left(\TN_{0,\infty,{\oHb}}\right)^{\otimes (-{\rm h})}
\otimes \left(\DT_{{\oHb}}\right)^{\otimes {\rm c}}\,.
\eqno (4.10)$$
such that the property (4.5) holds for any continuous map
$\xi:\;\oHb\to\oHb$ such that $\xi (0)=0$, $\xi(\infty )=\infty$, 
and the restriction $\xi\big|_{\Hb\cup U_0\cup U_\infty}$ is a 
holomorphic embedding, for some open neighbourhoods 
$U_0,U_\infty\subset\C P^1$ of points $0$, $\infty$
in $\C P^1$:  
$$\xi^*\big({\mbox{\boldmath${\lambda}$}}_{\oHb,0,\infty}\big)
={\mbox{\boldmath${\lambda}$}}_{\oHb,0,\infty}.$$
\medskip

{\bf 4.2.2. Trivialisations of line bundles on $Int_{0,\infty}$.}
Group $\R^\times_{>0}=Aut(\oHb,0,\infty)$ acts by dilations on
${\oHb}$ and hence on $Int_{0,\infty}$ and line bundles
$\TN_{{\oHb},0,\infty}$ and $\DT_{{\oHb},o,\infty}$. We construct a 
$\R^\times_{>0}$-equivariant trivialisation of both these 
bundles. 
By the definition of determinant line, we have a canonical
isomorphism
$$\dt_{{\cI},\oHb}\simeq \dt_{\Hb}\left/ \dt_{\Hb\setminus {\cI}}\right.\,\,\,.$$
Observe that for any interval ${\cI}\in Int_{0,\infty}({\oHb})$
the complement ${\Hb}\setminus {\cI}$ is isomorphic
to the disjoint union $\left({\Hb}\setminus {\cI}\right)^{\rm{left}}\sqcup
\left({\Hb}\setminus {\cI}\right)^{\rm{right}}$
of two copies of an open disk. Therefore the ratio of canonical vectors
gives a trivialisation 
$$v_{\cI}^{\rm{det}}:=v_{\Hb}\otimes\left(
v_{\left({\Hb}\setminus {\cI}\right)^{\rm{left}}}\otimes
v_{\left({\Hb}\setminus {\cI}\right)^{\rm{right}}}\right)^{\otimes(-1)}$$
of bundle $\DT_\oHb$, obviously invariant under $\R^\times_{>0}$ action.

Next, the tensor product of the unit tangent vector at $x=0$ to 
$\C P^1$ and its image under inversion at $y=\infty$ is a vector 
$$v^{\rm{tan}}\in \TN_{0,\infty,{\oHb}}$$ 
invariant under the action of $\R^\times_{>0}$.
Further, any $\R^\times_{>0}$-invariant measure 
${\mbox{\boldmath${\lambda}$}}_{\oHb ,0,\infty}$ on 
$Int_{0,\infty}$ with values in bundle (4.10)
gives an ordinary (scalar)\\ $\R^\times_{>0}$-invariant 
measure ${\mbox{\boldmath${\nu}$}}^{{\rm c},{\rm h}}$ on 
$Int_{0,\infty}$, after
division by the following section of this line bundle:
$${\cI}\in Int\,(\oHb )\mapsto 
(v^{\rm{tan}})^{\otimes (-{\rm h})}
\otimes (v_{{\cI}}^{\rm{det}})^{\otimes{\rm c}}.$$
The last measure should satisfy a certain condition,
called the restriction covariance property and discussed below.
\medskip

{\bf 4.2.3. Restriction covariance property for measures on $Int_{0,\infty}$.}
Let $\alpha:\;\Hb\hookrightarrow\Hb$
be an embedding of the open half-plane 
into itself, which extends by continuity to
a continuous map $\oHb\hookrightarrow\oHb$, denoted again by 
$\alpha$, such that
${\alpha}(0)=0$, ${\alpha}(\infty )=\infty$, and 
${\alpha}$ can be continued to a holomorphic map 
to $\C P^1$ near points $0$ 
and $\infty$. Then $\alpha$ induces an open embedding 
$$\alpha_*:\;Int_{0,\infty}\hookrightarrow
Int_{0,\infty}\,.\eqno (4.11)$$

%\vskip 5 truemm
%\begin{center}
%\includegraphics[width=.43\textwidth]{Fig19.eps}
%\end{center}
%\vskip 5 truemm

\vskip 5 truemm
\begin{figure}[ht]
\centering
\includegraphics[height=55mm]{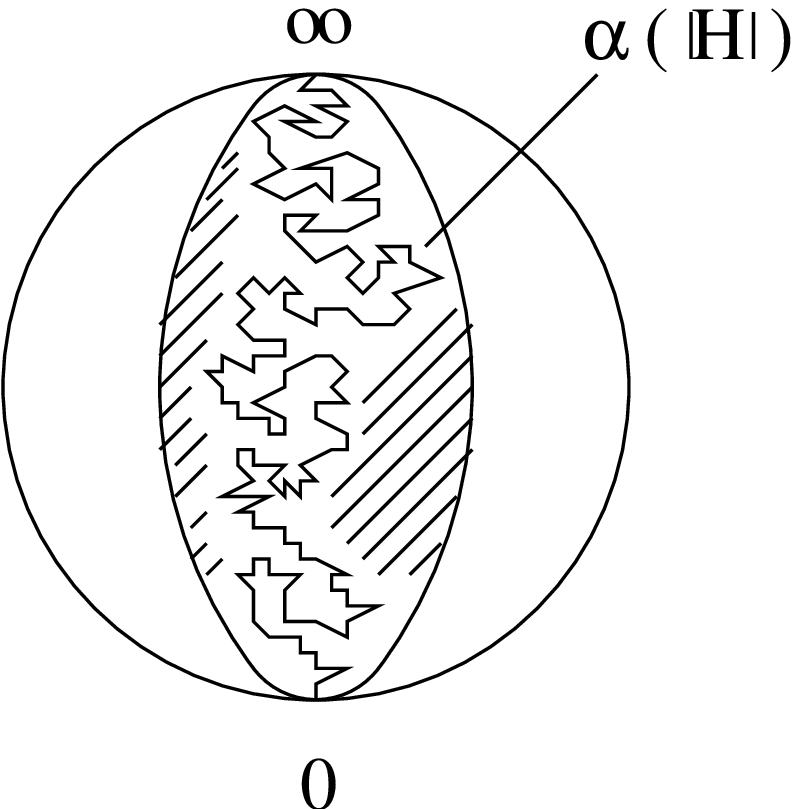}
\caption{Interval in $\alpha(\Hb)\subset \Hb$}
\end{figure}
\vskip 5 truemm

Given $\alpha$ as above, there are defined a positive 
constant, $q^{\rm{tan}}_\alpha$, and
a positive continuous function, $q^{\rm{det}}_\alpha ({\cI} )$, 
${{\cI}}\in Int_{0,\infty}$. 
Constant $q^{\rm{tan}}_\alpha$ is given by the product 
$$q^{\rm{tan}}_\alpha=q^{\rm{tan}}_{\alpha,0}\;q^{\rm{tan}}_{\alpha,\infty}.
\eqno (4.12)$$
Here numbers $q^{\rm{tan}}_{\alpha,0}$, $q^{\rm{tan}}_{\alpha,\infty}$ $>0$ 
are determined from the Taylor expansions at $0$ and $\infty$:
$$\alpha(z)=q^{\rm{tan}}_{\alpha,0}\;z+O(z^2),\;\;z\to 0;\;\;\;
\frac{1}{\alpha (z)}=q^{\rm{tan}}_{\alpha,\infty}\;\frac{1}{z}+
O\left(\frac{1}{z^2}\right),\;\;z\to \infty .\eqno (4.13)$$

Next, function $q_\alpha^{\rm{det}}({{\cI}})$ on $Int_{0,\infty}$ 
is defined as follows. Given ${\cI}\in Int_{0,\infty}$ and map 
$\alpha$, we construct a neutral collection $\mS_{\alpha,{\cI}}$
consisting of six spheres $\Sigma_1$, $\Sigma_2$, $\Sigma_3$, 
$\Sigma_4$, $\Sigma_5$ and $\Sigma_6$ that are the doubles of 
six open disks $D_1$, $D_2$, $D_3$, $D_4$, $D_5$ and  
$D_6$, correspondingly. Namely, these disks will be 
$$\alpha(\Hb),\alpha (\left({\Hb}\setminus {\cI}\right)_{\rm{L}}),
\alpha (\left({\Hb}\setminus {\cI}\right)_{\rm{R}}),\Hb,
\left({\Hb}\setminus \alpha ({\cI})\right)_{\rm{L}},
\left({\Hb}\setminus \alpha ({\cI})\right)_{\rm{R}}\eqno (4.14)$$
taken with weights $\mu_1=+1$, $\mu_2=-1$, $\mu_3=-1$, 
$\mu_4=-1$, $\mu_5=+1$ and $\mu_6=+1$. As before, subscripts
${\rm L}$ stand for left and ${\rm R}$ for right.

The non-Hausdorff surface $S^{\rm{nH}}$ containing these six
spheres is the union of the doubles of disks $D_1,\dots,D_6$ 
glued all along the domain that is the pullback to the double covering
of the union of two thin strips on the left and on the right of 
${\cI}\subset\oHb$ (see Figure 20).

%\vskip 5 truemm
%\begin{center}
%\includegraphics[width=.9\textwidth]{Fig20.eps}
%\end{center}
%\vskip 5 truemm

\vskip 5 truemm
\begin{figure}[ht]
\centering
\includegraphics[height=80mm]{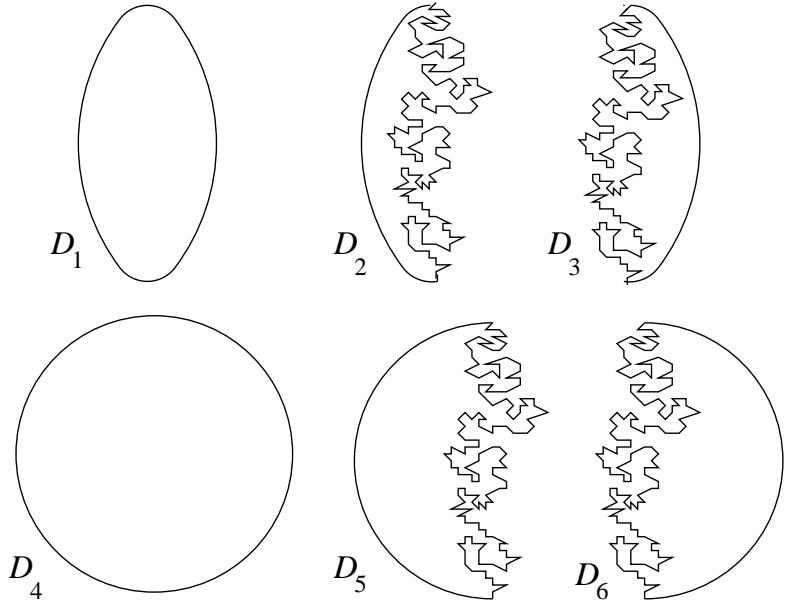}
\caption{Six disks $D_1,\dots,D_6$}
\end{figure}
\vskip 5 truemm

Schematically, one can draw surface $S^{\rm{nH}}$ for collection
$\mS_{\alpha,{\cI}}$ as drawn on Figure 21. 

%\vskip 5 truemm
%\begin{center}
%\includegraphics[width=.7\textwidth]{Fig21.eps}
%\end{center}
%\vskip 5 truemm

\vskip 2mm
\begin{figure}[ht]
\centering
\includegraphics[height=45mm]{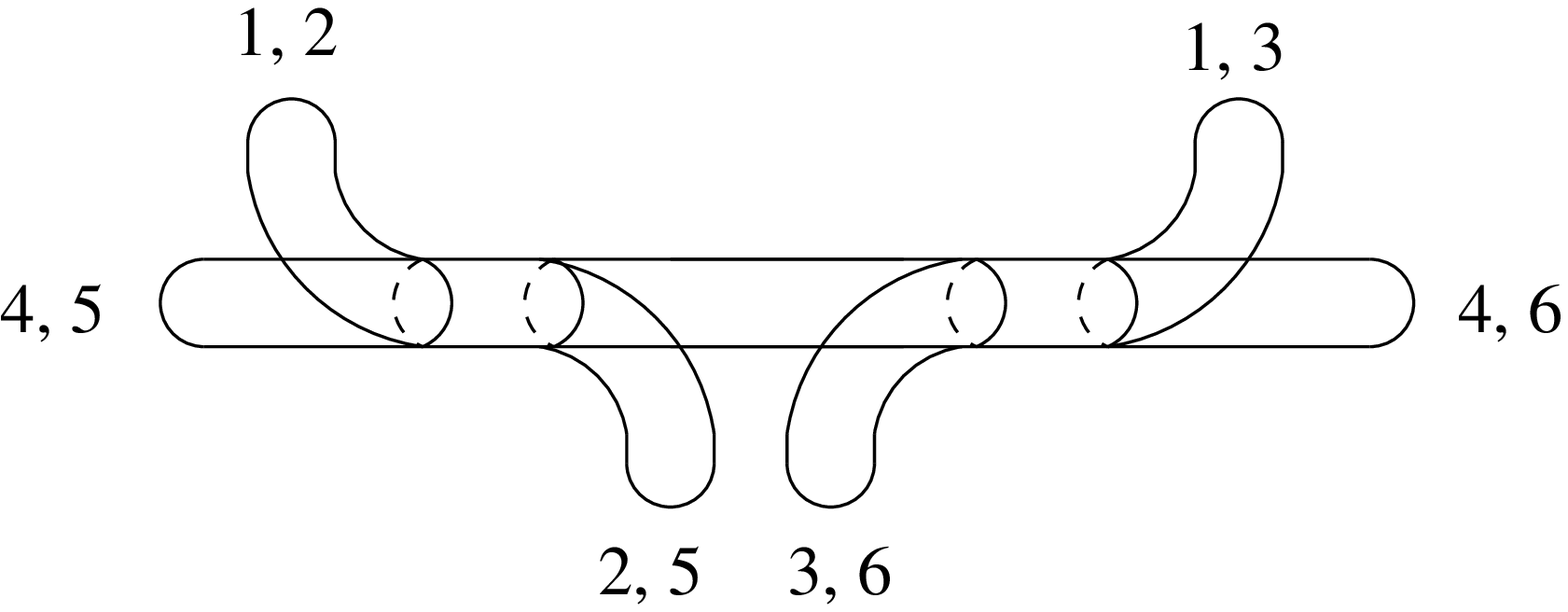}
\caption{Non-Hausdorff surface $S^{nH}$}
\end{figure}
\vskip 2mm
 Here sphere $S_i$ connects two half-spherical 
caps which have value $i$ among the pair of indeces attached to
them. (So, sphere $S_4$ is the horizontal one.)

Value $q_\alpha^{\rm{det}}({\cI})$ is then defined as follows:
$$q_\alpha^{\rm{det}}({\cI})=\left(v_{\mS_{\alpha,{{\cI}}}}\left/
\operatornamewithlimits{\otimes}_{k=1}^6v_{S_k}^{\otimes \mu_k}
\right.\right)^{1/2}.\eqno (4.15)$$ 
\medskip

{\bf Definition 4.2.} We call a (scalar) measure ${\mbox{\boldmath{$\nu$}}}$ 
on $Int_{0,\infty}$ $({\rm c},{\rm h})$-{\it restriction covariant},
or briefly, restriction covariant (RC) if, for any embedding
$\alpha:\;\Hb\hookrightarrow\Hb$ 
as above, the pullback $\alpha^*\big({\mbox{\boldmath${\nu}$}}
\big|_{\alpha_*(Int_{0,\infty})}\big)$ of the restriction 
${\mbox{\boldmath${\nu}$}}\big|_{\alpha_*(Int_{0,\infty})}$
of measure ${\mbox{\boldmath${\nu}$}}$ to the image  
$\alpha_*(Int_{0,\infty})$ (which is an open subset 
in $Int_{0,\infty}$) is absolutely continuous with respect to 
${\mbox{\boldmath${\nu}$}}$ and has the Radon-Nikodym 
derivative
$$\frac{{\rm d}\big[\alpha^*\big(
{\mbox{\boldmath${\nu}$}}\big)\big]\big|_{\alpha_*(Int_{0,\infty})}}{
{\rm d}{\mbox{\boldmath${\nu}$}}}\;({\cI} )=
\left(q^{\rm{tan}}_\alpha\right)^{\rm h}\left(q^{\rm{det}}_\alpha ({\cI} )
\right)^{\rm c},\;\;{\cI}\in Int_{0,\infty}.\quad\blacksquare\eqno (4.16)$$

\medskip

By definition, measure ${\mbox{\boldmath${\nu}$}}^{{\rm c},{\rm h}}$ 
identified in subsection 4.2.2 is $({\rm c},{\rm h})$-RC. 
Summarising the arguments produced in section 4.2, we 
obtain the following lemma
\medskip

{\bf Lemma 4.1.} {\sl There is a
one-to-one correspondence between LCC assignments  
$(\Sigma,x,y)\mapsto{\mbox{\boldmath${\lambda}$}}_{\Sigma,x,y}$  
and scalar RC measures 
${\mbox{\boldmath${\nu}$}}^{{\rm c},{\rm h}}$ on $Int_{0,\infty}$ 
invariant under the  
antiholomorphic involution $\sigma_{\,\oHb}:z\mapsto -\overline{z}$.}
\medskip

Invariance of ${\mbox{\boldmath${\nu}$}}^{{\rm c},{\rm h}}$
under $\sigma_{\,\oHb}$ (also valid by definition of this measure 
in subsection 4.2.2)
is needed here for independence of
${\mbox{\boldmath${\lambda}$}}_{\Sigma,x,y}$ of the orientation 
of $\Sigma$ near interval ${\cI}\subset \Sigma$.
 
Note that the assertion of Lemma 4.1 remains correct regardless
of condition $\theta\in (0,1]$. However, we need this condition
in the course of constructing and RC measure 
${\mbox{\boldmath${\nu}$}}^{{\rm c},{\rm h}}$.

\subsection{A reminder on SLE processes}

{\bf 4.3.1. The space of hulls and the canonical time parametrisation.}
Here we follow works [Sc1], [LSW] and their sequel,  where a 
one-parameter family of random processes SLE$_\kappa$, 
$0<\kappa<+\infty$ was 
introduced and investigated in great detail. For recent reviews 
of progress in this direction, see [Sc2], [W2], [W5] and the 
bibliography therein.

Define a {\it hull} as a closed subset $\cK\subset \Hb$ with a 
contractible complement
$\Hb\setminus \cK$ and such that $\infty$ does not lie in the closure 
of $\cK$ in $\oHb$.
\medskip 
  
{\bf Remark 4.2.} Our definition of a hull slightly differs from the 
standard one, see the aforementioned references.
In the standard definition, a hull is the closure 
of a hull in our sense, in $\oHb$. The advantage
of our definition is that there is a canonical one-to-one 
correspondence between hulls and certain holomorphic mappings, see below.
$\blacksquare$ 
\medskip 
 
For every hull $\cK$ there exists a unique uniformisation 
of its complement $\Hb\setminus \cK$. It is a bijection 
$$\gamma_\cK:\Hb\setminus{\cK}\simeq\Hb,
\eqno (4.17)$$ 
admitting a holomorphic extension to a neighborhood of 
point $\infty\in\C P^1$ (which, for simplicity, we 
denote by the same symbol $\gamma_\cK$) such that
$$\gamma_\cK({\bar z})=\overline{\gamma_\cK(z)},\;\gamma_\cK(\infty)
=\infty,\;\gamma_\cK (z)=z+o (1)\;
\hbox{as $|z|\to\infty$, $\;z\in \Hb$.}$$

The space $Hull$ of hulls is endowed with the following Hausdorff 
separable topology (and the associated Borel structure).
A sequence of hulls $\cK_n$ is convergent to $\cK$ iff (i) all
Taylor coefficients of $1/\gamma_{\cK_n}(z)$ at $z=\infty$ 
converge to those of $1/\gamma_{\cK}(z)$, and (ii) $\exists$ a neighbourhood $U_\infty$ 
of point $\infty$ such that $\cK\cap U_\infty=\emptyset$ $\forall$ $n$. 

We introduce a continuous function $Time: Hull\to [0,+\infty)$ by
$$Time(\cK)=2 \gamma^{(-1)}_\cK\eqno (4.18)$$
where $\gamma^{(-1)}_\cK$ is the first non-trivial 
coefficient\footnote{In [W4], number 
$Time(\cK)/2=\gamma_\cK^{(-1)}$ is called the {\it capacity} of hull $\cK$
(from infinity).} 
of the Taylor expansion of $\gamma_\cK$ at $z=\infty$ (i.e.,
the coefficient in front of $1/z$):
$$\gamma_\cK (z)=z+\frac{\gamma^{(-1)}_\cK}{z}+\dots .\eqno (4.19)$$
The inequality $\gamma^{(-1)}_\cK\geq 0$ (in fact, $\gamma^{(-1)}_\emptyset =0$ 
and $\gamma^{(-1)}_\cK > 0$ for $\cK\neq\emptyset$) is well
known in the theory of conformal embeddings. See, e.g., [W2]. Function
$Time$ defines a foliation of space $Hull$ into its level sets $Time^{-1}(t)$,
which we repeatedly use below.

For a given real-valued continuous function $w =(w_s)_{s\geq 0}$, 
taking $s\in [0,+\infty)$ to $w_s\in\R$, with $ w_0=0$, there exists 
a unique solution $g_t(z)(=g_t(z;( w_s)))$ of the Loewner equation
$$\frac {\partial g_t(z)}{\partial t}= \frac{2}{g_t(z)- w_t}\,,
\;\;t>0,\;\;z\in \Hb,\eqno (4.20)$$
with the initial condition 
$$g_0(z)=z,\;\;z\in \Hb.\eqno (4.21)$$ 
This solution determines a family of hulls 
$\cK_t\left(=\cK_t\big((w_s)_{s\geq 0}\big)\right)$,
with $\cK_0=\emptyset$, via the identification
$$g_t(z)=\gamma_{\cK_t}(z),\;\;z\in\Hb.\eqno (4.22)$$ 
In what follows, we repeatedly use identification (4.22), 
without stressing it every time again. It follows immediately 
from the Loewner equation that
$$Time(\cK_t)=t.\eqno (4.23)$$

Furthermore, with respect to the above topology on $Hull$, for any 
given real-valued continuous function $( w_s)$ such that $w_0=0$, 
the solution $g_t(z)$ of the Loewner equation determines a continuous 
path $(\cK_t)_{t\geq 0}$ in $Hull$, with $\cK_0=\emptyset$. We will 
call $(\cK_t)_{t\geq 0}$ a path (or a trajectory) 
driven by $w=(w_s)_{s\geq 0}$. On the other hand, $w$ is called a driving 
function (for path $(\cK_t)$). 

The {\it chordal process} SLE$_\kappa$ is the (Borel) probability measure 
on continuous paths $(\cK_t)_{t\geq 0}$ in $Hull$, with $\cK_0=\emptyset$,
generated by the standard Brownian motion
$(B_s)_{s\geq 0}$ with diffusion coefficient $\kappa>0$, 
by means of the above construction (i.e., via the random function 
$g_t(z,(B_s))$ emerging via (4.20)-- (4.22). 
We denote this probability measure by ${\mbox{\boldmath${\mu}$}}^\kappa$.
In short, SLE$_\kappa$ is a random path $({\mbox{\boldmath${\cK}$}}_t
)_{t\geq 0}$ in $Hull$ driven by Brownian motion
$(B_s)$ with diffusion coefficient $\kappa>0$: ${\mbox{\boldmath${\cK}$}}_t
={\mbox{\boldmath${\cK}$}}_t\big((B_s)\big)$. 
The scaling property of the Brownian motion
implies the scale covariance of process SLE$_\kappa$. Namely, 
$\forall$ $\lambda>0$ the dilation of time $t\mapsto \lambda t$
corresponds to the dilation of the hull $\cK_t\mapsto{\sqrt\lambda}
\cK_t$:
$$({\mbox{\boldmath${\cK}$}}_{\lambda t})
\sim ({\sqrt\lambda} {\mbox{\boldmath${\cK}$}}_t).\eqno (4.24)$$
Formally, it means that two probability measures obtained
from ${\mbox{\boldmath${\mu}$}}^\kappa$ 
by the above dilations, coincide. 

It is convenient to slightly generalise the above set-up
and introduce a Borel subset 
${\wh{Hull}}$ of the Cartesian product 
$Hull\times \R$ whose points are pairs $(\cK,x)$, 
or, equivalently, $(\gamma_\cK,x)$, such that
$$\hbox{either }\;(\cK,x)=(\emptyset, 0)\;\hbox{ or }\;
{\overline\cK}\cap\partial\oHb=\{0\}\;\hbox{and}\;\gamma^{-1}_\cK(x)\in
\partial{\overline\cK}.\eqno (4.25)$$
Here and below, $\gamma^{-1}_\cK(x)$ stands for the 
embedding $\Hb\hookrightarrow\Hb\setminus{\cK}$,
inverse to $\gamma_{\cK}$.

The reason for introducing ${\wh{Hull}}$ 
is that if we start the SLE$_{\kappa}$ process 
at a point from ${\wh{Hull}}$, it stays in ${\wh{Hull}}$. 
More precisely, given $(\cK ,x)\in{\wh{Hull}}$, 
for any real-valued continuous function $w =( w_s)_{s\geq 0}$ with 
$w_0=x$, we can define a path 
$\big(\cK_t, w_t\big)_{t\geq 0}$ in $\wh{Hull}$, with $\cK_0=\cK$.
Namely, we set $\gamma_{\cK_t}(z)=g_t(z,\cK)$ where 
$g_t(z,\cK)$ satisfies Loewner equation (4.20) driven by $( w_s)$,
with the initial condition 
$$g_0(z,\cK)=\gamma_\cK(z),\;\;z\in\Hb\setminus\cK,\eqno (4.26)$$ 
instead of (4.21). We again call
$\big(\cK_t, w_t\big)$ a path driven by $ (w_s)$, and starting from
$(\cK,x)$. From the above 
definitions (and independence of increments in Brownian motion)
it follows that SLE$_{\kappa}$ generates a time-homogeneous
Markov process on ${\wh{Hull}}$. Namely, the process starting from
point $(\cK,x)$ 
is represented by a random path $({\mbox{\boldmath${\cK}$}}_t,B_t+x)$
driven by the shifted Brownian motion $(B_s+x)_{s\geq 0}$. 

We will call the above Markov process on ${\wh{Hull}}$ an
{\it extended} SLE$_\kappa$ process. Correspondingly, 
${\wh{Hull}}$ is called the {\it extended phase space} of
the extended SLE$_{\kappa}$ process.

We will also denote by $Time$ the pullback of the time 
function from $Hull$ to ${\wh{Hull}}$.

The infinitesimal generator of process  SLE$_{\kappa}$ in coordinate 
$(\gamma_\cK, x)$ on ${\wh{Hull}}$ is given by
$$\frac{\kappa}{2}\left(\frac{\partial}{\partial x}\right)^2
+ \frac{2}{\gamma_\cK- x}\;\frac{\delta}{\delta\gamma_\cK}\eqno (4.27)$$
where vector field $\diy{\frac{2}{\gamma_\cK- x}\frac{\delta}{\delta 
\gamma_\cK}}$ is defined by
$$\begin{cases}\dot{x}=0,\\
\dot{\gamma_\cK}=\diy{\frac{2}{\gamma_\cK- x}}.\end{cases}\eqno (4.28)$$
\medskip
    
{\bf Definition 4.3.} There is a convenient algebra $\bA$ of measurable
functions on $\widehat{Hull}$ (separating all points)
consisting of polynomials in $ w$ and all non-trivial
Taylor coefficients $\gamma_\cK^{(-1)}$, $\gamma_\cK^{(-2)}$, $\ldots$. 
We endow $\bA$ with a graduation by associating weights
$${\rm{weight}}(x)=1,\,\,{\rm{weight}}(\gamma_\cK^{(-n)})=n
\mbox{ for }n\ge 1$$
to its generators. Algebra $\bA$
has a natural exhaustive increasing filtration
by finite-dimensional linear subspaces 
$\bA_0\subset\bA_1\subset
\dots\subset \bA$, where $\bA_n$
consists of linear combinations of monomials of weight $\le n$.
$\blacksquare$ 
\medskip

It is easy to see that the generator of the extended 
SLE$_{\kappa}$ process preserves
finite-dimensional spaces $\bA_n$, hence the action 
of the evolution operator on $\bA$ is well-defined.
\medskip
    
{\bf 4.3.2. Hulls and intervals for $\kappa\le 4$.}
From now on we assume that $0<\kappa\leq 4$. The reason is that,
as was shown in [RS], if $\kappa\in (0,4]$ (and only if this 
condition holds), then with ${\mbox{\boldmath${\mu}$}}^\kappa$-probability $1$ the path $(\cK_t)$ 
of the SLE$_\kappa$ process satisfies the following property.  
Sets $\cK_t\cup \{0\}$, $t>0$, are intervals embedded in $\oHb$
and increasing with $t$: $\cK_{t_1}\subset\cK_{t_2}$ for $0<t_1<t_2$. 
Next, the `tip' of the interval $\cK_t$ approaches point $\infty$, 
in the limit $t\to+\infty$, 
By continuity, process SLE$_\kappa$, with $0<\kappa\leq 4$, gives rise to a 
probability measure on $Int_{0,\infty}$ which we denote by  
${\mbox{\boldmath${\mu}$}}^\kappa_\infty$. Like before, we can 
associate this probability measure
with a random interval ${\mbox{\boldmath${\cI}$}}$ in $Int_{0,\infty}$.
Scaling covariance of 
SLE$_{\kappa}$ (see (4.24)) implies a similar property
of ${\mbox{\boldmath${\cI}$}}$ in $Int_{0,\infty}$. 

To analyse properties of probability measure 
${\mbox{\boldmath${\mu}$}}^\kappa_\infty$ on
$Int_{0,\infty}$, it is convenient to introduce the space
$SInt$ of finite {\it semi-intervals} (in $\oHb$).
A finite semi-interval is denoted by ${\cJ}$ and is defined
an equivalence class of homeomorphic embeddings of the unit segment 
$$\iota:\;[0,1]\hookrightarrow{\oHb},\;\;\hbox{with
$\iota(0)=0$ and $\iota((0,1])\subset\Hb$,} 
\eqno (4.29)$$
modulo the action of the group of orientation-preserving 
homeomorphisms $[0,1]\to [0,1]$ preserving point $0$.
Obviously, a finite semi-interval is a particular case of a 
hull; viewed in 
this way, $SInt$ is a Borel subset in $Comp ({\oHb})$,
and we consider it as a topological space, with the induced
topology. However, $SInt$ is not closed
and not locally compact.

Note that $SInt$ can also be naturally identified with 
a subspace of 
${\widehat{Hull}}$. The reason is that the real number 
$x$ giving the second entry of the coordinate 
$(\cK,x)$ in ${\widehat{Hull}}$ 
can be uniquely determined from the first entry, $\cK$ 
(which is, in general, a hull, but under condition $0<\theta\leq 
1$, a semi-interval). In fact, if ${\cJ}\in SInt$ is a semi-interval
and ${\cJ}=\iota ([0,1])$, then 
$$x(=x({\cJ}))=\gamma_{\iota (0,1]}(\iota (1)).\eqno (4.30)$$

At the same time, the union $\{0\}\sqcup SInt$
can be treated as the path space of the SLE$_\kappa$ process. More
precisely, semi-intervals ${\cJ}\in\sqcup SInt$
can be parametrised by means of function
$Time$ and will then represent `stopped trajectories' SLE$_\kappa$.
This picture can be extended to intervals ${\cI}\in Int_{0,\infty}$:
points of such an interval will be parametrised by $[0,+\infty]$.
Furthermore, for any ${\cI}\in Int_{0,\infty}$ of the form
${\cI}=\iota ([0,1])$, the map $[0,1]\to [0,+\infty ]$ given by
$$\tau\mapsto Time\;(\iota ([0,\tau]))\eqno (4.31)$$
is a homeomorphism. The function $t(\tau)=Time (\iota ([0,\tau]))$ 
provides a convenient {\it canonical} parametrisation 
of the interval ${\cI}$ by $[0,+\infty]$. As a result, we associate
with  ${\mbox{\boldmath${\mu}$}}^\kappa_\infty$  
a family of a probability measures ${\mbox{\boldmath${\mu}$}}^\kappa_t$  
on the level set $Time^{-1}(t)\subset SInt\subset {\wh{Hull}}$, where
$$ Time^{-1}(t)=\{{\cK}:\;\;Time ({\cK})=t\},\;\;0<t<\infty.\eqno (4.32)$$

\section{The SLE-measures, II}

\subsection{The restriction martingale}

Let $\alpha:\Hb\hookrightarrow\Hb$ be
an embedding, as in subsection 4.2.3, such that
$q^{\rm{tan}}_{\alpha,\infty}=1$. 
We associate with $\alpha$ an open embedding
$$\alpha_*: {\wh{Hull}}\hookrightarrow {\wh{Hull}}$$
in the following fashion: $\forall$ $({\wt\cK},{\wt x})\in{\wh{Hull}}$, 
$$\alpha_*({\wt\cK},{\wt x})=({\cK}, {x}) .\eqno (5.1)$$
Here ${\cK}$ is the closure $\overline{\alpha({\wt\cK} )}$ 
of the image of ${\wt\cK}$ under $\alpha$ in $\Hb$. 

Next, in order to determine ${x}\in\R$, we introduce the (partially 
defined holomorphic) mapping $h(=h_{{{\wt\cK}},{\cK}})$: 
$\Hb\hookrightarrow\Hb$, by
$$h=\gamma_{{\wt\cK}}\circ\alpha^{-1}\circ\gamma^{-1}_{\cK}.
\eqno (5.2)$$
It is easy to see that both $h$ and the inverse mapping 
$h^{-1}=\gamma_{\cK}
\circ\alpha\circ\gamma^{-1}_{{\wt\cK}}$ can be extended
continuously to an invertible real analytic map (with strictly 
positive derivative) in a neighbourhood 
of ${\wt x}\in\R$, $\forall$ $({\wt\cK},{\wt x})\in{\wh{Hull}}$. 
We then define ${x}$ in (5.1) by 
$${x}=h^{-1}({\wt x}).\eqno (5.3)$$ 

Therefore, we obtain two coordinate systems, 
$({\wt\cK},{\wt x})$ and $({\cK}, {x})$, on ${\wh{Hull}}$,
related by (5.1). In what follows we will treat $h$ as a function 
on ${\wh{Hull}}$ with values in (partially defined) holomorphic 
mappings $\Hb\hookrightarrow \Hb$. 

Embedding $\alpha_*$ generates (by restriction) a similar 
embedding $SInt\hookrightarrow SInt$.

Let us introduce a new random process SLE$_{\kappa,\alpha}$ whose phase 
space is the same space $\{0\}\sqcup SInt\subset{\wh{Hull}}$ 
as for the original process SLE$_\kappa$. The time function $Time^\alpha$ 
for SLE$_{\kappa,\alpha}$ is equal to $Time\circ \alpha_*$. We then 
introduce process SLE$_{\kappa,\alpha}$ as the result of the time redefinition
(from $Time$ to $Time^\alpha $) of process SLE$_\kappa$.
  
So, $\forall$ $t\in (0,+\infty)$ we have two probability measures
on the level set $Time^{-1}(t)$ (see (4.32)).
The first measure, ${\mbox{\boldmath${\mu}$}}_t
\big(={\mbox{\boldmath${\mu}$}}^\kappa_t\big)$, 
is generated by the process SLE$_\kappa$.
The second measure, ${\mbox{\boldmath${\mu}$}}_{t,\alpha}
\big(={\mbox{\boldmath${\mu}$}}^\kappa_{t,\alpha}\big)$, 
is the pushforward of the measure 
generated by SLE$_{\kappa,\alpha}$ under map $\alpha_*$. 
  
We associate with pair $(\alpha ,{\cK})$, where ${\cK}=\overline
{\alpha ({\wt{\cK}})}$ for some ${\wt{\cK}}\in Hull$,   
the  neutral collection $\mF_{\alpha,{\cK}}$
consisting of four spheres $S_1$, $S_2$, $S_3$, $S_4$, identified as
the doubles of four open disks
$$\alpha(\Hb),\,\,\alpha (\Hb)\setminus{\cK},\,\, \Hb, \,\,
\Hb\setminus{\cK}\eqno (5.4)$$
taken with weights $+1,-1,-1,+1$.
The glueing of these spheres is defined similarly to that in 
subsection 2.5.1. 

On ${\wh{Hull}}$ we consider the function $r^{\rm{det}}_\alpha$ which is 
defined, in coordinate $({\cK},{x})$, by
$$r^{\rm{det}}_\alpha({\cK})=\left(
v_{\mF_{\alpha,{\cK}}}\left/\operatornamewithlimits{\otimes}_{k=1}^4
v_{S_k}^{\otimes w_k}\right)^{1/2}\right.,
\eqno (5.5)$$
and depends on hull ${\cK}$ but not on ${x}$. 
The relation of this function to function $q^{\rm{det}}_\alpha$ 
defined in (4.15) on $Int_{0,\infty}$ will be explained in section 
5.2 (see Eqn (5.24)).

Next, set:
$$r({\cK}, {x})=\left(h'({x})\alpha'(0)\right)^{\rm h}
\left[r^{\rm{det}}_\alpha ({\cK})\right]^{\rm c},\eqno (5.6)$$
where $({\cK}, {x})\in\alpha_*({\wh{Hull}})$, and parameters ${\rm h,c }$ are defined
 by (4.6) with $\theta:=\kappa/4$.
\medskip

{\bf Theorem 2.} {\sl $\forall$ $t>0$,
measure ${\mbox{\boldmath${\mu}$}}_{t,\alpha} $ is 
absolutely continuous with
respect to ${\mbox{\boldmath${\mu}$}}_t$. The Radon-Nikodym 
derivative 
$$r_t:=\frac{{\rm d}{\mbox{\boldmath${\mu}$}}_{t,\alpha}}{
{\rm d}{\mbox{\boldmath${\mu}$}}_t}$$ 
coincides with the restriction to $Time^{-1}(t)$ 
of function $r$ defined in Eqn {\rm{(5.6)}}.
Moreover, the extension of function $\alpha_*r$ by $0$
outside the image $\alpha_*({\wh{Hull}})$ gives 
a martingale for process SLE$_\kappa$.}
\medskip

{\it Proof} : The proof of Theorem 2 is based on 
Propositions 3 and 4 below. Here we perform a series
of formal calculations with second order differential
operators on ${\wh{Hull}}$ related to the generators
of processes SLE$_\kappa$ and SLE$_\kappa^\alpha$. 
These can be converted into assertions 
about processes in the same way as in [W2], [W3].
(The fact that the SLE$_\kappa$-process is specified by
its generator on ${\wh{Hull}}$ is helpful here.)

Consider a positive function $H$ on ${\wh{Hull}}$ given,
in coordinate $({\cK},{x})$, by
$$H({\cK},{x})=h'({x})=\frac{\partial{\wt x}}{\partial{x}}.
\eqno (5.7)$$
Further, for a function $F$ on ${\wh{Hull}}$
(like $\;r$, $r^{-1}$, $H^2$, and so on), we denote by the same symbol 
$F$ the operator of multiplication by $F$. 

We claim that 
\medskip

{\bf Proposition 3.} {\sl The following operator identity holds true:
$$\begin{array}{l}r^{-1} \circ \left(\diy{\frac{\kappa}{2}} 
\left(\diy{\frac{\partial}{\partial{x}}}\right)^2+ 
\diy{\frac{2}{\gamma_{\cK}- {x}}}\;
\diy{\frac{\delta}{\delta \gamma_{\cK}}}\right)\circ r\\ 
\qquad\qquad{}= H^2\circ
\left(\diy{\frac{\kappa}{2}}\left(\diy{\frac{\partial}{\partial 
{\wt x}}}\right)^2+\diy{\frac{2}{\gamma_{{\wt\cK}}-{\wt x}}}\;\diy{\frac{\delta
}{\delta\gamma_{{\wt\cK}}}}\right).\end{array}\eqno (5.8)$$}
\medskip

Proposition 3  guarantees that $r$ determines a 
positive local martingale, and hence a semi-martingale, 
for SLE$_\kappa$.

The RHS of (5.8) gives the generator of SLE$_\kappa^\alpha$, 
as follows from Proposition 4: 
\medskip

{\bf Proposition 4.} {\sl In the above notation, one has the following 
functional identity on ${\wh{Hull}}$: 
$$\left(\frac{2}{\gamma_{{\wt\cK}}-{\wt x}}\;\frac{\delta}{
\delta\gamma_{{\wt\cK}}}\right)\big(Time^\alpha\big)
=\frac{1}{H^2}\,.\eqno (5.9)$$}
\medskip

The assertion of Theorem 2 then follows from Propositions 3 
and 4, by applying Girsanov's formula and the fact that
$r({\wt\cK},x)\to r(\emptyset,0)=1$ as pair $({\wt\cK},x)$ approaches
$(\emptyset,0)$ in the topology on ${\wh{Hull}}$.
 \medskip

{\it Proof of Proposition} 3. The first summand in the LHS of (5.8) is
the following operator:
$$r^{-1}\circ\left[\frac{\kappa}{2}\left(\frac{\partial}{\partial{x}}
\right)^2\right]\circ r=\frac{\kappa}{2}\circ H^{-{\rm h}}\circ
\left(\frac{\partial}{\partial{x}}\right)^2\circ H^{{\rm h}}.
\eqno (5.10)$$
The reason is that the other factors figuring in the formula for $r$ (see
(5.6)) do not depend on ${x}$. In what follows we will
denote by $H'$ and $H''$ the result of application of 
operators $\diy{\frac{\partial}{\partial{x}}}$ and
$\left(\diy{\frac{\partial}{\partial{x}}}\right)^2$ to 
function $H$. Then for the RHS of (5.10) we have the formula
$$\begin{array}{l}
\diy{\frac{\kappa}{2}}\circ H^{-{\rm h}}\circ
\left(\diy{\frac{\partial}{\partial{x}}}\right)^2\circ H^{{\rm h}}\\
=2\theta \left[\left(\diy{\frac{\partial}{\partial{x}}}\right)^2+
2{\rm h}\left(\diy{\frac{H'}{H}}\right)
\circ\diy{\frac{\partial}{\partial{x}}}+
{\rm h}({\rm h}-1)\left(\diy{\frac{H'}{H}}\right)^2
+{\rm h}\left(\diy{\frac{H''}{H}}\right)\right]
.\end{array}\eqno (5.11)$$

For the second summand in the LHS of (5.8) we have the
following operator representation
$$r^{-1}\circ\left(\frac{2}{\gamma_{\cK}-{x}}\;\frac{\delta
}{\delta\gamma_{\cK}}\right)
\circ r 
=\frac{2}{\gamma_{\cK}-{x}}\;\frac{\delta}{\delta\gamma_{\cK}}
+\left[\frac{2}{\gamma_{\cK}-{x}}\;\frac{\delta}{\delta\gamma_{\cK}}
\right](\log\;r ).\eqno (5.12)$$

Our next goal is to calculate the zero degree term 
in the RHS of (5.12). To this end, we consider the 
vector field 
$${V}=\diy{\frac{2}{\gamma_{\cK}-{x}}\;
\frac{\delta}{\delta{\gamma_{\cK}}}}.\eqno (5.13)$$ 
and calculate the action of field ${V}$ on 
the function
$$(\log\;r)({\cK},{x})={\rm h}\log\;\alpha'(0)
+{\rm h}\log\;h'({x})+{\rm c}\log\;r^{\rm{det}}_\alpha
({\cK}).\eqno (5.14)$$
Observe that the first summand in the RHS of (5.14) is constant;
hence we can discard it in future calculations.
\medskip

{\bf Lemma 5.1.} {\sl Vector field ${V}$ acts on 
function $h=h_{{\cK}}$ as
$${\dot h}(z)=h'({x})^2\frac{2}{h(z)-h({x})}-
h'(z)\frac{2}{z-{x}}.\eqno (5.15)$$}
\medskip

{\it Proof of Lemma} 5.1. We have the following identity (cf (5.2)):
$$h\circ\gamma_{\cK}=\gamma_{\cK}\circ\alpha^{-1}.$$ 
Next, applying field ${V}$, we obtain the identity
$${\dot h}\circ\gamma_{\cK} +\frac{2}{\gamma_{\cK}- 
{x}}\cdot h'\circ 
\gamma_{\cK}= \dot{\gamma_{{\wt\cK}}}\circ \alpha^{-1}.$$
It is clear, geometrically, that $\dot{\gamma_{{\wt\cK}}}$ is proportional
to $\diy{\frac{2}{\gamma_{{\wt\cK}}-{\wt x}}}$, with 
${\wt x}=h({x})$, as we perform a 
Schiffer variation here. Hence, we obtain, for given $({\cK},{x})$,
that
$${\dot h}(z)+\frac{2}{z-{x}}h'(z)={\rm{const}}\frac{2}{h(z)
-h({x})}.$$
The proportionality coefficient 
is equal to $h'({x})^2$ as follows from the condition that 
$h(z)$ is non-singular at $z={x}$. This completes the proof
of Lemma 5.1. $\Box$
\medskip

{\bf Lemma 5.2.} {\sl Vector field ${V}$ acts on function $\log\;H$:
$({\cK},{x})\mapsto\log\;h'({x})$
as follows:
$${V}(\log\;H)=-\frac{4}{3}\frac{H''}{H}+\frac{1}{2}\left(\frac{H'}{H}
\right)^2.$$}
\medskip

{\it Proof of Lemma} 5.2. Expand $h$ 
near point ${x}$:
$$h(z)=a_0+a_1(z-{x})+a_2(z-{x})^2+a_3(z-{x})^3+\ldots$$
where coefficients $a_i$ are functions on ${\wh{Hull}}$: 
$$a_0={\wt x},\;a_1=H,\;a_2=\frac{H'}{2},\;a_3=\frac{H''}{6},\ldots .$$
By Lemma 5.1, we have:
$$\begin{array}{cl}{\dot h}(z)&=\diy{\frac{2h'({x})^2}{h(z)-h({x})}-
\frac{2h'(z)}{z-{x}}}\\
\;&=\diy{\frac{2a_1^2}{a_1(z-{x})+a_2(z-{x})^2+a_3(z-{x})^3
+\ldots}}\\ \;&\;\\
\;&\qquad\qquad{}-\diy{\frac{2[a_1+2a_2(z-{x})
+3a_3(z-{x})^2+\ldots ]}{z-{x}}}.\end{array} \eqno (5.16)$$
The coefficient at $(z-{x})$ in the RHS equals
$${\dot H}=-2a_3+\frac{2a_2^2}{a_1}-6a_3=-\frac{4}{3}H''+\frac{1}{2}
\frac{(H')^2}{H}.$$
This completes the proof of Lemma 5.2. $\Box$ 
\medskip

{\bf Lemma 5.3.} {\sl Vector field ${V}$ 
acts on function $\log\;q^{\rm{det}}_\alpha$ as follows:
$$\big[V(\log\;r^{\rm{det}}_\alpha )\big]({\cK},x)
=-\frac{1}{6}{\cS}_h({x})
=-\frac{1}{6}\left(\frac{H''}{H}-\frac{3(H')^2}{2H^2}\right).
\eqno (5.17)$$}  
\medskip

{\it Proof of Lemma} 5.3: follows immediately from Proposition 1
in subsection 3.3.2.
$\Box$
\medskip
 
We now can calculate the LHS of (5.8):
$$\begin{array}{l}r^{-1} \circ \left(\diy{\frac{\kappa}{2}} 
\left(\diy{\frac{\partial}{\partial{x}}}\right)^2+ 
\diy{\frac{2}{\gamma_{\cK}- {x}}}\;
\diy{\frac{\delta}{\delta \gamma_{\cK}}}\right)\circ r\\ 
=2\theta\left(\diy{\frac{\partial}{\partial{x}}}\right)^2+
4{\rm h}\theta\left(\diy{\frac{H'}{H}}\right)\circ\diy{\frac{\partial
}{\partial{x}}}+{V}.\end{array}$$

The next task is to calculate the RHS in (5.8) in coordinate
$({\cK},{x})$. The first summand is 
calculated by using the 
functional identity $\diy{\frac{\partial}{\partial{\wt x}}=
H^{-1}\frac{\partial}{\partial{x}}}$:
$$H^2\circ\frac{\kappa}{2}\left(\frac{\partial}{\partial{\wt x}}\right)^2=
\frac{\kappa}{2}\left[\left(\frac{\partial}{
\partial{x}}\right)^2-\frac{H'}{H}\frac{\partial}{\partial{x}}
\right].\eqno (5.18)$$
\medskip

{\bf Lemma 5.4.} {\sl Vector field 
$${\wt V}=\diy{\frac{2}{\gamma_{{\wt\cK}}-{\wt x}}\;
\frac{\delta}{\delta{\gamma_{\wt\cK}}}}$$ 
acts in coordinate $({\cK},{x})$ as
$$H^{-2}\circ {V}+3\frac{H'}{H^3}\frac{\partial}{\partial{x}}.
\eqno (5.19)$$}
\medskip

{\it Proof of Lemma} 5.4. An argument similar to that in
the proof of Lemma 5.2, shows that 
$${\wt V}=H^{-2}{V}+\Phi\frac{\partial}{\partial{x}},$$
where $\Phi$ is a function on ${\wh{Hull}}$. This function $\Phi$ is calculated
by using the identity ${\wt V}({\wt x})={\wt V}(h({x}))=0$. 
This identity yields that 
$$\left[H^{-2}{V}(h)\right](z)\Big|_{z={x}}
+(\Phi H)(z)\Big|_{z={x}}=0.$$
The value $[{V}(h)](z)\big|_{z={x}}$ is the zeroth
coefficient in the RHS of (5.16):
$$[{V}(h)](z)\big|_{z={x}}=-2a_2-4a_2=-3H'.$$
This proves Lemma 5.4. $\Box$
\medskip

Combining Eqns (5.10)--(5.19), we obtain the assertion of 
Proposition 3. $\Box$
\medskip

{\it Proof of Proposition} 4. Taking into account 
the above facts, the proof is concise. We have to calculate 
${\wt V}(Time^\alpha)$. The result follows directly from Lemma 5.4
as $\diy{\frac{\partial}{\partial{x}}}(Time^\alpha )=0$
and ${V}(Time^\alpha )=1$. This concludes the proof of 
Proposition 4. $\Box$
\medskip

Theorem 2 has now been proved. $\Box$
\medskip

{\bf Remark 5.1.} In [W3], Werner consctructed a local martingale
for the SLE$_\kappa$ process given by the formula
$$r({\cK}_t,{w}_t)=\big[h_t({w}_t)\big]^{\rm h}
\exp\;\left[\frac{\rm c}{6}\int\limits_0^t{\cS}_{h_s}({w}_s)
{\rm d}s\right],\eqno (5.20)$$
where $({\cK}_s,{w}_s)_{s\geq 0}$ is a path of the 
SLE$_\kappa$ process starting at $(\emptyset,0)$ and $h_s$
stands for the mapping $h$ associated with $({\cK}_s,{w}_s)$.
It follows from Lemma 5.3 that (5.20) coincides with 
$r({\cK}_t,{w}_t)$, modulo the constant factor 
$\alpha'(0)^{\rm h}$. An advantage of our formula (5.6)
is that it refers to the final point of the path 
$({\cK}_s,{w}_s)$, at $s=t$. $\blacksquare$ 

\subsection{End of proof of Theorem 1}

To complete the proof of Theorem 1, it remains
to check that measures ${\mbox{\boldmath${\mu}$}}^\kappa_\infty$ 
on $Int_{0,\infty}$
have the RC property; see Lemma 4.1. We will check this 
property in the special case where embedding $\alpha$
is such that the closure $\overline{\alpha (\Hb)}$ of
$\alpha (\Hb)$ in $\oHb$
contains either $[0,+\infty]$ or $[-\infty,0]$. The general 
case will follow by composition of two embeddings with
the above property.

Set $\cA=\Hb\setminus\alpha (\Hb)$; it is a hull touching
$\partial\oHb$ either strictly to the left or strictly to the 
right of $0$. 
\medskip

{\bf Proposition 5.} {\sl For $\mu^\kappa_\infty$-almost every trajectory
$({\wt\cK}_s,{\wt x}_s)_{s\geq 0}$ 
avoiding $\cA$, and the associate trajectory
$({\cK}_t,x_t)_{t\geq 0}$, where 
$$({\cK}_t,x_t)=
\alpha_*({\wt\cK}_s,{\wt x}_s)\;\hbox{ and }\;t=Time^\alpha (
{\wt\cK}_s,{\wt x}_s),\eqno (5.21)$$ 
the Radon-Nikodym derivative $r_t({\cK}_t,{x}_t)$ (cf. Theorem {\rm 2})
has a limit as $t\to \infty$ (and $s\to\infty$). Namely,
$$\lim_{t\to \infty} r_t({\cK}_t,{x}_t)=
\left(q_\alpha^{\rm{tan}}\right)^{\rm h} \left(q_\alpha^{\rm{det}}\right)^{\rm c} 
({\wt\cK}_{\infty}).\eqno (5.22)$$
Here we use the fact that 
$${\wt\cK}_\infty=\lim_{s\to\infty}{\wt\cK}_s\eqno (5.23)$$ 
is an element of $Int_{0,\infty}$ (see subsection {\rm{3.3.2}}),
and constant $q_\alpha^{\rm{tan}}$ and function $q_\alpha^{\rm{det}}$ are 
defined in {\rm{(4.12)}} and {\rm{(4.15)}}, respectively.}
\medskip

{\it Proof of Proposition} 5. By definition, 
$q_\alpha^{\rm{tan}}({\wt\cK}_\infty)$ 
coincides with $h'_0( 0)=\diy{\frac{1}{\alpha'(0)}}$.
\medskip

{\bf Lemma 5.5.} {\sl In the assumptions of the Proposition 
{\rm 1} one has
$$\lim_{t\to \infty} h'_t({x}_t)=1.\eqno (5.24)$$}
\medskip

{\it Proof of Lemma} 5.5. As was mentioned in Remark 4.2, 
with probability one the trajectory $({\wt\cK}_s,{\wt x}_s)$ is
a growing family of truncations ${\cI}_s$ of an interval
${\cI}\in Int_{0,\infty}$. It is easy to see that for any 
interval ${\cI}\in Int_{0,\infty}$ 
avoiding $\cA$ there exists
a function $b (s)>0$, $s>0$, such that the uniformisation
$$b (s)\gamma_{{\cI}_s}:\;\Hb\setminus{\cI}_s\to
\Hb\eqno (5.25)$$
has the following properties. (i) $b (s)\gamma_{{\cI}_s}$ 
maps $\infty$ to $\infty$ and the tip of 
${\cI}_s$ to $0$, and (ii) $b (s)\gamma_{{\cI}_s}$ maps
hull $\cA$ to a domain $\cA_s$ lying in 
an $\epsilon_s$-neighborhood of point $1\in \oHb$ or
point $-1\in \oHb$, depending on the position
of $\cA$, where $\lim\limits_{s\to \infty} \epsilon_s=0$. 

Obviously, the uniformising coordinate $w(z)$ on $\Hb\setminus\cA_s$, 
normalised so as
$w(z)=z+O(1)$ and $w(0)=0$, approaches, together with
its first derivative $w'(z)$, to the standard coordinate on 
$\Hb$ at $z=0$ as $s\to \infty$. This proves Lemma 5.5.
$\Box$

Finally, by Proposition 2 from section 3.4, we have
$$\lim_{t\to \infty} r^{\rm{det}}_\alpha({\cK}_t)
=q^{\rm{det}}_\alpha({\wt\cK}_\infty)\,.\eqno (5.26)$$
As was mentioned earlier,
Eqn (5.26) establishes the relation between (4.15) and (5.5).

Thus the limit (5.21) is established. This completes the
proof of Proposition 5.    $\Box$
\medskip

Now we are ready to finish the proof of Theorem 1.
\medskip
   
{\bf Proposition 6.} {\sl Consider the probability measure 
${\mbox{\boldmath${\mu}$}}^\kappa_\infty$ 
on $Int_{0,\infty}$ generated by process SLE$_\kappa$. Then 
${\mbox{\boldmath${\mu}$}}^\kappa_\infty$ is 
${\rm{(c,h)}}$-RC.}
\medskip
   
{\it Proof of Proposition} 6.  Proposition 5 implies that
measure ${\mbox{\boldmath${\mu}$}}^\kappa_\infty$ is invariant 
under any embedding $\alpha$
such that $q^{\rm{tan}}_{\alpha,\infty}=1$, in the notation from 
subsection 3.2.2. The invariance of ${\mbox{\boldmath${\mu}$}}^\kappa_\infty$ 
under dilations follows from the scaling covariance property 
of SLE$_\kappa$; see (4.24). The assertion of Proposition 6 then follows.     
$\Box$
\medskip

The invariance of measure ${\mbox{\boldmath${\mu}$}}^\kappa_\infty$ 
under the complex 
conjugation $z\mapsto -{\bar z}$ is obvious. This completes the 
proof of Theorem 1.  $\Box$

\subsection{Concluding remarks}

\medskip

{\bf Remark 5.2.} The first remark is that for ${\rm c}=0$,
the assignment 
$(\Sigma ,x,y)\mapsto{\mbox{\boldmath${\lambda}$}}_{\Sigma ,x,y}$  
is unique, up to a scalar factor. This can be verified by using
an argument similar to that from [W4]. $\blacksquare$ 
\medskip

{\bf Remark 5.3.} One can show that 
the assignment 
$(\Sigma ,x,y)\mapsto{\mbox{\boldmath${\lambda}$}}_{\Sigma ,x,y}$  
constructed in sections 4.1--4.5 is covariant under the exchange 
$x\leftrightarrow y $ of the endpoints.
It follows from the time reversal symmetry of SLE$_{\kappa}$ process
established in [W1]. $\blacksquare$ 
\medskip
 
{\bf Remark 5.4.} By using our construction of the LCC assignment\\
$(\Sigma ,x,y)\mapsto{\mbox{\boldmath${\lambda}$}}_{\Sigma ,x,y}$,
we can define a multi-interval assignment
$$(\Sigma ,\ux,\uy)
\mapsto{\mbox{\boldmath${\lambda}$}}_{\Sigma ,\ux,\uy}\eqno (5.27)$$
satisfying the corresponding LCC property: for any embedding 
$\xi:\;\Sigma \hookrightarrow\Sigma'$,
$$\xi^*\big({\mbox{\boldmath${\lambda}$}}_{\Sigma',\xi (\ux),\xi (\uy)}\big)
={\mbox{\boldmath${\lambda}$}}_{\Sigma,\ux,\uy},\eqno (5.28)$$
Here, $\ux$ and $\uy$ are two disjoint ordered 
collections of distinct 
points from $\pSigma$ and $\xi (\ux)$ and $\xi (\ux)$ are their images 
under $\xi$:  
$$\begin{array}{c}\ux=(x_1,\ldots ,x_n),\;\;\uy=(y_1,\ldots ,y_n),\\
\xi(\ux )=\big(\xi (x_1),\ldots ,\xi(x_n)\big),\;\;
\xi (\uy )=\big(\xi (y_1),\ldots ,\xi (y_n)\big).\end{array}\eqno (5.29)$$
Further, measure ${\mbox{\boldmath${\lambda}$}}_{\Sigma ,\ux,\uy}$
is supported by $n$-tuples of disjoint intervals\\ $({\cI}_1,\ldots ,{\cI}_n)
\in\diy{\operatornamewithlimits{\times}_{i=1}^n}Int_{x_i,y_i}(\Sigma )$, with 
values in the tensor product
$$\operatornamewithlimits{\otimes}_{i=1}^n
\left(\TN^{\otimes (-{\rm h})}_{\Sigma ,x_i,y_i}
\otimes\DT^{\otimes{\rm c}}_{\Sigma ,x_i,y_i}\right).\eqno (5.30)$$
of corresponding bundles (4.4). 

Namely, the set of $n$-tuples of disjoint intervals 
$({\cI}_1,\ldots ,{\cI}_n)$ is an open subset, $Int^{rm{disj}}_{\ux,\uy}(\Sigma )\subset
\diy{\operatornamewithlimits{\times}_{i=1}^n}Int_{x_i,y_i}(\Sigma )$, and 
${\mbox{\boldmath${\lambda}$}}_{\Sigma ,\ux,\uy}$
is the restriction of the product-measure 
$\diy{\operatornamewithlimits{\times}_{i=1}^n}
{\mbox{\boldmath${\lambda}$}}_{\Sigma ,x_i,y_i}$ on $Int^{\rm{disj}}_{\ux,uy}(\Sigma )$.
If set $Int_{\ux,\uy}(\Sigma )$ is non-empty then assignment
(5.27) is non-zero.

Again, in the case ${\rm c}=0$, it is possible to check that
such an assignment is unique, up to a scalar factor. However, for
a general ${\rm c}=0$ the uniqueness of the LCC assignment remains open.
$\blacksquare$ 
\medskip

{\bf Remark 5.5.} Now consider assignments
$(\Sigma ,x,y)\mapsto{\mbox{\boldmath${\lambda}$}}_{\Sigma ,x,y}$  
where one of the two endpoints lies
strictly in the interior $\Sigma\setminus \pSigma$. 
The line bundle where the measure should take its  
values is modified for the corresponding endpoint.
Suppose for definiteness that $x\in\Sigma\setminus\pSigma$
and $y\in\pSigma$. Then we replace, in (4.3), (4.4), the 
factor $\left(|T_x\pSigma |\right)^{\otimes {(-{\rm h})}}$ by
$$\big|\det T_x\Sigma\big|^{\otimes (-{\rm h})}
=\left(|\wedge^2 T_x\Sigma |\right)^{\otimes {(-{\rm h})}}\,.\eqno (5.31)$$
Constructions from sections 4.1--4.3 can be extended to cover 
this case, but instead of chordal, one will have to use 
radial SLE$_{\kappa}$ processes; see [BF], [LSW], [W2]. 
Again it will yield an LCC assignment, which for ${\rm c}=0$ 
is unique up to a scalar factor.

In the case where two endpoints lie in the interior 
$\Sigma\setminus\pSigma$, the question of existence 
and uniqueness of an LCC assignment remains open. $\blacksquare$  
\medskip

{\bf Remark 5.6.} It is possible to define LCC assignments
$\Sigma\mapsto{\mbox{\boldmath${\lambda}$}}_{\Sigma, {\rm{free}}}$
on spaces of intervals 
$\diy{\operatornamewithlimits{\cup}_{x,y\in\pSigma:\;x\neq y}}   
Int_{x,y}(\Sigma )$ with non-fixed endpoints. 
The measure ${\mbox{\boldmath${\lambda}$}}_{\Sigma, {\rm{free}}}$  
will take values in the line bundle with a fiber at point ${\cI}
\in Int_{x,y}(\Sigma )$ equal to 
$$\big(|T_x\pSigma|\otimes|T_t\pSigma|\big)^{\otimes (1-{\rm h})}
\otimes \dt_{{\cI},\Sigma}^{\otimes{\rm c}}.\eqno (5.32)$$
The reason is that there is a canonical (`tautological') measure
${\mbox{\boldmath${\tau}$}}_{\pSigma}$ 
on $\pSigma$ with values in $|T\pSigma |$. Measure 
${\mbox{\boldmath${\lambda}$}}_{\Sigma, {\rm{free}}}$ is the
product of measure
$${\mbox{\boldmath${\tau}$}}_{\pSigma}\times
{\mbox{\boldmath${\tau}$}}_{\pSigma}\;\hbox{ on }\;\big(\pSigma\times
\pSigma )\setminus {\rm{diag}}\big(\pSigma\times\pSigma\big)$$ 
and the family of measures 
$${\mbox{\boldmath${\lambda}$}}_{\Sigma, x,y}\;\hbox{ on }\;
Int_{x,y}(\Sigma ),\;\hbox{ where }\;x,y\in\pSigma,\;x\neq y.
\quad \blacksquare $$

\section{Applications to statistical physics}

This section follows some parts of a talk given by one of us
 at the Arbeitstagung (Bonn, 2003), see [K2].

\subsection{Phase boundaries}

It is believed that the conformal field theory (CFT) helps to
describe a large-scale behaviour of lattice
models near phase transition points. In particular, the
CFT (and its massive perturbations by relevant fields) are
credited with predictions of asymptotics of correlators 
of local observables. However, there is a different part 
of the picture, not reduced directly to local observables
and related to statistics of phase boundaries. See, e.g., [C1]. 
 
The basic example here is the two-dimensional Ising model on
the square lattice, with the zero magnetic field and at a 
temperature $T=T_{\rm{crit}}-\delta T$ with small $\delta T
>0$. Here, in the thermodynamic limit we 
will have with probability $1/2$ the `sea' of 
spins $+1$ with `islands' of spins $-1$, or vice versa.
Typical `large' islands will have 
size $\simeq (\delta T)^{-\mu}$ for some critical
exponent $\mu>0$. Inside islands of, say spins $-1$,
the system is `confused' about the 
global phase, and one expects that there will be yet smaller 
`second-order' islands of spins $+1$, etc.
Passing to the limit $\delta T\to 0$ and rescaling 
simultaneously the distance on $\R^2\supset \Z^2$ by factor 
$(\delta T)^{\mu}$, one obtains, hypothetically,  a 
random collection of closed pairwise disjoint Jordan
curves on $\R^2$, called phase boundaries (or domain walls). 
This collection is, with probability $1$, everywhere dense, 
but there will be `very few' curves of a large size $\gg 1$. 
Furthermore, there will be  many curves 
of size $\simeq 1$ covering a positive part of 
the total area. 

This picture is {\it not} conformally invariant and should be
associated in general with a massive perturbation of a CFT 
with two vacua.

Next, consider the behavior of 
phase boundaries at small distances, i.e. rescale 
again the distance in $\R^2$. By general heuristic 
arguments, one can show that the limiting
distribution of collections of phase boundaries is not degenerate, 
i.e. there are many curves of size (diameter) 
$\simeq 1$, and the distribution is now {\it scale 
invariant}. One can also expect that this distribution
is also {\it conformally invariant}.

Many people, e.g. the late colleagues Roland Dobrushin and 
Claude Itzykson, asked about how to derive from a CFT the description 
of the probabilistic ensemble of loops. 
A strong motivation for works in this direction was provided
by recent spectacular development connected with the SLE processes. 
In this context, a hypothetical picture of the phase boundaries
was outlined in the last chapter in [F] and 
in an earlier presentation [FK]. In these publications,
a description was given, of a probability measure on intervals, 
which connect two phase changing points on the 
boundary of a surface. We will discuss this approach in
section 6.2. We note that a possibility of a connection between
the subjects of the CFT and the SLE was earlier discussed in [BB].
          
\subsection{The Malliavin measures and the CFT}

In this section we describe a new approach to the ensemble of phase 
separating loops based on the Malliavin measures. We  remind some basic 
facts about the CFTs in two dimensions. The basic parameter 
characterising a CFT is a {\it central charge} ${\rm c}\in \R$. Next, 
with any surface $\Sigma$ there is associated a partition 
function $Z_\Sigma\in \dt_\Sigma^{\otimes{\rm c}}\otimes_\R \C$. The usual 
axiomatics of the CFT assumes that the theory is unitary and  
oriented towards the quantum field theory on surfaces with 
Lorentzian metrics. (Recently, there appeared non-unitary versions
of the CFT, with discrete spectrum and logarithmic operator
product expansion (OPE).)
To our knowledge, there is yet no systematic 
approach proposed to CFTs based on the probability theory,
despite the common belief that the simplest unitary CFT with 
${\rm c}=1/2$ must describe the large scale behavior of the
Ising model at the critical temperature. A recent work  
[SW], [W5] indicates that for any value of the central charge 
${\rm c}\in (0,1]$ there should exists a probabilistic CFT which 
has a natural Markov property and gives a random field of disjoint 
Jordan loops describing (hypothetically) a picture of phase boundaries 
in a stochastic particle system. This differs sharply from the
the unitary CFTs, where all theories with ${\rm c}<1$ were classified, 
and only a discrete set of values of ${\rm c}=1-6/(k(k+1))$, 
$k=2,3,\dots$, is allowed.
   
In a probabilistic model of CFT one should expect $Z_\Sigma$ to 
be a positive point of $\dt_\Sigma^{\otimes {\rm c}}$. 
In a sense, $Z_\Sigma$ is a regularised value of 
the partition function for a lattice approximation.

Similarly, in probabilistic CFT models with {\it boundary conditions} 
(in short, boundary CFTs (BCFTs); see [C2]) 
one should have positive points $Z_{\Sigma,\omega}\in 
\dt_\Sigma^{\otimes {\rm c}}$ where $\Sigma$ is a surface
of finite type and $\omega$ is a specified boundary condition. To 
be concrete, let us focus from now on on the continuous limit
of the critical Ising model. In this case, a boundary condition 
$\omega$ (in the microscopic description) 
is a locally constant map assigning values $+$ or $-$ to each 
connected component of the boundary $\partial \Sigma$.
(In the case where $\Sigma$ is a surface without a boundary,
we have a positive point $Z_{\Sigma,{\rm{free}}}\in
\dt_\Sigma^{\otimes {\rm c}}$ and speak about free boundary conditions.)
Consider a loop ${\cL}\in Loop (\Sigma )$, and also attach sign $+$ to one
side of ${\cL}$ in $\Sigma$, and the sign $-$
to the opposite side. We are interested in the 
probability that in the critical Ising model with 
boundary condition $\omega$ there will
be a phase separating loop close to ${\cL}$ (with 
phases near ${\cL}$ specified by our choices of 
signs).

Thus we will talk about `cooriented' loops, i.e., pairs ${\ocL}
=({\cL},\vartheta )$
where $\vartheta$ indicates the $\pm$ signs on both sides of ${\cL}$.
Denote by $\Sigma'(=\Sigma'_{\cL})$ the complement $\Sigma\setminus{\cL}$ 
with canonically attached boundaries so that the canonical conformal 
structure  in the interior of $\Sigma'$ extends smoothly to the 
boundary $\partial \Sigma'$. Given an `initial' boundary condition 
$\omega$ and an attachment $\vartheta$, 
we obtain a boundary condition $\omega'$ for $\Sigma'$.
(In the case of surface $\Sigma$ without a boundary, $\omega'$ 
is reduced to $\vartheta$.) There is 
a canonical isomorphism between the oriented real lines
$$\dt_{{\cL},\Sigma}\simeq \dt_\Sigma/\dt_{\Sigma'}.\eqno (6.1)$$
Hence the ratio 
$$\frac{Z_{\Sigma',\omega'}}{Z_{\Sigma,\omega}}\;\left(\hbox{ or }\;
\frac{Z_{\Sigma',\omega'}}{Z_{\Sigma,{\rm{free}}}}\;\right)
\eqno (6.2)$$ 
can be interpreted (as a function of ${\cL}$) as a section of the line 
bundle $\DT_\Sigma^{-\otimes {\rm c}}$ on $Loop (\Sigma )$. Therefore, 
the product
$$\rho_{\Sigma,\omega}\left(=\rho^{(1)}_{\Sigma,\omega}\right)
={\mbox{\boldmath${\lambda}$}}_{\Sigma}\,\frac{Z_{\Sigma',\omega'}
}{Z_{\Sigma,\omega}}
\left(\hbox{ or }
\rho_{\Sigma,{\rm{free}}}\left(=\rho^{(1)}_{\Sigma,{\rm{free}}}\right)
={\mbox{\boldmath${\lambda}$}}_{\Sigma}\,\frac{Z_{\Sigma',\omega'}
}{Z_{\Sigma,{\rm{free}}}}\right)
\eqno (6.3)$$
is a scalar measure on the space ${\oLoop}\;(\Sigma)$ of cooriented loops
in $\Sigma$ (it is a double covering of $Loop(\Sigma)$). 
Superscript $(1)$ in notation
$\rho^{(1)}_{\Sigma,\omega}$ (or $\rho^{(1)}_{\Sigma,{\rm{free}}}$) 
has a straightforward (and important) 
meaning which we explain below. For simplicity, we will not 
treat the case of a surface without boundary separately;
in this case the reader should substitute the subscript ${\rm{free}}$
in place of $\omega$. 
	 
Our prediction is that measure (6.3) is proportional
to the rate measure (more precisely, the first-order 
rate measure), 
of the random field of phase separating loops. Formally, for 
any Borel subset ${\obU}\subset\oLoop\;(\Sigma)$, the quantity
$$\zeta\int\limits_{Loop (\Sigma )}
{\mathbf 1}_{\obU}\;\rho_{\Sigma,\omega}=
\zeta\;\rho_{\Sigma,\omega}(\obU )\eqno (6.4)$$
gives the expected value of the random number of (equipped) loops falling 
in ${\obU}$. (Here and below, ${\mathbf 1}_{\bW}$ stands for the indicator
function of a subset ${\bW}$, in a given (topological) space). 
The constant $\zeta >0$ standing in front in (6.4) depends on the 
specification of the BCFT
(recall that assignment $\Sigma\mapsto{\mbox{\boldmath${\lambda}$}}_{\Sigma}$
is determined up to a scalar factor).

Formula (6.3) for the rate measure is very natural. Indeed, 
it expresses the infinitesimal probability of
having ${\ocL}=({\cL},\vartheta )$ as a phase-separating curve in the form 
of probability of the event 
specified by the requirement of having spin $+$ on one side 
of ${\cL}$ and spin $-$ on the other side, specified by $\vartheta$. 
The probability
of such an event (in the lattice approximation) is then written as 
the ratio of two sums of Boltzmann weights. The numerator 
is the sum of Boltzmann weights  over configurations with 
boundary conditions $\omega'$, and the denominator is that over 
configurations with boundary condition $\omega$.
   	 
This description can be generalised directly to the case 
of several cooriented 
loops, leading to a sequence of `higher-order' rate measures
$\rho^{(n)}_{\Sigma,\omega}$, $n=1,2,\ldots$.
Here $\rho^{(n)}_{\Sigma,\omega}$ is a scalar measure on 
$\big[{\oLoop}\;(\Sigma )\big]^{\times n}_{\rm{disj}}$,   
the set of ordered $n$-tuples of 
disjoint cooriented equipped loops $({\ocL}_1,\ldots ,{\ocL}_n)$.
$\Big($ $\big[{\oLoop}\;(\Sigma )\big]^{\times n}_{\rm{disj}}$ is an   
open subset in the Cartesian product $[Loop (\Sigma )]^{\times n}=
{\oLoop}\;(\Sigma )\times\ldots\times{\oLoop}\;(\Sigma )$.$\Big)$ 
The meaning of $\rho^{(n)}_{\Sigma,\omega,\omega'}$ is, 
as above, that $\forall$ (Borel) ${\obU}^{(n)}\subseteq
\big[{\oLoop}\;(\Sigma )\big]^{\times n}_{\rm{disj}}$, the quantity
$$\zeta^n\int\limits_{[Loop (\Sigma )]^{\times n}_{\rm{disj}}}
{\mathbf 1}_{{\obU}^{(n)}}\;
\rho^{(n)}_{\Sigma,\omega}=
\zeta^n\;\rho^{(n)}_{\Sigma,\omega}\big({\obU}^{(n)}\big)
\eqno (6.5)$$
gives the expected value of the random number of $n$-tuples 
$({\cL}_1,\ldots ,{\cL}_n)$ falling in ${\obU}^{(n)}$ where
${\obU}^{(n)}$ is a Borel subset in 
$\big[{\oLoop}\;(\Sigma )\big]^{\times n}_{\rm{disj}}$. Measure 
$\rho^{(n)}_{\Sigma,\omega}$ is invariant under the action,
on $\big[{\oLoop}\;(\Sigma )\big]^{\times n}_{\rm{disj}}$, of
the permutation group of the $n$th order. 

The sequence of measures $\rho^{(n)}_{\Sigma,\omega}$ 
would eventually lead to a {\it random point field}
$\urho_{\,\Sigma,\omega}$
on $Loop (\Sigma )$ 
whose sample realisation is a countable 
collection of disjoint Jordan cooriented loops
from ${\oLoop}\; (\Sigma )$, compatible with each other 
and with boundary condition $\omega$ and 
everywhere dense in $\Sigma$. (For brevity, we will 
refer simply to a sample realisation, having
in mind all above-listed properties.)
A way to identify $\urho_{\,\Sigma,\omega}$
is discussed below.
    
A consequence of this proposal is a collection of 
inequalities
on partition functions $Z_{\Sigma ,\omega}$ involving alternate
values of measures $\rho^{(n+k)}_{\Sigma,\omega}$. 
More precisely, consider the class ${\mf}$ of (Borel) subsets 
${\obV}\subset \oLoop (\Sigma )$ such that
the series
$$\sum_{k\geq 1}\zeta^k\rho^{(k)}_{\Sigma,\omega}
\left({\obV}^{\times k}_{\not =}\right)
<\infty.\eqno (6.6)$$
Then, $\forall$ ${\obV}\in{\mf}$ and $n\geq 0$, we will have: 
$$0\leq\pi_{\Sigma ,\omega} ({\obV},n)\leq 1,\eqno (6.7)$$
where
$$\pi_{\Sigma ,\omega} ({\obV},n)
=\frac{1}{n!}\sum_{k\geq 0}(-1)^k\frac{1}{k!}x\zeta^{n+k}
\rho^{(n+k)}\left({\obV}^{\times (n+k)}_{\not =}\right),\eqno (6.8)$$
and, for $n=k=0$, $\rho^{(0)}\left({\obV}^{\times (0)}_{\not =}\right)$
is set to be equal to $1$.

The quantity $\pi_{\Sigma ,\omega} ({\obV},n)$ has a transparent
probabilistic meaning: it  gives the probability that in
the sample realisation, there will be exactly $n$ disjoint cooriented loops
falling in set ${\obV}$. Furthermore, these quantities, for different
${\obV}$ and $n$, will satisfy 
obvious compatibility properties.

We predict that
in the BCFT corresponding to the critical Ising model 
(for ${\rm c}=1/2$), $\forall$ $\zeta >0$, surface $\Sigma$ 
and boundary condition $\omega$, there exists a unique 
probability distribution $\urho_{\,\Sigma ,\omega}$ on
the space of sample realisations compatible with $\omega$, with
the following properties. 

(i) $\forall$ $n\geq 0$ and set ${\obV}\in{\mf}$,
the $\urho_{\,\Sigma ,\omega}$-probability
$$\urho_{\,\Sigma ,\omega}\Big(
\hbox{sample realisation contains exactly $n$ loops ${\cL}\in{\obV}$}\Big)
=\pi_{\Sigma ,\omega} ({\obV},n).\eqno (6.9)$$
(ii) $\forall$ $n\geq 1$ and 
set ${\obU}^{(n)}\subset[{\oLoop}]^{\times n}_{\rm{disj}}$, the 
expected value (relative to $\urho_{\,\Sigma ,\omega}$)
$$\begin{array}{l}
\begin{array}{r}\Eb_{\urho_{\,\Sigma ,\omega}}
\Big(\hbox{the number of ordered $n$-tuples of disjoint cooriented loops,}
\quad{}\\
\hbox{from the sample realisation,
which fall in ${\obU}^{(n)}$}\Big)\end{array}\\
\qquad\qquad =\zeta^n
\rho^{(n)}_{\Sigma ,\omega}({\obU}^{(n)}).\end{array}\eqno (6.10)$$

By construction, $\urho_{\,\Sigma,\omega}$
satisfies the following Markov property. For any cooriented loop 
${\ocL}=({\cL},\vartheta )\in{\oLoop}\;(\Sigma)$, 
the distribution of the sample realisation,
conditional on the fact that it contains ${\ocL}$ is decomposed 
into a product of two marginal distributions, one on the sample realisations
inside ${\cL}$, the other on the sample realisations 
outside ${\cL}$, both collections being compatible with
the boundary condition $\omega'$ on $\Sigma'=\Sigma\setminus{\cL}$
induced by $\omega$ and $\vartheta$ as explained above.

\subsection{On a proposal by Friedrich and the SLE measures}

One can also consider a CFT with boundary conditions which 
change their nature at some points on the boundary $\pSigma$
for a given surface $\Sigma$.
E.g., one can divide $\pSigma$ into a finite number of intervals
and put on each of these intervals boundary condition $+$ or $-$. 
In this case there is no canonical way to define partition function,
and correlators depend on certain insertions at boundary changing 
points. Such insertions form an infinite-dimensional vector space 
$\H_{+-}$. Friedrich's proposal [F] is that there exists a canonical 
vector $\psi \in \H_{+-}$ (unique up to a positive scalar factor) 
which is the highest vector for the natural Virasoro action and 
satisfies the property
$$L_n\psi =0,\;n\geq 1,\;\;
L_0\psi={\rm h}\psi,\;\;(\theta (L_{-1})^2 -L_{-2})\psi=0,
\eqno (6.11)$$
where $\theta$ and ${\rm h}$ are determined by the central charge 
${\rm c}$ via Eqns (4.6). 
      
Vector $\psi$ plays a role of a `vacuum vector' in $\H_{+-}$ and has 
the lowest conformal dimension. The correlators 
$\langle\psi (x_1)\ldots\psi (x_{2n})\rangle$, where $x_1,\ldots ,
x_{2n}\in\pSigma$, are points of change of the boundary condition,
should be positive and equal to the renormalised partition
functions in the lattice approximation.

Suppose we are
given a two-dimensional connected closed smooth manifold 
$S$ with non-empty boundary $S$
and an ordered collection $\ux$\\ $ =(x_1,\ldots x_m)$ of $m$ points in
$\partial S$. Denote by ${\cM}_{S,\ux}$ the space of moduli 
of pairs $(\Sigma,\uy)$ diffeomorphic to $(S,\ux)$, 
where $\Sigma$ is a surface (endowed with a conformal structure), 
and $\uy$ are marked points on the boundary $\pSigma$.
This is a finite-dimensional orbifold. We will assume that 
we are in a hyperbolic case, with $m>2\chi (S)$, where
$\chi (S)$ is the Euler characteristic of $S$.
We define the oriented real line bundles $|T_i|$, $i=1,\ldots ,m$,
and $\DT$ on ${\cM}_{S,\ux}$ as follows. The fiber
of $|T_i|$ at point $[(\Sigma,\uy)]$ (the equivalence class
represented by $(\Sigma,\uy)$) is defined as $|T_{y_i}\pSigma |$.
The fiber of $\DT$ at point $[(\Sigma,\uy)]$ is $\dt_\Sigma$.
The element $(\theta (L_{-1})^2 -L_{-2})$ of the envelopping algebra 
of the Virasoro algebra gives rise to a collection of second-order 
hypolelliptic differential operators $\Delta_i$ on ${\cM}_{S,\ux}$,
$i=1,\ldots ,m$:
$$\begin{array}{l}
\Delta_i:\; \Gamma ({\cM}_{S,\ux},\otimes_{j=1}^m|T_j|^{\otimes
(-{\rm h})}\otimes \DT^{\otimes{\rm c}})\\
\qquad{}\to
\Gamma ({\cM}_{S,\ux},\otimes_{j=1}^m|T_j|^{\otimes
(-{\rm h})}\otimes \DT^{\otimes{\rm c}}\otimes |T_i|^{\otimes (-2)}).
\end{array}\eqno (6.12)$$
This fact can be deduced from the well-known Virasoro uniformisation
of moduli spaces (see [K1]), as explained in [K2], [F] and [FK]. 

It follows from (6.11) that the correlator 
$\langle\psi (y_1)\ldots \psi (y_{2n})\rangle$ (with an even 
number of points $m=2n$)
is a harmonic section
of line bundle $\otimes_{j=1}^{2n}|T_j|^{\otimes
(-{\rm h})}\otimes \DT^{\otimes{\rm c}})$ with respect to each operator
$\Delta_i$. Every $\Delta_i$ gives rise, after division
by $\langle\psi (y_1)\ldots \psi (y_{2n})\rangle$, 
to the generator of a
Brownian motion on ${\cM}_{S,\ux}$, defined modulo a time change.
Friedrich's remark is that, for $n=1$, the random path of
the Brownian motion on ${\cM}_{S,\ux}$, associated with
$\Delta_1$, corresponds to a self-avoiding curve growing in $\Sigma$, from
point $y_1\in\pSigma$, and eventually reaching $y_2\in\pSigma$. 
Such a random interval should correspond to a (random) phase 
boundary. 

Considerations from section 6.2 can be extended in a straightforward
way to incorporate both phase-separating loops and intervals. Namely,
one should replace the partition function $Z_{\Sigma,\omega}$ by
the correlator $\langle\psi (y_1)\ldots \psi (y_{2n})\rangle$ 
and use a joint measure whose marginals are the corresponding 
Malliavin and SLE measures. For the critical 
Ising model this would give a description of a `joint' ensemble
of phase separating lines combining loops and intervals. 

If we focus on intervals only then the corresponding prediction
will give the {\it same} measure as in
Friedrich's proposal. This can be deduced from Girsanov's formula.
In a sense, our construction gives a justification of Friedrich's
proposal, as our approach is physically more transparent.

\subsection{Operadic structure and quadratic identities
 for\\ partition functions}

The conjectured assignments $\Sigma\mapsto 
{\mbox{\boldmath${\lambda}$}}_{\Sigma}$ can be used to 
construct a nice algebraic structure called the 
modular operad [GK]. The prototype is the collection of homology 
groups $H_*(\overline{{\cal M}}_{g,n})$ of moduli stacks of stable 
curves with marked points, where $g,n\geq 0$ and $2-2g-n<0$. There 
exist polylinear operations on these spaces given by pushforward 
maps from the boundary strata of moduli stacks. 

In this section we assume that all surfaces
are oriented. Modifications needed in the non-oriented
case are straightforward. Without stressing it every time again,
we assume that we are given an ${\rm c}$-LCC assignment 
$\Sigma\mapsto 
{\mbox{\boldmath${\lambda}$}}_{\Sigma}$.
   
Let us define (for given $c\in (-\infty,+1]$) an 
infinite-dimensional real vector space
$V_{g,n}$ as the space of measurable sections 
of the line bundle $\dt^{\otimes {\rm c}}$
on the moduli space ${\cM}_{g,n}^{\rm{holes}}$
of puncture-free conformal structures on 
surfaces of genus $g$ with $n$ enumerated holes. 
Here we assume that $g\geq 1$, $n\geq 0$
or $g=0$, $n\geq 2$. Space $V_{g,n}$  contains a convex
cone $V_{g,n}^+$ consisting of non-negative sections.

Our goal is to define certain polylinear maps between
spaces $V_{g,n}$. These maps will be only partially defined
and preserve cones $V_{g,n}^+$. 

Suppose we are given a two-dimensional connected
$C^\infty$-manifold $S$, of finite topological type, 
and a collection of disjoint loops ${L}_1$, $\ldots$,
${L}_k$ in $S$. Let $S_1$, $\ldots$, $S_m$ be the connected 
components of $S\setminus(\sqcup_i{L}_i)$. 
We assume that none of $S_1$, $\ldots$, $S_m$ 
is a sphere or a disk. We associate with these topological
data a map 
$${\ma}_{S,L_1,\ldots ,L_k}:\;\otimes_{i=1}^mV_{g_,n_i}\to V_{g,n}.$$ 
Here $g_i$
and $g$ are the genera of, and
$n_i$ and $n$ the numbers of holes in,
$S_i$ and $S$, respectively. 
Namely, given sections $s_i\in V_{g_i,n_i}$ and a surface
$\Sigma$ representing point 
$[\Sigma]\in{\cM}_{g,n}^{\rm{holes}}$, the value of
the section ${\ma}_{S,L_1,\ldots ,L_k}(s_1\otimes\ldots\otimes s_k)$ 
at the point $[\Sigma ]$ is given by
$$\begin{array}{l}\big({\ma}_{S,L_1,\ldots ,L_k}
(s_1\otimes\ldots\otimes s_k)\big)([\Sigma])\\
=\diy{\int\limits_{{\cU}(S,L_1,\ldots ,L_k)}}
\otimes_{i=1}^ms_i([\Sigma\setminus(\sqcup_{j=1}^k{\cL}_j)])\;
{\rm d}\lambda_\Sigma({\cL}_1)\cdots
{\rm d}\lambda_\Sigma({\cL}_k)\end{array}\eqno (6.13)$$
Here, ${\cU}(S,L_1,\ldots ,L_k)$ consists
of disjoint $k$-tuples of loops 
$({\cL}_1,\ldots ,{\cL}_k)\in [Loop (\Sigma )]^k_{\rm{disj}}$
such that $(\Sigma,{\cL}_1,\ldots ,{\cL}_k)$ is 
homeomorphic to $(S_,L_1,\ldots ,L_k)$. (We use here an 
obvious identification of line bundles.)

In general, convergence of the integral in the RHS
of (6.13) is not guaranteed; 
in the case of non-negative sections $s_1$, $\ldots$, $s_m$,
the integral is finite or equal to $+\infty$.
 
The set of polylinear maps ${\ma}_{S,L_1,\ldots ,L_k}$, for varying
$S,L_1,\ldots ,L_k$, is closed under composition. In particular, 
in the case where $S$ is an open cylinder, $k=1$ and $L_1$ 
is a single-winding loop, we obtain a
bilinear operation $\star$ on $V_{0,2}$. Operation $\star$
is associative, as follows from the composition property.
This operation is also commutative: this follows from
the symmetry of the cylinder under swapping the 
boundary circles with each other.

The conclusion is that we obtain a partially defined
commutative associative product $\star$ on $V_{0,2}$,
depending on the value ${\rm c}$. By using 
the conformal parameter of a cylinder, and the canonical vector
$v_{\Sigma}$ from subsection 2.2.4, space
$V_{0,2}$ can be identified with 
the set of measurable functions on the interval $(0,1)\simeq 
{\cM}_{2,0}^{\rm{holes}}$.  

An easy combinatorial argument shows,  
heuristically, that
our prediction for phase boundaries implies 
the following identity.
Let $Z_{++}=Z_{--}$ and $Z_{+-}=Z_{-+}$ be the elements of 
$V_{0,2}$ corresponding to the partition functions of the 
critical Ising model on a cylinder with boundary conditions 
$+,+$ on both boundary components ($+$ on one 
components and $-$ on another component respectively). Then one has
$$Z_{+-}=Z_{++}\star Z_{--}-Z_{+-}\star Z_{+-}.
\eqno (6.14)$$

The reason is that for a configuration of $\pm$-spins on a cylinder, 
with the boundary condition $+$ on the left and $-$ on the right,
one should always have an {\it odd} number of single-winding
phase-separating loops. Moreover, the number of such loops
with attachment $+|-$ (plus to the left, minus to the right)
equals one plus the number of loops with attachment $-|+$ 
(plus to the right, minus to the left). See the figure below.

%\vskip 5 truemm
%\begin{center}
%\includegraphics[width=.7\textwidth]{Fig22.eps}
%\end{center}
%\vskip 5 truemm

\vskip 5 truemm
\begin{figure}[ht]
\centering
\includegraphics[height=2cm]{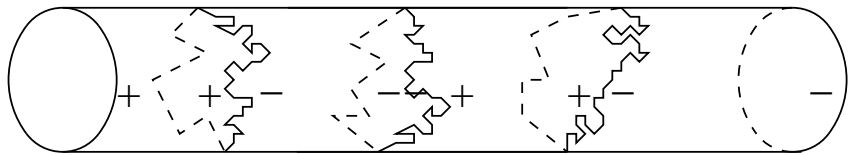}
\caption{Signle-winding phase-deparating loops}
\end{figure}
\vskip 5 truemm

The conjectured random point field on the cylinder
$\Sigma$ with the $+/-$ boundary condition 
should be supported by sample realisations containing finitely many
single-winding loops. The expected number of $+|-$ loops
equals 
$$\diy{\frac{(Z_{++}\star Z_{--})([\Sigma ])}{Z_{+-}([\Sigma ])}}.$$ 
Similarly, the expected number of $-|+$ loops
equals 
$$\diy{\frac{(Z_{+-}\star Z_{+-})([\Sigma ])}{Z_{+-}([\Sigma ])}}.$$ 
The identity (6.14) follows from the aforementioned
relation that the number of the $+|-$ loops is one
more than that of the $-|+$ ones. 
 
In this regard, we state the following problem:
\medskip 

{\sl Fix ${\rm c}\in (-infty,1]$ and consider the 
corresponding product $\star$ on functions on $(0,1)$. It is 
natural to expect that it has the representation
$$(f\star g)(r)=\int_0^1\int_0^1K(r;r_1,r_2)f(r_1)g(r_2)
\frac{{\rm d}r_1}{r_1}\frac{{\rm d}r_2}{r_2}\eqno (6.15)$$
Calculate kernel $K(r;r_1,r_2)$ in a closed form.}
\medskip 

Similar questions may be posed for general compositions
${\ma}_{S,L_1,\ldots ,L_k}$.

Equations of a type similar to (6.14) can be derived in 
the case of a surface $\Sigma$ of a higher genus. This results
in an infinite system of integral equations on partition functions
$Z_{\Sigma,\omega}$ in the critical Ising model. 
\medskip

{\bf Acknowledgements.} M.K. thanks Profs P. Malliavin and W. Werner
and Dr R. Friedrich for numerous fruitful discussions. Y.S. thanks IHES,
Bures-sur-Yvette, for hospitality during visits in 2004--2006;
participation in the IHES Euro-programme 2 is 
particularly acknowledged. Y.S. thanks DIAS, Dublin, for hospitality
during visits in 2004-2005. Y.S. thanks Mathematics Department, UC 
Davis, for hospitality during the Fall Quarter, 2005.

\end{document}